\documentclass[11pt]{article}

\usepackage{amsmath}

\usepackage[cp866]{inputenc}

\newcommand{\N}{N\raise.7ex\hbox{\underline{$\circ $}}$\;$}

\begin{document}

\thispagestyle{empty}

{\em International Conference on

Non-Euclidean Geometry and its  applications.

5 -- 9 July 2010, Kluj-Napoca (Kolozv\'{a}r), ROMANIA}

\vspace{2mm}

\begin{center}
{\bf ON THE NON-ABELIAN MONOPOLES  ON THE  BACKGROUND   OF SPACES WITH
CONSTANT CURVATURE
\\[3mm]
 V.M. Red'kov }
\end{center}

\begin{center}
Institute of Physics,  National Academy of Sciences of Belarus\\
redkov@dragon.bas-net.by
\end{center}

Procedure of constructing  the  BPS solutions  in SO(3) model  on
the background of 4D-space-time with the spatial  part as a model
of constant curvature: Euclid, Riemann, Lobachevsky,  is
reexamined.  It is shown that among possible  solutions
$W^{k}_{\alpha}(x)$ there exist just three  ones which  in a
one-to-one correspondence can be associated with respective
geometries, the known non-singular BPS-solution in the flat
Minkowski space  can be understood as a result of somewhat
artificial combining the Minkowski space model with a possibility
naturally linked up with the Loba\-chevsky geometry. A special
solution $W^{k}_{(triv)\alpha} (x)$ in three spaces is described,
which can be  understood as result of embedding the Abelian
monopole potential into the non-Abelian model.

The problem  of  Dirac  fermion  doublet  in
the~external BPS-monopole potential  in  these curved spaces
is examined on the base of generally covariant tetrad formalism by Tetrode-Weyl-Fock-Ivanenko.
In the frame of  spherical  coordinates, and (Schr\"{o}dinger's) tetrad basis, and special unitary    basis in isotopic space,
 an analog of Schwin\-ger's  one in Abelian case,
 there arises a Schr\"{o}dinger's structure for extended
operator ${\bf J} = {\bf l} + {\bf S} + {\bf T}$. Correspondingly,
instead of monopole harmonics, the  language of conventional Wigner $D$-functions is  used,
radial equations are founds in all three models, and solved in the case of trivial $W^{k}_{(triv)\alpha} (x)$
in Lobachevsky and Riemann models.
 In the  particular case  $W^{k}_{(triv)\alpha} (x)$,
the~doublet-monopole  Hamiltonian
is invariant under additional one-para\-metric group.
This symmetry
results in a freedom in choosing a~discrete operator
$\hat{N}_{A}$  entering the complete set of
quantum variables.

\vspace{3mm}


\section{ Introduction}

While  there  not  exists  at  present  definitive  succeeded
experiments concerning  monopoles, it is nevertheless true  that
there exists a~veritable jungle  of  literature  on  the~monopole
theories. Moreover, properties of more  general
monopoles, associated with large gauge groups now thought to be
relevant in  physics.
As evidenced  even  by   a~cursory examination of some popular surveys,
 the~whole monopole area covers and touches
quite a~variety of fundamental problems. The~most outstanding of them are:
the~electric charge quantization, $P$-violation in purely  electromagnetic
processes, scattering on the~Dirac string, spin from monopole
and spin from isospin, bound states in fermion-monopole system
and violation of the Hermiticity property,
fermion-number breaking in the~presence of a~magnetic monopole and
monopole catalysis of baryon decay.

The~tremendous volume of publications on monopole topics
(and there is no hint  that  its  raise  will  stop)  attests
the~interest which  they  enjoy among theoretical  physicists,  but
the~same token, clearly  indicates  the~unsettled  and  problematical
nature of those objects:  the~puzzle of monopole seems to be
one of  the~still yet unsolved problems of particle  physics.

Many
physicists have contributed to investigation of the~monopole-based theories.
The~wide scope of the~field and the~prodigious number of investigators
associated with various of its developments make it all but hopeless to list
even the~principal contributors. The list of references given  in the end
is not complete and the paper  does not pretend to be
a~survey in this matter, most of references  may be useful to the~readers who wish some supplementary material or
are interested in more technical developments beyond  the~scope of the~present
treatment.

In general, there are several ways of approaching  the monopole problems.
As known, together with geometrically topological  way of exploration into
them,  another  approach   to   studying  such
configurations is possible; namely, that  concerns  any physical
manifestations  of monopoles when they are considered as external  potentials.
Moreover,
from the~physical standpoint, this latter method can  thought  of  as
a~more visualized one in comparison with  less  obvious  and
direct  topological  language;  in the present treatment the  accent  is made  just on this aspect.

 The~basic   frame   of   the present
 investigation  is  the~study of a Dirac particle isotopic doublet in the~external monopole potentials
 on the background of curved spaces, these are  4D-spaces with 3-spatial geometry of constant curvature:
 Euclid $E_{3}$, Riemann  $S_{3}$, and Lobachevsky $H_{3}$.

For convenience of the readers, some remarks about the~approach  used in the~work are to be given.

The technical  and geometrical novelty is that, in the paper, the~tetrad generally
relativistic method of Tetrode -- Weyl -- Fock -- Ivanenko for
describing a~spinor particle will be exploited
(the first publications were
  for
describing a~spinor particle will be exploited are
\cite{1928-Tetrode, 1929-Weyl(1), 1929-Weyl(2), 1929-Weyl(3),
1929-Fock(1), 1929-Fock(2), 1929-Fock(3), 1929-Fock(4), 1929-Fock(5}).
Choosing  this method is  not an~accidental. It is matter that,
the~use of a~special spherical tetrad
in the~theory of a~spin $1/2$ particle had led Schr\"{o}dinger
 \cite{1932-Schrodinger, 1938-Schrodinger(1), 1938-Schrodinger(2)}
 to a~basis of remarkable features.
This Schr\"{o}dinger's basis had been used with great efficiency by Pauli
in his  investigation  \cite{1939-Pauli} on   the~problem of allowed spherically
symmetric wave  functions in quantum mechanics; also see  M\"{o}glich \cite{1938-Moglich} cited in Pauli's paper.
  In particular,
the~following explicit expression for (spin $1/2$ particle's)
angular momentum operator  had been found

\begin{eqnarray}
J_{1}= l_{1} +  i \sigma ^{12}{ \cos \phi \over{ \sin\theta}}
\; , \qquad
J_{2}= l_{2} +  i \sigma ^{12}{ \sin \phi \over{\sin\theta}}
\; , \qquad J_{3} = l_{3} \; ;
\label{i.1}
\end{eqnarray}

\noindent    such a structure for $J_{i}$ typifies this frame in bispinor
space. This Schr\"{o}dinger's basis had been used with great efficiency by Pauli
in his  investigation \cite{1939-Pauli} on   the~problem of allowed spherically
symmetric wave  functions in quantum mechanics.  For our purposes,
just several simple
rules extracted from the~much more comprehensive Pauli's analysis will be
quite   sufficient (those are almost mnemonic working regulations).
They can be explained on the~base of $S=1/2$ particle case.
indeed,  using Weyl's representation of Dirac  matrices where
$\sigma^{12} = {1 \over 2}\;(\sigma_{3} \oplus \sigma_{3})$   and taking into account the~explicit form for
$\vec{J}^{2}, J_{3}$ according to (\ref{i.1}), it is readily verified that the~most
general bispinor functions  with fixed quantum numbers $j,m$ are to be
\begin{eqnarray}
\Phi _{jm}(t,r,\theta ,\phi ) =
                              \left | \begin{array}{l}
          f_{1}(t,r) \; D^{j}_{-m,-1/2} \\
          f_{2}(t,r) \; D^{j}_{-m,+1/2} \\
          f_{3}(t,r) \; D^{j}_{-m,-1/2} \\
          f_{4}(t,r) \; D^{j}_{-m,+1/2}
                               \end{array} \right |
\label{i.2}
\end{eqnarray}

\noindent where $D^{j}_{mm'}(\phi ,\theta ,0)$ designates the Wigner's $D$-functions
\cite{1927-Wigner, 1931-Wigner}
(the~notation
and subsequently required formulas  according to  \cite{1974-Warshalovich-Moskalev-Hersonskii} are adopted).
One should take notice of the low right indices
$-1/2$ and $+1/2$ of $D$-functions  in (\ref{i.2}),  which correlate with the~explicit diagonal structure of
the~matrix $\sigma^{12} = {1 \over 2}\; ( \sigma_{3} \oplus \sigma_{3})$.
The~Pauli criterion allows only half integer values for $j$.

So, one may
remember several very simple  facts of $D$-functions theory and then produce,
almost automatically, proper wave functions. There may exist a~generalized analog of such a~representation
for $J_{i}$-operators, that  might be successfully used whenever
in a~linear problem there exists a~spherical symmetry.

In particular, the~case of
electron in the~external Abelian monopole field, together with
the~ problem of selecting the~allowed wave functions as well as
the~Dirac charge quantization condition \cite{1931-Dirac},  completely come under
that Shr\"{o}dinger-Pauli method. In particular,  components of the generalized
conserved momentum can be expressed as follows

\begin{eqnarray}
j^{eg}_{1} = l_{1} + (i\sigma ^{12} - eg) { \cos \phi
\over { \sin \theta }} \; ,
\nonumber
\\
j^{eg}_{2} = l_{2} + (i\sigma ^{12} - eg) { \sin\phi
\over {\sin \theta }} \; ,  \;
j^{eg}_{3} = l_{3}\; ,
\label{i.3}
\end{eqnarray}

\noindent where $e$ and $g$  are an~electric  and magnetic charges respectively,
respectively. In accordance with the~above rules,
the~corresponding electron-monopole wave functions can be
constructed like in the purely electron pattern (\ref{i.2}) but with a
single change
$
D^{j}_{-m,\pm1/2} \;\; \Longrightarrow \;\;
D^{j}_{-m,eg\pm1/2}.
$
\noindent The~Pauli criterion produces two results: first,
$\mid eg \mid = 0, 1/2, 1, 3/2,\ldots $ (what is called the~Dirac charge
quantization condition; second, the quantum number $j$
 may take the values $\mid eg \mid -1/2 , \mid eg \mid +1/2,
\mid eg \mid +3/2, \ldots$ that  selects the~proper spinor particle-monopole
functions.

So, it seems rather a~natural step yo use  some generalized
Schr\"{o}dinger's basis at analyzing the~problem of  particles
 in the Abelian and non-Abelian monopole fields.

There exists additional reasons justifying the~interest to just
the~aforementioned approach: the~Shr\"{o}dinger's tetrad basis and
Wigner's $D$-functions are deeply  connected with what is called
 the~formalism of spin-weight harmonics:
 Goldberg -- Macfarlane -- Newman -- Rohrlich -- Sudarshan
\cite{1967-Goldberg-Macfarlane-Newman-Rohrlich-Sudarshan},
  developed in
the~frame of the~Newman-Penrose method of light (or isotropic))
tetrad by Newman and Penrose \cite{1962-Newman}; see also Frolov \cite{1977-Frolov},
Alexeev  and  Khlebnikov  \cite{1978-Alexeev}, Penrose and Rindler
\cite{1984-Penrose-Rindler}.
On relationships between  spinor monopole
harmonics  of Wu and Yang \cite{1976-Wu-Yang, 1977-Wu-Yang.} and
 spin-weight  see in:
Dray \cite{1985-Dray, 1986-Dray}, Gal'tsov -- Ershov
\cite{1988-Gal'tsov-Ershov},
also see Krolikowski --  Rzewuski -- Turski
\cite{1976-Krolikowski-Rzewuski, 1986-Krolikowski-Turski, 1986-Turski}.
Also see \cite{1956-Gel'fand-Minlos-Shapiro},
Lochak \cite{1959-Lochak}, Halbwachs -- Hillion -- Vigier
\cite{1959-Halbwachs-Hillion-Vigier(1), 1959-Halbwachs, 1959-Hillion-Vigier.,
1959-Halbwachs-Hillion-Vigier(2)}, Pandres \cite{1965-Pandres}.
The~present work   follows the~notation used in \cite{1988-Red'kov(1)}.

There is still more reason for  special attention just to
the~Scr\"{o}dinger's  basis on the~background of non-Abelian monopole matter.
As will be seen subsequently,  that basis  can be associated  with the~unitary
isotopic gauge in the~non-Abelian monopole problem.

\vspace{3mm}

The main guideline of the present paper  is  as follows.

\vspace{3mm}

In {\bf Sections 2 -- 5 (PART I)},  we examine   constructing   the  BPS
solutions  in $SO(3)$ model  on the background of 4D-space-time with the spatial  part as a model  of constant
curvature: Euclid, Riemann, Lobachevsky.  It is shown that among possible
 solutions $W^{k}_{\alpha}(x)$ (constructed in conformally flat coordinates)
there exist just three  ones which  in a one-to-one
correspondence can be associated with respective geometries,
the known non-singular BPS-solution in the flat
Minkowski space  can be understood as a result of somewhat
artificial combining the Minkowski space model with a
possibility naturally linked up with the Lobachevsky geometry.
Besides,
a special
solution $W^{k}_{(triv)\alpha} (x)$ in three spaces is described,
which can be  understood as result of embedding the Abelian
monopole potential into the non-Abelian model
 (first, such a~specific non-Abelian solution was found out in \cite{1975-Bais-Russel}).

In {\bf Sections 6 -- 10 (PART II)},
we look into the problem of particle in monopole  background.
Firstly we  we consider main points of spin $1/2$  quantum particle
in the presence of the Abelian external field.

In {\bf Section 11 -- 15 (PART III)}  we   consider  a doublet  of Dirac particles in non-Abelian  monopole potentials.
Because the above mentioned   special solution $W^{k}_{(triv)\alpha} (x)$ in three spaces
can be  understood as
result of embedding the Abelian monopole potential into the non-Abelian model
we assume that such a trivial potential is presented in the
well-known monopole solutions by t'Hooft and Polyakov \cite{1974-Polyakov, 1974-t'Hooft, 1975-Julia-Zee}
we  establish explicitly that constituent structure.
The use of the~spherical coordinates and special gauge transformation
enables us to  introduce heuristically
useful concepts of three gauges:  Cartesian, Dirac and Schwinger's; both later are
unitary ones in isotopic space. The use of Schwinger's isotopic gauge enables us
to reduce the non-Abelian doublet-monopole problem to the above Schr\"{o}dinger's type.
 The~Pauli criterion  allows here
all positive integer values for $j : j = 0, 1, 2, 3,\ldots$

As known, an important case in theoretical  investigation  is
the electron-monopole system at the minimal value of  the  quantum
number $j$; so, the~case $j = 0$  should  be  considered  especially
carefully, and we  do  this.  In  the~chosen  frame,  it  is
the~independence on $\theta ,\phi $-variables that  sets  the~wave
functions  of minimal $j$  apart  from  all  other  particle
multiplet states. Correspondingly, the~relevant  angular term in
the~wave equation will be  effectively eliminated.

The~systems of radial equations found by separation of
variables ($4$ and $8$ equations in the cases of $j = 0$ and $j > 0$,
respectively) are simplified by searching a~suitable
operator that can be diagonalized simultaneously with
$ \vec J ^{2} , J_{3}$.
The~usual  space  reflection ($P$-inversion) operator
for a~bispinor doublet field has to be followed by a~certain discrete
transformation in the~isotopic space, so that  a~required quantity
could be constructed.
The problem of discrete symmetry in presence of monopole
has been studied intensively in the literature, but previous results are not  general as much as possible.

As a~result we find out  that there are two
different possibilities depending on what type of external monopole
potential is taken. So, in case of the~non-trivial potential,
the~composite reflection operator with required properties is

\begin{eqnarray}
\hat{N}^{S.} \; = \;  \hat{\pi
} \otimes \hat{P}_{bisp.} \otimes  \hat{P} , \qquad   \hat{\pi } = +
\sigma _{1}
\label{i.7}
\end{eqnarray}

\noindent here, the quantities  $ \hat{\pi }$ and $\hat{P}_{bisp.}$ represent
fixed matrices acting in the isotopic and bispinor space, respectively,
and changing simultaneously with any variations of relevant bases.
 A~totally  different situation occurs in case of the~simplest
monopole potential.  Now, a~possible additional operator, suitable
for separating the~variables, depends on an~arbitrary
real  numerical parameter $A$ (for  some detail in case of complex-valued $A$ see in \cite{2000-Red'kov(2)}):

\begin{eqnarray}
\hat{N}^{S.}_{A} = \hat{\pi }_{A} \otimes \hat{P}_{bisp.} \otimes
\hat{P} , \qquad     \hat{\pi }_{A} = e^{iA \sigma _{3}} \sigma _{1} \; .
\label{1.8a}
\end{eqnarray}

\noindent The same quantity  $A$ appears  also  in  expressions
for the~corresponding  eigenfunction
(the eigenvalues $ N_{A} = \delta  (-1)^{j+1} ; \delta  = \pm  1$):

\begin{eqnarray}
\Psi ^{A}_{\epsilon jm\delta}(x) = T_{+1/2} \otimes  F(x) \; + \;
                       \delta\; e^{iA}\; T_{-1/2} \otimes  G(x)  \; .
\label{1.8b}
\end{eqnarray}

Further
the~fermion doublet just in this simplest monopole
field. In the~first place,  we have constructed
a~remaining operator from a~supposedly  complete set:  $\{\; \hat{H},
\;\vec{J}^{2},\; J_{3}, \hat{N}_{A}, \; \hat{K} = ?  \}$ .  That
$\hat{K} $  is determined as a~natural extension of the~well-known
(Abelian) Dirac operator to the~non-Abelian case. Correspondingly,
the~set of  radial equations is eventually reduced to a~set of two
ones; which can be solved in hyper geometrical functions in all three spaces.
The spectrum of energy in the space $S_{3}$ is discrete.

On simple comparing the~non-Abelian doublet functions
with the~Abelian ones, we arrive at an~explicit factorization of
the~doublet functions by Abelian ones and isotopic basis vectors.
The~relevant decompositions have been found  for
the~composite states with all values of $j$, including the~minimal
one $j_{min.}=0$ too.

We are especially interested in the question: where does
the above ambiguity come from? It is quite easily understandable that
this possibility is closely connected with the fact of decoupling of
two isotopic components in the wave equation.
The situation can be formulated in terms of an additional hidden
symmetry: there are two operators, $t_{3}$ and
$\hat{N}_{A}$, commuting with the Hamiltonian but not commuting with
each other.
In formal mathematical  terms,  the~origin of the above freedom
in discrete symmetry operations lies in the~existence of
an~additional (one parametric) operation $U(A)$ that leaves
the~doublet-monopole Hamiltonian invariant. Just this operation
$U(A)$ changes $\hat{N}_{A=0}$ into $\hat{N}_{A}$.  Different values
for $A$ lead to the~same whole functional space; each fixed $A$
governs only the basis states $\Psi^{A} (x)$ of
it, and the~symmetry operation acts transitively on those states:
$\Psi ^{A}(x) = U(A) \Psi ^{A'=0}(x)$.

Additionally,  we
 draw an analogy between this isotopic symmetry and more familiar
chiral symmetry transformation  ($\gamma ^{5}$ symmetry in
massless Dirac field theory \cite{1980-Berestetzkiy-Lifshitz-Pitaevskiy}).
The~role of the~Abelian $\gamma^{5}$-matrix is
taken by the~isotopic $\sigma_{3}$-matrix: its form in the Schwinger's isotopic gauge is
 $U^{S.}(A) = \mbox{exp}\; (A/2) \; \mbox{exp} \; (i {A \over 2} \sigma_{3})$.

Also some technical details touching the discrete operation
$\hat{N}_{A}$  are given; in  particular,
the  form of that  transformation in  the
Cartesian isotopic gauge is  calculated:

\begin{eqnarray}
U_{C.}(A) = e^{+iA/2}
\exp \;  (\; -i\; {A\over 2} \;\vec{\sigma }\; \vec{n}_{\theta ,\phi }\; ) \; ;
\label{i.10a}
\end{eqnarray}

\noindent correspondingly, the discrete operator looks

\begin{eqnarray}
\hat{N}^{C.}_{A} \; =\; (-i) \; \mbox{exp}  (\; - i \; A \; \vec{\sigma }\;
 \vec{n}_{\theta ,\phi }  )
\otimes \hat{P}_{bisp.} \otimes \hat{P}  \; .
\label{1.10b}
\end{eqnarray}

\noindent  The~explicit coordinate dependence  in  Cartesian gauge
results from the~non-commutation $\sigma_{3}$ with a~gauge transformation
involved into transition from Shwinger's to Cartesian isotopic basis.
In the~analogous Abelian situation, the form of the~chiral transformation
remains the~same because $\gamma^{5}$ and the~relevant gauge matrix (that
belongs to the~bispinor local representation of the group $SL(2.C)$) are
commutative with each other.

 It  may be  stressed   that
these symmetry operations occur only in the case of special
monopole potential; instead, for the 't Hooft-Polyakov
potential as well as for the free isotopic doublet case no such
additional symmetry occur.

\newpage

\begin{center}

{\bf PART I}
\end{center}

\section{  BPS-monopole, radial equations}

In the literature, a $SU(2)$-monopole problem in the limit of
Bogomolny -- Prasad -- Sommerfield for Minkowski flat and  curved space-time backgrounds
has attracted great  interest

Polyakov \cite{1974-Polyakov},
t'Hooft \cite{1974-t'Hooft},
Julia and Zee \cite{1975-Julia-Zee},
Prasad and Sommerfield \cite{1975-Prasad-Sommerfield},
Bais and Russel \cite{1975-Bais-Russel},
Wang \cite{1975-Wang},
Nieuwenhuizen et al \cite{1976-Nieuwenhuizen-Wilkinson-Perry},
Benguria et ak\l \cite{1977-Benguria-Cordero-Teitelboim},
Witten \cite{1976-Witten},
Ray \cite{1978-Ray},
Goddard and Olive \cite{1978-Goddard-Olive},
Cervero and Jacobs \cite{1978-Cervero-Jacobs},
Boutaleb et al \cite{1978-Boutaleb-Joutei--Chakrabarti--Comtet},
Actor \cite{1979-Actor},
Harnad et al \cite{1979-Harnad-Shnider-Vinet},
\cite{1979-Harnad-Tafel-Shnider},
Maison \cite{1981-Maison},
Clement \cite{1981-Clement},
Gu \cite{1981-Gu},
Schigolev \cite{1982-Schigolev, 1984-Schigolev},
Kamata \cite{1982-Kamata},
Henneaux \cite{1982-Henneaux},
Hitchin \cite{1982-Hitchin},
Kasuya \cite{1982-Kasuya},
Hitchin \cite{1983-Hitchin},
Melnikov and Shigolev \cite{1984-Melnikov-Shigolev},
Comtet et al \cite{1984-Comtet-Forgacs-Horvathy},
Deser \cite{1984-Deser},
Atiyah \cite{1984-Atiyah(2)},
Chakrabarti \cite{1985-Chakrabarti},
Nash \cite{1986-Nash},
Gibbons and Manton \cite{1986-Gibbons-Manton},
Chakrabarti \cite{1987-Chakrabarti},
Atiyah and Hitchin \cite{1988-Atiyah-Hitchin},
Garland and Murray \cite{1989-Garland-Murray},
Pajput and Rashmi \cite{1989-Pajput-Rashmi},
Ershov and Gal'tsov \cite{1990-Ershov-Gal'tsov},
Yaffe \cite{1990-Yaffe},
Yang \cite{1990-Yang},
Bartnik \cite{1990-Bartnik},
Austin and Braam {1990-Austin-Braam},
Pedersen and Tod \cite{1991-Pedersen-Tod},
Ortiz \cite{1992-Ortiz},
Balakrishna and Wali \cite{1992-Balakrishna-Wali},
Breitenlohner er al \cite{1994-Breitenlohner-Forgacs-Maison},
Hitchin et al \cite{1995-Hitchin-Manton-Murray},
Volkov \cite{1996-Volkov},
Jarvis and Norbury \cite{1997-Jarvis-Norbury(1), 1997-Jarvis-Norbury(2)},
Kraan and van Baal \cite{1998-Kraan-van Baal},
Kimyeong Lee and Changhai Lu \cite{1998-Kimyeong Lee and Changhai Lu},
Houghton et al \cite{1999-Houghton-Manton-Romao},
Volkov and Gal'tsov \cite{1999-Volkov-Gal'tsov},
Norbury et al \cite{2001-Norbury, 2003-Murray-Norbury-Singer, 2004-Norbury},
Meng \cite{2004-Meng},
Landweber \cite{2005-Landweber},
Gibbons and Warnick \cite{2006-Gibbons-Warnick},
Weinberg and Yi \cite{2007-Weinberg-Yi},
Harland \cite{2009-Harland}.

In a space-time with a metrics tensor $g_{\alpha \beta}(x)$ let us
consider the Yang - Mills - Higgs system. Lagrangian of that
system is given by
\begin{eqnarray}
L  = {1 \over 2}  g^{\alpha \beta}(x)  D_{\alpha}
\Phi^{a}  D _{\beta} \Phi ^{a}
- {1\over 4}\; g^{\alpha \rho}(x) g^{\beta\sigma}(x) F^{a}_{\alpha
\beta} F^{a}_{\rho \sigma}
-  {\lambda \over 4}  (\Phi^{2} - V^{2})^{2} \; .
\nonumber
\end{eqnarray}

\noindent
Three scalar fields $\Phi^{a}(x)\;$ are supposed to be real; correspondingly,
the Lagrangian is invariant under local  $SO(3.R)$ group transformations.
The  operator $D_{\alpha}$ is
$$
   D_{\alpha} \Phi^{a} =
\partial_{\alpha} \Phi^{a} \;  + \; e \; \epsilon _{abc}\; W_{\alpha}^{b} \;
\Phi^{c} \;  .
$$
The
 $W_{\alpha}^{b}$ stands for the Yang-Mills isotriplet.
Antisymmetric  generally covariant strength tensor is given by
$$
F^{a}_{\alpha \beta} =  \partial_{\alpha} W^{a}_{\beta} \; - \;
\partial_{\beta} W^{a}_{\alpha} \; + \; e \; \epsilon _{abc} \;
W_{\alpha}^{b}  \;   W_{\beta}^{c} \; .
$$
In accordance with the variational principle
one can derive equations
\begin{eqnarray}
{1 \over \sqrt{-g}} \; \partial _{\alpha} \; \sqrt{-g} \;
D^{\alpha}\; \Phi^{a}\; + \;  e \; \epsilon_{abc} \;
W_{\alpha}^{b} D^{\alpha} \Phi^{c} =  -  \lambda ( \Phi^{2} \;-\;
V^{2}) \; \Phi^{a} \; ,
\nonumber
\\
{1 \over \sqrt{-g}} \; \partial _{\alpha} \; \sqrt{-g}\; F^{\alpha
\beta}_{a} \; + \; e \; \epsilon_{abc}\; W^{b}_{\alpha} \;
F_{c}^{\alpha \beta}  = - e \; \epsilon _{abc} \; \Phi^{b} \;
D^{\beta} \; \Phi^{c} \; .
\label{1.5b}
\end{eqnarray}

In the following, all analysis will be done  for three (curved) space models:
 Euclid's -- $E_{3}$, Riemann's -- $S_{3}$, and Lobachevsky's -- $H_{3}$;
 conformally flat coordinates  will be used (we employ dimensionless variables  $x^{\alpha} / \rho \Longrightarrow
 x^{\alpha}$, where $\rho$ is a curvature radius):
\begin{eqnarray}
dS^{2} =  (dx^{0})^{2} -
 {(dx^{1})^{2}  +   (dx^{2})^{2}   +   (dx^{3})^{2})   \over \Sigma^{2} }\; .
\label{1.6}
\end{eqnarray}

\noindent
To $E_{3}$-model there corresponds  $\Sigma  = 1$, to
$$
S_{3} - \Sigma = 1 + r^{2} /4 , \qquad H_{3} -
\Sigma  = 1 - r^{2} /4 \; ,  \qquad r^{2} =
(x^{1})^{2} +  (x^{2})^{2}+  (x^{3})^{2}.
$$
Starting with the  the known dyon substitution
\begin{eqnarray}
\Phi^{a}(x) = x^{a} \; \Phi (r) \; , \;\; W^{a}_{0}(x) = x^{a} \;
f(r) \; , \;\;  W^{a}_{i}(x) = \epsilon _{iab} \; x^{b} \; K(r)\;
, \label{1.7}
\end{eqnarray}

\noindent after simple calculation, we  get the  radial equations for
$\Phi ,  f, K $  --  below
only the situation in  absence  of self-interactions
between components of scalar triplet will be examined (the Bogomolny-Prasad-Sommerfield
limit)
\begin{eqnarray}
\Phi''\; + \; {4\over r} \; \Phi' \;-\; 2 e \Phi \; (2 + e r^{2}K)
\; K \;-\;  {\Sigma ' \over  \Sigma} \; ( \Phi ' + {\Phi \over
r})\; =   0 \; ,
\nonumber
\\
f''\; + \; {4\over r} \; f' \;-\; 2 e f \; (2 + e r^{2} K) \; K
\;-\;  {\Sigma ' \over  \Sigma} \; ( f ' + {f \over r})\; =   0 \; ,
\nonumber
\\
K'' + {4 K' \over r} + e\; {(f^{2} - \Phi^{2}) \; (1 + e r^{2}K)
\over \Sigma ^{2} }
\nonumber
\\
- \; e K^{2} \; (3 + e r^{2}K) + {\Sigma'
\over \Sigma}\; (K' + {2 K \over r}) = 0 \; .
\label{1.14}
\end{eqnarray}

\section{ Solutions in flat space}

Now let us turn to    eqs. (\ref{1.14}) specified for the flat Minkowski space (in next Sections we will extend the solving procedure to $H_{3}$ and $S_{3}$ models).
As $\Sigma = 1$ eqs. (\ref{1.14}) take  the form
\begin{eqnarray}
\Phi''\; + \; {4\over r} \; \Phi' \;-\; 2 e \Phi \; (2 + e r^{2}K)
\; K  =  0 \; ,
\nonumber
\\
f''\; + \; {4\over r} \; f' \;-\; 2 e f \; (2 + e r^{2} K) \; K =
0 \; ,
\nonumber
\\
K'' + {4 K' \over r} + e\; (f^{2} - \Phi^{2}) \; (1 + e r^{2}K) \;
- \; e K^{2} \; (3 + e r^{2}K)  = 0 \; .
\label{2.1'}
\end{eqnarray}

\noindent It is known that  the dyon system (\ref{1.7}) can be solved on the base of
solution  for a purely monopole system:
\begin{eqnarray}
\Phi^{a}(x) = x^{a} \; \Phi (r) \; , \;\; W^{a}_{0}(x) = 0 \; ,
\;\;
 W^{a}_{i}(x) = \epsilon _{iab} \; x^{b} \; K(r)\; ,
\label{2.2a}
\end{eqnarray}

\noindent  when  radial equations are
\begin{eqnarray}
\Phi''\; + \; {4\over r} \; \Phi' \;-\; 2 e \Phi \; (2 + e r^{2}K)
\; K  =  0 \; ,
\nonumber
\\
K'' + {4 K' \over r} -  e \; \Phi^{2} \; (1 + e r^{2}K) \; - \; e
K^{2} \; (3 + e r^{2}K)  = 0 \; .
\label{2.2b}
\end{eqnarray}

\noindent  Indeed, turning  to eqs.  (\ref{2.1'}) and setting
 $f = c \; \Phi$ , where $ c $ is a constant, one  comes to
\begin{eqnarray}
f = c \; \Phi \; , \;\; {d^{2} \over dr^{2}} \Phi \; + \; {4 \over
r} \; {d \over dr} \Phi\; - \; 2e \Phi \; (2 + e r^{2} K) \; K = 0
\; ,
\nonumber
\\
{1 \over 1 -c^{2}} \; ( {d^{2} \over dr^{2}} K \; + \; {4 \over r}
\; {d \over dr} K ) \; - e \Phi^{2} \; (1 + e r^{2} K)  \; - \; {e
K^{2} \over 1 - c^{2} } \; (3 + e r^{2} K ) = 0 \; .
\nonumber
\end{eqnarray}

\noindent   From these,  having introduced a new radial variable and  a new function  $\tilde{K}$:
\begin{eqnarray}
r \; \rightarrow \; (1-c^{2})^{1/4} \; r = \tilde{r} \; , \qquad
{K(r) \over \sqrt{1 -c^{2}}}  = \tilde{K}((1-c^{2})^{1/4}r) \; ,
\nonumber
\end{eqnarray}

\noindent one obtains a system of the above type (\ref{2.2b}).
Therefore the dyon functions have been reduced  to monopole ones:
\begin{eqnarray}
\Phi(r) = \tilde{\Phi}((1-c^{2})^{1/4}\;r) \; , \;\; f(r) = c \;
\Phi (r) \; ,
\nonumber
\\
K(r) = \sqrt{1 -c^{2}} \; \tilde{K}((1-c^{2})^{1/4}\; r)    \; .
\label{2.3}
\end{eqnarray}

Bearing  this in mind,  we will examine only the purely
monopole equations  (\ref{2.2b}).
For further work instead of  $\Phi (r)$  and  $K(r)$ in  (\ref{2.2b}) it is
convenient to use new functions $f_{1}$ and $f_{2}$:
\begin{eqnarray}
1 \; +\;  e\; r^{2}\; K  = r\; f_{1} (r) \; ,
\qquad   1 \;+ \; e \;
r^{2}\; \Phi  = r \; f_{2} (r) \; ;
\label{2.4}
\end{eqnarray}

\noindent correspondingly eqs.  (\ref{2.2b})   transform into
\begin{eqnarray}
2 \; (\; f'_{2} \; + \; f_{1}^{2} \; ) \; + \; (\; f_{2}'' \; - \;
 2\; f_{1}^{2}\; f_{2} \; ) = 0 \; ,
\nonumber
\\
2\; (\; f_{1}' \; + \; f_{1} \; f_{2} \; ) \; + r \; (\; f_{1}''
\; - \; f_{1}\;f_{2}^{2} \; - \; f_{1}^{3} \; )  \; .
\label{2.5}
\end{eqnarray}

\noindent One can solve these equations by satisfying  four eqautions
\begin{eqnarray}
f'_{2} \; + \; f_{1}^{2}  = 0 \; , \qquad   f_{2}'' \; - \; 2\;
f_{1}^{2}\; f_{2} = 0 \; ,
\nonumber
\\
 f_{1}' \; + \; f_{1} \; f_{2}  =0 \; , \qquad   f_{1}'' \; -
\; f_{1}\;f_{2}^{2} \; - \; f_{1}^{3}  = 0 \; .
\label{2.6a}
\end{eqnarray}

\noindent
Second and fourth equations  are consequences
of the  first and third, so  we have only two  independent ones
\begin{eqnarray}
f'_{1} = -f_{1} \; f_{2}  , \;\; f'_{2} = - f_{1}^{2}  , \quad \mbox{or}\quad
 f_{2} = - {f_{1}' \over f_{1}}  , \;\;
\left ( {f_{1}' \over f_{1}} \right )' = f_{1}^{2} \; ;
\label{2.6b}
\end{eqnarray}

\noindent the task  reduces to a single differential equation
\begin{eqnarray}
\left ( {f_{1}' \over f_{1}} \right )' = f_{1}^{2} \; .
\label{2.6c}
\end{eqnarray}

\noindent  From whence one gets
\begin{eqnarray}
(\ln f_{1})'' = f_{1}^{2},
\qquad {d \over dr }\; \left [\; (\ln f_{1})'\; \right ]^{2} = {d \over
dr}\; f_{1}^{2}\; .
\nonumber
\end{eqnarray}

\noindent
From this it follows
\begin{eqnarray}
 \int {d\; f_{1} \over f_{1} \sqrt{c +
f_{1}^{2}}} = \pm \; (r \; + \; \mbox{const} )   \; .
\nonumber
\end{eqnarray}

\noindent  Depending on the sign of the constant $c$
we have three types of solutions:
\begin{eqnarray}
c = 0 \; , \qquad   \qquad  \qquad \qquad f_{1} = \pm
\; {A \over Ar +B } \;\; , \;\; f_{2} = {A \over Ar + B} \; ;
\nonumber
\\
c < 0 \; ,  \qquad  \qquad f_{1} = \pm \; {A \over
\mbox{sh}\; (A r + B) } \;\; ,\;\; f_{2} = {A  \over \mbox{tanh}\;
(A r + B)} \; ;
\nonumber
\\
c > 0 \; ,  \qquad  \qquad  f_{1} = \pm \; {A \over
\sin \;(A r + B) } \;\; ,\;\; f_{2} = {A  \over \mbox{tan}\; (A r +
B)} \; ;
\nonumber
\\
\label{2.7}
\end{eqnarray}

\noindent  where   $A$  and   $B$  are arbitrary  constants.
Turning  back to  (\ref{2.4}),  we get
\begin{eqnarray}
 K(r) = {1 \over e\; r^{2} } \; (\;  r\; f_{1}  -1 \; ) \; , \qquad
  \Phi (r) = {1 \over e\; r^{2} } \; ( \;  r \; f_{2}  - 1) \; ;
\label{2.8a}
\end{eqnarray}

\noindent in  usual unites,
 $A$  is measured in $(\mbox{meter})^{-1}$, and  $B$ is dimensionless.
Thus,  we arrive at six different solutions:
\begin{eqnarray}
K^{\pm}_{1} = {1 \over e\; r^{2} } \; [ \;  { \pm \;Ar \over Ar+B}
-1\;  ]\; ,\qquad \Phi_{1}(r) = {1 \over e\; r^{2} } \; [ \;  { A
r\over Ar+B} -1 \; ]\; ,
\nonumber
\\
K^{\pm}_{2} = {1 \over e\; r^{2} } \; [ \;  { \pm \;Ar \over
\mbox{sh}\; (Ar+B)} -1 \; ]\; ,\qquad \Phi_{2}(r) = {1 \over e\;
r^{2} } \; [  \;  { Ar \over  \mbox{tanh} \;(Ar+B)} -1 \;  ]\; ,
\nonumber
\\
K^{\pm}_{3} = {1 \over e\; r^{2} } \; [ \;  { \pm \;Ar \over
\mbox{sin}\; (Ar+B)} -1 \; ]\; ,\qquad \Phi_{3}(r) = {1 \over e\;
r^{2} } \; [  \;  { Ar \over  \mbox{tan} \;(Ar+B)} -1 \;  ]\; .
\nonumber
\\
\label{2.8b}
\end{eqnarray}

Here it should be noted that in going from (\ref{2.6b}) to
(\ref{2.6c}) we have missed one simple solution
(which is to be interpreted as Abelian Dirac's nonopole being
placed into background of the non-Abelian  theory)
\begin{eqnarray}
f_{1}(r) = 0    \;\; , \;\; f_{2}(r)  = C \; , \qquad \mbox{or}
\nonumber
\\
K = - \;  {1 \over er^{2}} \; , \qquad \Phi (r)  = {1 \over
er^{2}} \; (  C \; r   -1 \; ) \; .
\label{2.9b}
\end{eqnarray}

\noindent It should be  noted that  if  $ f_{1}= e r^{2} K + 1
=0 $, the initial equations  (\ref{2.2b})
 become just one linear and  other nonlinear equations:
\begin{eqnarray}
\Phi''\; + \; {4\over r} \; \Phi' \;+\; 2  \Phi \;
 \; {1 \over r^{2}}   =  0 \; ,
\qquad
K'' + {4 K' \over r} - \;2 e
 K^{2}   = 0 \; .
\nonumber
\end{eqnarray}

\noindent
The nonlinear one is satisfied by the function
 $K = -1 /er^{2}$; whereas a general solution  $\Phi (r)$  is a  linear combination
\begin{eqnarray}
\Phi =  {c_{1} \over  r}  + {c_{2} \over  r^{2}} \; .
\label{2.11a}
\end{eqnarray}

\section{Some technical details for curved models}

In curved  models $H_{3}$ and $S_{3}$, analogously to the flat space $E_{3}$,
there exists possibility to construct dyon functions in terms of purely
monopole's ones  (all details are omitted).  By this reason, further we will examine  only  the  purely monopole  case:
\begin{eqnarray}
\Phi''\; + \; {4\over r} \; \Phi' \;-\; 2 e \Phi \; (2 + e r^{2}K)
\; K \;-\;  {\Sigma ' \over  \Sigma} \; ( \Phi ' + {\Phi \over
r})\; =   0 \; ,
\label{3.1a}
\nonumber
\\
K'' + {4 K' \over r} - e  \Phi^{2}  {(1 + e r^{2}K) \over
\Sigma ^{2} }  -
 e K^{2}  (3 + e r^{2}K) + {\Sigma' \over
\Sigma}  (K' + {2 K \over r}) = 0 \; .
\nonumber
\\
\label{3.1b}
\end{eqnarray}

\noindent Instead of
$K(r)$  and  $\Phi (r)$     let us  introduce $A(r)$ and $B(r)$:
\begin{eqnarray}
 1 + e\; r^{2} K  = A(r) \;\; , \qquad
 e \; r^{2} \Phi   = B(r) \;\; , \;\;
\label{3.2a}
\end{eqnarray}

\noindent    then  eqs. (\ref{3.1b}) transform into
\begin{eqnarray}
B''\; - \;{2\; B\;A^{2} \over r^{2}} \;+ \;{\Sigma' \over \Sigma}
\; ({B \over r } \;- \; B') = 0 \; ,
\label{3.2b}
\\
A'' \; - \; { A \; B^{2} \over  r^{2} \Sigma^{2} } \; + \; {A(1
-A^{2} )\over r^{2}}\; + {\Sigma ' \over \Sigma } \; A' = 0 \; .
\label{3.2c}
\end{eqnarray}

\noindent  For $A(r)$ and $B(r)$ let us use  substitutions
\begin{eqnarray}
A = c \; f_{1}(R) \; \; , \;\; B = a \; f_{2}(R) \;+\; b  \; ;
\label{3.3a}
\end{eqnarray}

\noindent
where $a(r),  b(r),  c(r), R(r)  $ stand for  some yet unknown functions
of  $r$, whereas   $f_{1}(R)$ and  $f_{2}(R)$ are assumed
to obey two relationships (see (\ref{2.6b}))
\begin{eqnarray}
{d \over dR }\; f_{1} = - f_{1} \; f_{2} \;\; , \qquad  {d \over
dR }\; f_{2} = - f_{1}^{2} \; ,
\nonumber
\end{eqnarray}
\noindent so that  $f_{1}, \;f_{2}$  coincide with those listed in (\ref{2.6b}).
Initial  functions look as follows:
\begin{eqnarray}
K (r) = {1 \over er^{2}} \;  \left [ \;  c(r) f_{1}(R) -1 \;
\right ]   , \quad
\Phi  (r) = {1 \over er^{2}} \; \left  [ \;
\; a(r) f_{2} (R) + b(r) \; \right ] \; ;
\label{3.3b}
\end{eqnarray}

\noindent limiting transition to the case of the flat space should be
\begin{eqnarray}
c(r) \Longrightarrow r\;, \qquad a(r) \Longrightarrow r \;, \qquad
b(r) \Longrightarrow -1\; , \qquad  R(r) \Longrightarrow r \; .
\label{3.3c}
\end{eqnarray}

Substituting (\ref{3.3a}) into (\ref{3.2b})  we arrive at
\begin{eqnarray}
a'' \; f_{2} \;-\; (2 a' R' + a R'') \;f_{1}^{2} \;
\nonumber
\\
+ \; 2a\;
(R')^{2} \; f_{1}^{2} \; f_{2} \; + \; b'' - {2 \over r^{2}}\; (a
f_{2} + b )\; c^{2} f_{1}^{2} \:
\nonumber
\\
+ \; {\Sigma' \over \Sigma }\; [ \; {  a f_{2} + b \over r}
\; - \; (a' f_{2} - a R'\; f_{1}^{2} + b' ) \; ]  = 0 \; .
\label{3.4}
\end{eqnarray}

\noindent  Setting  factors at  $1, \; f_{2},
\;f_{1}^{2},\;f_{1}^{2}\;f_{2}$  equal to zero, we get four equations:
\begin{eqnarray}
1\; :   \qquad \qquad \qquad b'' + {\Sigma' \over \Sigma}( {b\over
r} - b') = 0\; ,
\nonumber
\\
f_{2} \; :  \qquad  \qquad  \qquad a'' + {\Sigma' \over \Sigma}(
{a\over r} - a') = 0\; ,
\nonumber
\\
f_{1}^{2} \; :  \qquad -2 a'\;R' - a \; R'' -
{2bc^{2} \over r^{2}} +{\Sigma' \over \Sigma}\; a R' = 0 \; ,
\nonumber
\\
f_{1}^{2}f_{2} \; :    \qquad \qquad  \qquad 2a\; (R')^{2} - {2 a
c^{2} \over  r^{2} }    = 0 \; .
\label{3.5}
\end{eqnarray}

\noindent Analogously, substituting  (\ref{3.3a}) into  (\ref{3.2c}),  we get
\begin{eqnarray}
c'' f_{1} \;- \;( 2c' \;R' + c \; R'') \;f_{1}\; f_{2} \; + \;
 c\; (R')^{2} \;f_{1}^{3} \; + \; c\;(R')^{2} \; f_{1} \; f_{2}^{2} \;
\nonumber
\\
- \;{c \over r^{2}} \;f_{1} \; (a^{2} f_{2}^{2} \;+\; 2 a b \;
f_{2}+b^{2})\; {1 \over \Sigma ^{2}} \; + {c f_{1} \over r^{2}} \;
(1 - c^{2}\;f_{1}^{2})\;
\nonumber
\\
+ \; {\Sigma '  \over \Sigma } \; (c'
f_{1} - cR'\; f_{1}\; f_{2}) = 0 \; ,
\label{3.6}
\end{eqnarray}

\noindent from  where it follow four equations:
\begin{eqnarray}
f_{1} \; : \qquad \qquad \qquad c'' - {c b^{2} \over r^{2}
\Sigma^{2}} + {c \over r^{2}} + {\Sigma ' \over \Sigma} \; c' = 0
\; ,
\nonumber
\\
f_{1}f_{2} \; : \qquad \qquad \qquad -2 c'\; R' - c
\; R'' - { 2abc \over r^{2} \Sigma^{2}}  -{\Sigma' \over \Sigma
}\; c R' = 0 \; ,
\nonumber
\\
f_{1}^{3} \; : \qquad \qquad \qquad \qquad c\; (R')^{2} - {c^{3}
\over r^{2} } = 0 \; ,
\nonumber
\\
 f_{1} f_{2}^{2} \; :     \qquad   \qquad  \qquad \qquad c (R')^{2} - {c a^{2} \over
r^{2} \Sigma^{2}} = 0 \; .
\label{3.7}
\end{eqnarray}

\noindent Collecting eqs. (\ref{3.5}) and  (\ref{3.7}) together, we get the system
\begin{eqnarray}
(R')^{2} = {c^{2} \over r^{2}} \; , \;\; (R')^{2} = {a^{2} \over
r^{2} \Sigma^{2}} \; ,
\label{3.8a}
\\
a'' + {\Sigma' \over \Sigma }\;( { a \over r } - a' )  = 0 \; , \;
b'' + {\Sigma' \over \Sigma }\;( { b \over r } - b' )  = 0 \; ,
\label{3.8b}
\\
-2 a'\;R' - a \; R'' - {2bc^{2} \over r^{2}} +{\Sigma' \over
\Sigma}\; a R' = 0 \; ,
\label{3.8c}
\\
-2 c'\; R' - c \; R'' - { 2abc \over r^{2} \Sigma^{2}}  -{\Sigma'
\over \Sigma }\; c R' = 0 \; .
\label{3.8e}
\end{eqnarray}

\noindent   It is readily seen that in the system,  eq. (\ref{3.8c})
can be  derived from others. Indeed,  let us  multiply
eq. (\ref{3.8e}) by  $c$,  then
\begin{eqnarray}
-2 c c' R' - c^{2} R'' - {2abc^{2} \over r^{2} \Sigma^{2}} -
{\Sigma' \over \Sigma}\; c^{2} R' = 0 \; ;
\nonumber
\end{eqnarray}

\noindent in turn, from eqs.  (\ref{3.8a})  it  follows
\begin{eqnarray}
c^{2} = {a^{2} \over \Sigma^{2}} \; , \qquad \Longrightarrow
\qquad cc' = {a a' \over \Sigma ^{2}} - {a^{2} \over \Sigma^{2}}
\; {\Sigma'\over \Sigma} \; .
\nonumber
\end{eqnarray}

\noindent Therefore, previous relation can be transformed to the form
\begin{eqnarray}
-2R'\;{a \over \Sigma^{2}} \; (a' - a \; {\Sigma' \over \Sigma}) -
{a^{2} \over \Sigma^{2}} \; R'' - {2abc^{2} \over r^{2}\Sigma^{2}}
- {\Sigma' \over \Sigma} \; {a^{2} \over \Sigma^{2}}\;  R' = 0 \;
.
\nonumber
\end{eqnarray}

\noindent From the latter it follows
\begin{eqnarray}
-2R'\; (a' - a \; {\Sigma' \over \Sigma}) - a\; R'' - {2bc^{2}
\over r^{2}} - {\Sigma' a  \over \Sigma} \;   R' = 0 \; .
\nonumber
\end{eqnarray}

\noindent  which coincides with  (\ref{3.8c}).
Therefore, independent equations are
\begin{eqnarray}
(R')^{2} = {c^{2} \over r^{2}} \;\; , \qquad  c^{2} = {a^{2} \over
\Sigma^{2} } \; ,
\label{3.9a}
\\
a'' + {\Sigma' \over \Sigma }\;( { a \over r } - a' )  = 0 \; ,
\qquad b'' + {\Sigma' \over \Sigma }\;( { b \over r } - b' )  = 0
\; , \label{3.9b}
\\
-2 c' \; R' - c \; R'' - { 2abc \over r^{2} \Sigma^{2}}  -{\Sigma'
\over \Sigma }\; c\; R' = 0 \; .
\label{3.9c}
\end{eqnarray}

\noindent Eq. (\ref{3.9c}) can be simplified. Indeed, let us multiply it by
  $cR'$:
\begin{eqnarray}
-(c^{2})' (R')^{2} \;-\; {1 \over 2} \; c^{2}\;[(R')^{2}]' \; - \;
{2abc^{2} \over r^{2}\Sigma^{2}} \; R' \; - \; {\Sigma '  \over
\Sigma }\; c^{2}\;(R')^{2}  = 0 \; ,
\nonumber
\end{eqnarray}

\noindent and  allow  for expressions for $c^{2}$ and  $(R')^{2}$  according to
(\ref{3.9a}):
\begin{eqnarray}
 - \; {a^{2} \over r^{2} \Sigma^{2}} \; {d \over dr}\;
({a^{2} \over \Sigma^{2}}) \; - \; {1\over 2} \; {a^{2} \over
\Sigma^{2}}\; {d \over dr}\;{a^{2} \over r^{2}\Sigma^{2}} \; -\;
{2ab \over r^{2}\Sigma^{2}} \; {a^{2} \over \Sigma^{2}} \; R' \; -
\; {\Sigma' \over \Sigma} \; {a^{2} \over \Sigma^{2}} \; {a^{2}
\over r^{2}\Sigma^{2}} = 0 \;  .
\nonumber
\end{eqnarray}

\noindent After simple calculation, we get to
\begin{eqnarray}
b = {1 \over 2R'} \;  ( -3a' \;+\;  2 \;{ \Sigma' \over \Sigma} \;
a \; + \; {a \over r} \;  ) \; .
\nonumber
\end{eqnarray}

\noindent Further, bearing  in mind the identity
\begin{eqnarray}
R' = \delta  \;{ a  \over   r \Sigma }  \;, \qquad \delta  = \pm 1
\; ,
\nonumber
\end{eqnarray}

\noindent we arrive at
\begin{eqnarray}
2ab = \delta \; r \Sigma \; ( \; -3a' \; + \; 2 {\Sigma' \over
\Sigma}
 \; a \; + \;{a \over r} \; ) \; ;
\nonumber
\end{eqnarray}

\noindent
Thus, the radial system will take the form
\begin{eqnarray}
a'' + {\Sigma' \over \Sigma }\;( { a \over r } - a' )  = 0 \; ,
\qquad  b'' + {\Sigma' \over \Sigma }\;( { b \over r } - b' )  = 0
\; ,
\label{3.11a}
\\
c  = \epsilon \;  {a  \over \Sigma  }  \;\; , \qquad \qquad   R' =
\delta \; {a \over r \Sigma}
 \; , \label{3.11b}
\\
2ab = \delta \; r \Sigma \; ( \; -3a' \; + \; 2 {\Sigma' \over
\Sigma}
 \; a \; + \;{a \over r} \; ) \; ;
\label{3.11c}
\end{eqnarray}

\noindent  here  $\delta ^{2} =1, \epsilon^{2} = 1$, and these two parameters are independent.
The quantity $\epsilon$ may be  excluded by inserting it   into $a(r)$, solution of the linear differential equation.
So, we have more simple system:
\begin{eqnarray}a'' + {\Sigma' \over \Sigma }\;( { a \over r } - a' )  = 0 \; ,
\qquad  b'' + {\Sigma' \over \Sigma }\;( { b \over r } - b' )  = 0
\; ,
\label{3.11a}
\\
c  =   {a  \over \Sigma  }  \;\; , \qquad \qquad   R' =  \delta \;
{a \over r \Sigma}
 \; ,
  \label{3.11b}
\\
2ab = \delta \; r \Sigma \; ( \; -3a' \; + \; 2 {\Sigma' \over
\Sigma}
 \; a \; + \;{a \over r} \; ) \; .
\label{3.11c}
\end{eqnarray}

\noindent
The way to solve the task is to be as follows:
first, one can find general expressions for $a(r), b(r)$; then
determine $c(r)$ and $R(r)$ from  (\ref{3.11b});  and finally one should substitute  $a(r)$ and $b(r)$ into
eq. (\ref{3.11c}).

\section{Radial solutions in the Riemann and  Lobachevsky
 models}

First, let us examine the case of Riemann space model.
Equations for  $a(r)$ and $b(r)$ are the same $(a, \; b = g)$:
\begin{eqnarray}
{d^{2} \over dr^{2}}\; g \;-\;{2r \over 4 \rho^{2} \;(1
+r^{2}/4\rho^{2})} {d \over dr}\;g\; +\; {2 \over 4\rho^{2} (1
+r^{2}/4 \rho^{2})}\; g = 0 \; ;
\label{4.1a}
\end{eqnarray}

\noindent  in this section we will use usual unites for $r$. General solutions are
\begin{eqnarray}
a = a_{1} \; r \; + \; a_{2} \; (1-r^{2} /4\rho^{2}) \; ,
\;\;
b = b_{1} \; r \; + \; b_{2} \; (1-r^{2} /4\rho^{2}) \;\; ;
\label{4.1b'}
\end{eqnarray}

\noindent $a_{1},\; a_{2},\;b_{1},\; b_{2}$ are   constants.
Correspondingly, from (\ref{3.11b}) for  $c(r)$ and  $R(r)$ we have
\begin{eqnarray}
c(r) =     a_{1} \; { r \over 1 +
r^{2}/4\rho^{2}} \; + \; a_{2} \; {1-r^{2} /4\rho^{2} \over 1 +
r^{2}/4 \rho^{2}} \; ,
\label{4.2a}
\\[2mm]
R(r) =    \delta \;  ( \; a_{1} \; 2\rho \; \arctan{r \over
2\rho} + a_{2}\;\ln {r/\rho \over 1 + r^{2}/4\rho^{2}}\; )
\;+ \; C\; .
\label{4.2b}
\end{eqnarray}

\noindent  Substituting    $a(r),\;b(r)$  (\ref{4.1b'}) into
eq. (\ref{3.11c}),
 we get the system of algebraic relations:
\begin{eqnarray}
2 a_{2} b_{2} = \delta\;  a_{2} \; \; , \qquad   a_{1} b_{2} +
a_{2} b_{1} = - \delta\; a_{1} \; ,
\nonumber
\\
\qquad  a_{1} b_{1} - {1\over 2} \; b_{2}a_{2} {1 \over \rho^{2} }= \delta\;  {5\over 4}
 \; a_{2} {1 \over \rho^{2}} \; .
\label{4.4b}
\end{eqnarray}

\underline{First, let  $a_{2} \neq 0$.} From first relation in (\ref{4.4b})  it follows
$b_{2} = \delta /2$, and two remaining ones take on the form
\begin{eqnarray}
b_{1} \; a_{2} = - {3\over 2} \; \delta \; a_{1} \;\; , \;\; a_{1}
\; b_{1} = + {3\over 2} \; \delta  \; a_{2} {1 \over \rho^{2}}
\;\;  ,
\nonumber
\end{eqnarray}

\noindent  from where it follows
\begin{eqnarray}
{a_{2} \over a_{1}} = - \; {a_{1} \over a_{2}}  \; \rho^{2}  \;
\Longrightarrow \;\; ({a_{2} \over a_{1}} )^{2} = - \rho^{2} \; \;
,
\nonumber
\end{eqnarray}

\noindent and therefore , $ a_{2} =   \pm \; i\; \rho \;   a_{1} \;$;
at this for  $b_{1}$  we have complex values:
$ \;b_{1} =  \pm { 3i \over 2}\; {\delta  \over \rho }\;$.
Thus, we arrive at the complex-valued solution:
\begin{eqnarray}
a = a_{1} \;  r \; \pm  \; i\;a_{1} \; \rho \;  (1 - {r^{2} \over
4 \rho^{2}}) \;  ,
\qquad  b =  \pm  { 3i \delta \over 2}  \;
{r \over \rho}  \; + {\delta \over 2} \; (1 - {r^{2} \over
4\rho^{2}  })  \; .
\label{4.5a}
\end{eqnarray}

\noindent
Correspondingly,  $c(r)$ and  $R(r)$ are
\begin{eqnarray}
c(r) =    a_{1} \; { r \over 1 + r^{2}/4\rho^{2}} \;  \;\pm  \;
i\;a_{1} \; \rho \;  \; {1-r^{2} /4\rho^{2} \over 1 + r^{2}/4
\rho^{2}} \; ,
\nonumber
\\
R(r) =   \delta \;  ( \; a_{1} \; 2\rho \; \arctan{r \over
2\rho}  \;\pm  \; i\;a_{1} \; \rho  \;\ln {r/\rho \over 1 +
r^{2}/4\rho^{2}}\;  ) \;+ \; C\; .
\label{4.5b}
\end{eqnarray}

\noindent
In the limit   $\rho \rightarrow \infty$, they behave
\begin{eqnarray}
a = \pm  \; i\;a_{1} \; \rho \; \;  , \qquad
 b =   + {\delta \over 2} , \qquad
c(r) =    \pm  \; i\;a_{1} \; \rho \;   \; ,
\nonumber
\\
R(r) =   \delta \;
 \left [ \; a_{1} \; r  \;\pm  \; i\;a_{1} \; \rho  \;\ln {r  \over  \rho } \;  \right
] \;+ \; C\; .
\label{4.5b}
\end{eqnarray}

\noindent
This solution is complex-valued and it has no physical meaning in the limit
of the flat space, in the following this solution  will not be considered.

\underline{Now, let   $a_{2} = 0$}, then  eqs.  (\ref{4.4b})  give
$a_{1} \; b_{2} = - \delta \; a_{1} \;\; , \;\; a_{1} \; b_{1} = 0
\; ,
$ from where we arrive at two solutions:
\begin{eqnarray}
{\bf I} \qquad a_{2} = 0 ,\;  a_{1} \neq 0 \; , \qquad b_{1} = 0 \;\; , \qquad
b_{2} = - \delta \;\;  ;
\nonumber
\\
 a (r) = a_{1} r \; , \qquad  \qquad  b (r)  = - \delta \;   (1 - {r^{2} \over 4 \rho^{2}}) \; ,
\nonumber
\\
c(r)  =    {  a_{1} r \over 1 + r^{2}/4\rho^{2}}\; , \qquad  R (r)
= \delta a_{1} \;  \;  ( 2 \rho\;  \mbox{arctan} \; {r \over
2\rho})  +  C \; .
\label{4.6a}
\end{eqnarray}
\begin{eqnarray}
{\bf II} \qquad a_{2} = 0 \; \; , \; a_{1} =0: \qquad
 b = b_{1} \; r \; + \; b_{2}\; (1 - {r^{2} \over 4 \rho^{2}
}) \; ,
\nonumber
\\
 a (r) = 0 \; , \; c (r)  = 0 \; , \; R  (r) = C \; .
\label{4.6b}
\end{eqnarray}

\noindent
In order to have a needed  behavior  in  the limit of the flat space, one
must consider only the following solutions:
\begin{eqnarray}
{\bf I}
 \qquad  a (r) = a_{1} r \; , \qquad  \qquad  b (r)  = -   (1 - {r^{2} \over 4 \rho^{2}}) \; ,
\nonumber
\\
c(r)  =    {  a_{1} r \over 1 + r^{2}/4\rho^{2} }\; , \qquad  R
(r)  = a_{1} \;  ( 2 \rho \; \mbox{arctan} \; {r \over 2 \rho})  +
C \; .
\label{4.7ab}
\end{eqnarray}
\begin{eqnarray}
{\bf II}
 \qquad  b = b_{1} \; r \;  + b_{2} (1 - {r^{2} \over 4 \rho^{2}
}) \; ,
\nonumber
\\
 a (r) = 0 \; , \; c (r)  = 0 \; , \; R  (r) = C \; .
\label{4.7}
\end{eqnarray}

\noindent
Respective expressions for    $K(r)$  and   $\Phi (r)$
look as
\begin{eqnarray}
 {\bf I}
\qquad K (r) = {1 \over er^{2}} \;   [ \;   {  a_{1} r \over 1 +
r^{2}/4\rho^{2}} \;
 f_{1}[  a_{1} \;  ( 2 \rho \; \mbox{arctan} \; {r \over 2 \rho})  +  C  ] -1 \;  ]  \; ,
\nonumber
\\
\Phi  (r) = {1 \over er^{2}} \;   [ \; \; a_{1} r \; f_{2} [
a_{1} \;  ( 2 \rho\; \mbox{arctan} \; {r \over 2 \rho })  +  C  ]
-   (1 - {r^{2} \over 4 \rho^{2} })  \;  ] \; ;
\nonumber
\\
\label{4.8a}
\end{eqnarray}
\begin{eqnarray}
{\bf II}
 \qquad K (r) =  - {1 \over er^{2}}   \; , \qquad \Phi  (r) = {1 \over
er^{2}} \; \left  [ b_{1} \; r \;  + b_{2} (1 - {r^{2} \over 4
\rho^{2} }) \right ] \; ;
\label{4.8b}
\end{eqnarray}

It is readily verified that solution of the type  {\bf I} (depending on  $f_{1},f_{2}$ there are three different
possibilities) has  a good behavior in flat space  limit.
Indeed, in the limit $\rho \; \longrightarrow \infty$ we get
\begin{eqnarray}
K (r) = {1 \over er^{2}} \;  \left [ \;     a_{1} r \;
 f_{1}(  a_{1} r  +  C  ) -1 \; \right ]  \; ,
 \nonumber
 \\
\Phi  (r) = {1 \over er^{2}} \; \left  [ \; \; a_{1} r \; f_{2} (
a_{1} \;r +  C )] -   1   \; \right ] \; .
 \label{(4.8c}
\end{eqnarray}

\noindent From this, choosing for instance  $f_{1}$  and   $f_{2}$
according to  (see  (\ref{2.7}) )
\begin{eqnarray}
f_{1}(x)  = \pm {\alpha  \over \sin ( \alpha x +  \beta ) } ,
\qquad  f_{2}(x)  =
 {\alpha \over \mbox{tan}\;  ( \alpha x + \beta ) } \; ,
\nonumber
\end{eqnarray}

\noindent  we get
\begin{eqnarray}
K (r) = {1 \over er^{2}} \;  ( \;
  { \pm  \alpha  a_{1} r\over \sin (\alpha  (  a_{1} r  +  C  )  + \beta) }     -1 \;  )  \; ,
\nonumber
\\
\Phi  (r) = {1 \over er^{2}} \;  (  \;
  {  \alpha  a_{1} r\over \mbox{tan} \; (\alpha  (  a_{1} r  +  C  )  + \beta) }     -1
\;  ) \; ;
\nonumber
\end{eqnarray}

\noindent from where, with the notation
$\alpha a_{1} = A \; , \; \alpha C + \beta = B \; $,
  we arrive at
\begin{eqnarray}
K (r) = {1 \over er^{2}} \;  ( \;
  { \pm  A r\over \sin ( A r +B )}     -1 \;  )  \; ,
\qquad \Phi  (r) = {1 \over er^{2}} \; ( \;
  {  A r\over \mbox{tan} \; ( Ar +B) }     -1
\; ) \; ;
\nonumber
\end{eqnarray}

\noindent which coincides with $K_{3}$ and $\Phi_{3}$,
according to  (\ref{2.8a}). In the same manner can be considered two other cases from  (\ref{2.7}).

Now, let us show that solution of the type {\bf II}  will give
a trivial monopole solution in the flat space limit. Indeed,
in this limit, eqs. (\ref{4.8b})   look as
\begin{eqnarray}
K (r) =  - {1 \over er^{2}}   \; , \qquad \Phi  (r) = {1 \over
er^{2}} \;  ( b_{1} \; r \;  + b_{2}  ) \; ,
\nonumber
\end{eqnarray}

\noindent  which coincides with    (\ref{2.11a}).
It is readily verified that for such  a trivial solution, Yang-Mills equations
become  just two independent differential equations (linear and nonlinear). Indeed,
let  $er^{2} K(r) +1 = 0$,  then equations become
\begin{eqnarray}
\Phi''\; + \; {4\over r} \; \Phi' \;+\; 2  \Phi \; {1 \over r^{2}}
-\;  {\Sigma ' \over  \Sigma} \; ( \Phi ' + {\Phi \over r})\; = 0
\; ,
\nonumber
\\
K'' + {4 K' \over r}  - 2 e K^{2}  + {\Sigma' \over \Sigma}\; (K'
+ {2 K \over r}) = 0 \; .
\nonumber
\end{eqnarray}

\noindent
Evidently, the nonlinear equation is satisfied  by $K(r)
= -1 /er^{2}$. In turn equation for $\Phi (r)$
\begin{eqnarray}
\Phi''\; + \; {4\over r} \; \Phi' \;+\; 2  \Phi \; {1 \over r^{2}}
-\;  {r/2 \rho^{2} \over  1  + r^{2} /4 \rho^{2} } \; ( \Phi ' +
{\Phi \over r})\; = 0 \; ,
\nonumber
\end{eqnarray}

\noindent has two independent solutions
\begin{eqnarray}
\Phi_{1} =  {1 \over r} , \qquad  \Phi_{1} =  {1 -r^{2} /4
\rho^{2} \over r^{2}} \; ,
\label{4.10b}
\end{eqnarray}

\noindent which are  in accordance with (\ref{4.8b}).

The case of Lobachevsky space is treated in the similar manner. The results are
\begin{eqnarray}
 {\bf I}
 \qquad K (r) = {1 \over er^{2}} \;  \left [ \;   {  a_{1} r \over 1 -
r^{2}/4 \rho^{2}} \;
 f_{1} \; [ \;  a_{1} \;  ( 2 \rho \; \mbox{arcth} \; {r \over 2\rho})  +  C  \; ] - 1 \; \right ]  \; ,
\nonumber
\\
\Phi  (r) = {1 \over er^{2}} \; \left  [ \; \; a_{1} r \; f_{2}\;  [\;
a_{1} \;  ( 2 \rho\; \mbox{arcth} \; {r \over 2\rho})  +  C \;  ] -
(1 + {r^{2} \over 4\rho^{2}})  \; \right ] \; ;
\nonumber
\\
\label{5.6a}
\end{eqnarray}
\begin{eqnarray}
{\bf II}
 \qquad
 K (r) =  - {1 \over er^{2}}   \; , \qquad \Phi  (r) = {1 \over
er^{2}} \; \left  [ b_{1} \; r \;  + b_{2} (1 + {r^{2} \over 4
\rho^{2} }) \right ] \; .
\label{5.6b}
\end{eqnarray}

Solution of the type  {\bf I}  is analogue of
the known monopole solution in flat space (\ref{2.8a}).
Let us show that solution of the type {\bf II}  will give
a trivial monopole solution  in the flat space limit. Indeed, its limit at $\rho \longrightarrow \infty$
looks as follows
\begin{eqnarray}
K (r) =  - {1 \over er^{2}}   \; , \qquad \Phi  (r) = {1 \over
er^{2}} \; \left  [ b_{1} \; r \;  + b_{2}  \right ] \; ,
\label{5.7}
\end{eqnarray}

\noindent  which coincide with   (\ref{2.9b}) -- (\ref{2.11a}).
It is readily verified that for such  a trivial solution, Yang-Mills equations
become  just two independent differential equations (linear and nonlinear). Indeed,
let  $er^{2} K(r) +1 = 0$,  then equations become
\begin{eqnarray}
\Phi''\; + \; {4\over r} \; \Phi' \;+\; 2  \Phi \; {1 \over r^{2}}
-\;  {\Sigma ' \over  \Sigma} \; ( \Phi ' + {\Phi \over r})\; = 0
\; ,
\nonumber
\\
K'' + {4 K' \over r}  - 2 e K^{2}  + {\Sigma' \over \Sigma}\; (K'
+ {2 K \over r}) = 0 \; ;
\nonumber
\end{eqnarray}

\noindent Nonlinear equation is satisfied by  $K(r) = -1 /er^{2}$. Linear equation
\begin{eqnarray}
\Phi''\; + \; {4\over r} \; \Phi' \;+\; 2  \Phi \; {1 \over r^{2}}
+\;  {r/2 \rho^{2}  \over  1  - r^{2} /4\rho^{2}} \; ( \Phi ' +
{\Phi \over r})\; = 0 \; ,
\nonumber
\end{eqnarray}

\noindent has  two independent solutions
\begin{eqnarray}
\Phi_{1} =  {1 \over r} , \qquad  \Phi_{1} =  {1 +r^{2} /4
\rho^{2}  \over r^{2}} \; .
\label{5.9b}
\end{eqnarray}

\vspace{5mm}

\begin{center}

{\bf PART II}
\end{center}

\section{ The Pauli criterion}

Let the $J^{\lambda }_{i}$   denote
\begin{eqnarray}
J_{1}=  l_{1} + \lambda\; {{\cos  \phi  }\over {\sin \theta}}\;  ,\qquad
J_{2}=  l_{2} + \lambda\; {{\sin  \phi  }\over {\sin \theta}}\;  ,\qquad J_{3} = l_{3} \; ,
\label{2.1}
\end{eqnarray}

\noindent where $l_{i}$  stand for the components of orbital momentum operator \cite{1974-Landau-Lifshitz}:
\begin{eqnarray}
 l_{1} = i\; (\sin \phi \partial_{\theta} + ctg \; \theta \cos \phi \partial _{\phi} ) \; ,
\nonumber
\\
l_{2} = i\; (- \cos \phi \partial_{\theta} + ctg \; \theta \sin \phi \partial _{\phi} )\; ,\;\;
l_{3} = -i \partial_{\phi} \; .
\nonumber
\end{eqnarray}

\noindent At arbitrary $\lambda$, as readily  verified,  those $J_{i}$
 satisfy the~commutation rules of the~Lie algebra
$SU(2)$.
As known,  all  irreducible  representations  of  such  an~ abstract
algebra are determined by a~set of weights
 $$j = 0, 1/2, 1, 3/2,... \; \; ;\;  \mbox{dim} \;j = 2j + 1 \;.
 $$

Given the~explicit expressions of  $J_{a}$   above,  we  will
find functions
               $\Phi ^{\lambda }_{jm}(\theta ,\phi )$
on which the~representation of  weight $j$ is realized. In agreement with
the~general approach \cite{1974-Landau-Lifshitz}, those solutions are to be established by
the~following relations
\begin{eqnarray}
J_{+} \; \Phi ^{\lambda }_{jj} \; = \; 0 ,\qquad
\Phi ^{\lambda }_{jm} \; = \; \sqrt{{(j+m)! \over (j-m)! \; (2j)! }} \; J^{(j-m)}_{-} \;
\Phi^{\lambda}_{jj} \; ,
\nonumber
\\
J_ {\pm}  =  J_{1} \pm i J_{2}  =
e^{\pm i\phi }\; (\; \pm { \partial \over \partial \theta } \; + \;
 i \cot \theta \; { \partial \over  \partial \phi} \; + \;
 { \lambda \over  \sin  \theta }\; )\; .
\label{2.2}
\end{eqnarray}

\noindent From the equations
$J_{+} \; \Phi ^{\lambda }_{jj} \; = \; 0 \;$ and
$\; J_{3} \; \Phi ^{\lambda }_{jj} \;  = \; j \; \Phi ^{\lambda }_{jj}$
it follows that
\begin{eqnarray}
\Phi ^{\lambda }_{jj}  = \;
N^{\lambda }_{jj} \;  e^{ij\phi} \; \sin^{j}\theta \;\;
{( 1 + \cos \theta  )^{+\lambda /2} \over ( 1 - \cos \theta )^{\lambda /2}}\; ,
\nonumber
\\
N^{\lambda }_{jj} \; =  \; {1 \over \sqrt{2\pi}} \; { 1 \over 2^{j} } \;
 \sqrt{{(2j+1) \over \Gamma(j+m+1) \; \Gamma(j-m+1)}} \; .
\nonumber
\end{eqnarray}

\noindent Further, employing (\ref{2.2}) we produce the~functions
$\Phi ^{\lambda }_{jm}$
\begin{eqnarray}
\Phi ^{\lambda }_{jm}  =  N^{\lambda }_{jm}  e^{im\phi}
 {1 \over \sin^{m}\theta }
{(1 - \cos \theta)^{\lambda/2} \over (1 + \cos \theta)^{+\lambda/2}}
\nonumber
\\
\times
({ d \over d \cos  \theta})^{j-m}  \; \left [\;  (1 + \cos  \theta ) ^{j + \lambda }
(1 - \cos  \theta ) ^{j-\lambda }  \;  \right ] \; ,
\label{2.3}
\end{eqnarray}

\noindent where
\begin{eqnarray}
N^{\lambda }_{jm}   =  {1 \over \sqrt{2\pi} 2^{j}} \;
 \sqrt{{(2j+1) \; (j+m)! \over
2(j-m)!  \Gamma(j + \lambda +1) \; \Gamma(j- \lambda +1)}}  \;\; .
\nonumber
\end{eqnarray}

\noindent The Pauli criterion tells us that the $(2j + 1)$ functions
$ \Phi ^{\lambda }_{jm}(\theta ,\phi ),$
so  constructed are  guaranteed  to be a~basis for a~finite-dimension
representation, providing that the function
$\Phi ^{\lambda }_{j,-j}(\theta ,\phi )$
found by this procedure obeys the~identity
\begin{eqnarray}
J_{-} \;\; \Phi ^{\lambda }_{j,-j} \; = \; 0 \; .
\label{2.4a}
\end{eqnarray}

\noindent After substituting the~function $\Phi ^{\lambda }_{j,-j}(\theta ,\phi )$
to the (\ref{2.4a}), the latter reads
\begin{eqnarray}
J_{-} \Phi ^{\lambda }_{j,-j}  = N^{\lambda }_{j,-j}
e^{-i(j+1)\phi } \; (\sin \theta)^{j+1}
{(1 - \cos  \theta )^{\lambda /2} \over (1 + \cos  \theta )^{\lambda /2}} \; \times
\nonumber
\\
\times \;
({d \over d \cos \theta})^{2j+1} \; [ (1 + \cos  \theta )^{j+\lambda }
(1 - \cos  \theta )^{j-\lambda } )  ] =  0 \; ,
\label{2.4b}
\end{eqnarray}

\noindent which in turn gives the~following restriction on
 $j$  and $\lambda $
\begin{eqnarray}
({d \over d \cos  \theta})^{2j+1} \; [\; (1 + \cos  \theta  )^{j+\lambda } \;
(1 - \cos  \theta  )^{j-\lambda } \; ] \;  = \; 0\;.
\label{2.4c}
\end{eqnarray}

\noindent But the~relation (\ref{2.4c}) can be satisfied  only  if  the~factor
 $P(\theta )$  subjected to the~operation of taking derivative  $( d/d \cos \theta ) ^{2j+1}$
is a~polynomial of degree $2j$  in $ \cos \theta$. So, we have (as a~result
of the~Pauli criterion)

\vspace{5mm}
\noindent
{\bf 1 }  {\em the} $\lambda$ {\em is allowed to take values}
$, +1/2,\; -1/2,\; +1,\; -1, \ldots$.
\vspace{5mm}

\noindent Besides, as the~latter condition is satisfied,
$P(\theta )$  takes different forms depending on
the $(j - \lambda)$-correlation:
\begin{eqnarray}
P(\theta )  =  (1 + \cos \theta )^{j+\lambda }
  (1 - \cos \theta )^{j - \lambda }  =
P^{2j}(\cos \theta ), \;\; if \;\;  j = \mid \lambda \mid,
\mid \lambda \mid +1,...
\nonumber
\end{eqnarray}

\noindent
or
\begin{eqnarray}
P(\theta ) \; = \;
{ P^{2j+1}(\cos \theta ) \over \sin \theta }, \qquad if \qquad
 j = \mid \lambda \mid +1/2, \mid \lambda \mid +3/2,...
\nonumber
\end{eqnarray}

\noindent so  the second necessary condition  resulting from
the~Pauli criterion  is

\vspace{5mm}
\noindent
{\bf 2 }  {\em given} $\lambda$ {\em according to {\bf 1},
    the number j is allowed to take values}
  $$j = \mid \lambda \mid, \mid \lambda \mid +1,...
  $$
\vspace{3mm}

Hereafter, these two conditions: {\bf 1}   and  {\bf 2} will  be referred   as  the~first
and     the~second   Pauli consequences respectively.
It should be noted  that  the~angular  variable $\phi $  is  not
affected (charged) by this Pauli condition; in other words, it is
effectively eliminated out of  this  criterion,  but
a~variable that  worked above is the~$\theta$. Significantly,  in
the contrast to this, the well-known procedure of deriving  the
Dirac  quantization  condition  from investigating continuity
properties of quantum mechanical wave functions,  such
a~working variable is the $\phi $.

If the~first and second Pauli consequences fail, then we face
rather unpleasant mathema\-ti\-cal and physical problems\footnote{Reader  is
referred  to  the Pauli article \cite{1939-Pauli} for more detail about
those peculiarities.
However, all these peculiarities may be ignored and then there arise new possibilities
 -- see Hunter et al \cite{1998-Hunter et al}-\cite{2005-Hunter-Schlifer} and references therein. }.

As a~simple illustration,  we  may  indicate
the~familiar case  when $\lambda= 0$; if in those circumstances,
 the~second  Pauli  condition had failed, then we would have the~integer and
half-integer  values  of the orbital angular momentum number
$l = 0, 1/2, 1, 3/2,\ldots\;$

As regards  the Dirac  electron  with  the  components  of  the  total
angular momentum in the form \cite{1980-Berestetzkiy-Lifshitz-Pitaevskiy}
\begin{eqnarray}
J_{1} =   l_{1} + { \cos \phi \over \sin  \theta }\; \Sigma _{3}\;,\qquad
J_{2} = l_{2}  +  { \sin \phi \over \sin  \theta } \; \Sigma _{3}\;,\qquad
 J_{3} = l_{3}
\nonumber
\end{eqnarray}

\noindent we are to employ the Pauli  criterion  in  the  constituent
form  ($\lambda $  changes into $\Sigma _{3}$):
\begin{eqnarray}
\Sigma _{3} =
\left | \begin{array}{cccc}
               +1/2 &   0 &  0  &  0 \\
                0   &-1/2 &  0  &  0 \\
                0   &   0 &+1/2 &  0 \\
                0   &   0 &  0  &-1/2
                           \end{array} \right |     .
\nonumber
\end{eqnarray}

\noindent In this case,  we obtain the allowable set $j = 1/2, 3/2, \ldots$.

Significantly that the  functions
$\Phi ^{\lambda }_{jm}(\theta, \phi )$ constructed  above  relate
directly  to  the  well-known Wigner $D$-functions (bellow we will use
the notation  according  to \cite{1974-Warshalovich-Moskalev-Hersonskii}):
\begin{eqnarray}
\Phi ^{\lambda }_{jm}(\theta , \phi ) \; =  \;
(-1)^{j-m} \; D^{j}_{-m, \lambda}(\phi, \theta, 0)\; .
\label{2.5}
\end{eqnarray}

\noindent Because of the detailed development of $D$-function  theory,
relation (2.5) will be  of vital importance  in the following.

Closing this paragraph, we draw attention to   that
the  Pauli  criterion
\begin{eqnarray}
J_{-} \Phi _{j,-j}(t,r,\theta ,\phi )\; =\; 0
\nonumber
\end{eqnarray}

\noindent
(here $\Phi ^{\lambda }_{j,-j}(\theta ,\phi )$
denotes  a  spherically symmetrical wave function)
affords a condition that is invariant relative to possible  gauge
transformations. The function $\Phi _{j,m}(t,r,\theta ,\phi )$
may be subjected  to any $U(1)$ transformation, but if  all the~components
 $J_{i}$ vary  in a~corresponding way too, then the
Pauli  condition provides the~same result on $J$-quantization. In contrast
to this, the common  requirement to be a~single-valued function of spatial
points is often applied to producing a~criterion on selection  of
allowable wave functions in quantum mechanics; but that is  not
invariant under gauge transformations.

\section{  Electron in a spherically symmetric
                 geometric background   and Wigner $D$-functions}

Below we  review briefly  some  relevant facts about the
tetrad formalism. In  the presence of an  external  gravitational
field, the ordinary  Dirac equation
\begin{eqnarray}
(\; i \gamma ^{a} \partial _{a} \; -
\;m \; ) \Psi (x)\; = \; 0
\nonumber
\end{eqnarray}

\noindent
is generalized into  \cite{1988-Red'kov(1), 1998-Red'kov(1)}
\begin{eqnarray}
[\; i \gamma ^{\alpha }(x) ( \partial _{\alpha }  +
\Gamma _{\alpha }(x) \; )  -  m \; ] \; \Psi (x)  =  0\; ,
\label{3.1}
\end{eqnarray}

\noindent where $e^{\alpha }_{(a)}(x)$  is a~tetrad:
\begin{eqnarray}
\gamma ^{\alpha }(x) \; = \;
\gamma ^{a} e^{\alpha }_{(a)}(x)   ,
\qquad
e^{\alpha }_{(a)}(x)e^{\beta }_{(b)}(x) \eta^{ab} = g^{\alpha \beta}(x) \; ;
\nonumber
\end{eqnarray}

\noindent $ \Gamma _{\alpha }(x)$ is the bispinor connection:
\begin{eqnarray}
\Gamma _{\alpha }(x) \; = \;
 {1 \over 2} \sigma ^{ab} \; e^{\beta }_{(a)} \;
\nabla _{\alpha } (e^{\alpha } _{(b)\beta }) \; ;
\nonumber
\end{eqnarray}

\noindent $\nabla _{\alpha }$ is  the
covariant derivative symbol. In the spinor basis \cite{1980-Berestetzkiy-Lifshitz-Pitaevskiy}
\begin{eqnarray}
\sigma ^{a} = (I, \; + \sigma ^{k}) \; , \;\;
 \bar{\sigma}^{a} = (I, \; -\sigma ^{k}\;) \; ,\qquad
\gamma ^{a} =
\left | \begin{array}{cc}
               0   &  \bar{\sigma}^{a} \\
            \sigma ^{a}  &  0
\end{array} \right |,
\nonumber
\\
\psi (x)  =
\left | \begin{array}{c}
                       \xi(x)  \\     \eta (x)
\end{array} \right | \; , \;\;
\xi(x)  =  \left | \begin{array}{c}
                 \xi ^{1}  \\  \xi ^{2}
\end{array} \right | \; , \;\;
\eta (x) =
\left | \begin{array}{c}
            \eta_{\dot{1}} \\ \eta_{\dot{2}}
\end{array} \right | \; , \;\;
\nonumber
\end{eqnarray}

\noindent we  have  two equations
\begin{eqnarray}
i \sigma ^{\alpha }(x) \; [\;  \partial_{\alpha } \; + \;
\Sigma _{\alpha }(x)\; ]\; \xi (x) \; = \; m \; \eta (x) \; ,
\nonumber
\\
i\; \bar{ \sigma ^{\alpha}}(x) \; [\;  \partial_{\alpha} \; + \;
\bar{\Sigma} _{\alpha }(x) \;]  \; \eta (x) \;  =  \; m \;   \xi (x)\; ;
\label{3.2}
\end{eqnarray}

\noindent the symbols $\sigma ^{\alpha }(x), \bar{\sigma} ^{\alpha }(x),
 \Sigma _{\alpha }(x), \bar{\Sigma} _{\alpha }(x)$  denote respectively
\begin{eqnarray}
\sigma ^{\alpha }(x)  =  \sigma ^{a}\; e^{\alpha }_{(a)}(x) \; , \;\;
\Sigma _{\alpha }(x)  = {1 \over 2} \Sigma ^{ab} e^{\beta }_{(a)}
\nabla _{\alpha }(e_{(b)\beta }) \;, \;\;
\Sigma ^{ab} = {1 \over 4}
(\bar{\sigma} ^{a} \sigma ^{b}  -  \bar{\sigma} ^{b} \sigma ^{a}) \; ,
\nonumber
\\
\bar{\sigma} ^{\alpha }(x)  =  \bar{\sigma}^{a} \; e^{\alpha }_{(a)}(x)\; , \;\;
\bar{\Sigma} _{\alpha }(x)  = {1 \over 2}
\bar{\Sigma} ^{ab}e^{\beta }_{(x)} \nabla _{\alpha }(e_{(b)\beta } ) \; , \;\;
\bar{\Sigma} ^{ab} = {1 \over 4} (\sigma ^{a} \bar{\sigma} ^{b}  -
\sigma ^{b} \bar{\sigma} ^{a} ) \;  .
\nonumber
\end{eqnarray}

\noindent Setting $m$ equal to zero, we obtain the~Weyl equations for neutrino
$\eta (x)$  and anti-neutrino $\xi (x)$, or Dirac's equation for a~massless
particle.

The form of equations (\ref{3.1}) -- (\ref{3.2}) implies   quite
definite their symmetry properties. It is common, considering the
Dirac equation in the same  space-time,  to  use  some   different
tetrads $e^{\beta }_{(a)}(x)$ and $e^{'\beta}_{(b)}(x)$, so that we have
the~equation  (3.1)
and analogous one with a~new tetrad mark. In other words, together
with (3.1) there exists an equation on $\Psi'(x)$  where the quantities
$\gamma^{'\alpha }(x)$ and $\Gamma'_{\alpha}(x)$, in comparison with
$\gamma^{\alpha}(x)$ and $\Gamma_{\alpha}(x)$, are  based
on another tetrad $e^{'\beta }_{(b)}(x)$ related to $e^{\beta }_{(a)}(x)$
 through  some  local Lorentz matrix
$
e_{(b)}^{'\beta} (x) \;  = \; L^{\;\;a}_{b}(x) \; e^{\beta }_{(a)}(x) $.
It may be shown that these two Dirac equations on  functions
$\Psi (x)$ and $\Psi'(x)$ are related to each other by a~quite definite
bispinor transformation
\begin{eqnarray}
\xi'(x) \; = \; B(k(x)) \; \xi (x) \; , \qquad
\eta'(x) \; = \; B^{+} (\bar{k}(x))\; \eta (x) \; .
\label{3.3a}
\end{eqnarray}

\noindent Here, $B(k(x)) = \sigma ^{a} k_{a}(x)$ is a~local matrix from
the $SL(2.C)$ group; 4-vector $k_{a}$ is the well-known parameter
on this group  (for instance, see   Wightman \cite{1960-Wightman},  Macfarlane \cite{1962-Macfarlane},
Fedorov \cite{1979-Fedorov}, Red'kov \cite{2009-Red'kov}).  The
matrix $L^{\;\;a}_{b}(x)$  can be expressed  as  a~function  of
arguments $k_{a}(x)$  and $k^{*}_{a}(x)$:
\begin{eqnarray}
L^{\;\;a}_{b}(k, k^{*})  =  \bar{\delta}^{c}_{b}  \;
[ \; - \delta ^{a}_{c}  k^{n}  k^{*}_{n}  +
k_{c}  k ^{a*}  +
k^{*}_{c} \; k^{a}  +
i \; \epsilon ^{\;\;anm}_{c}  k_{n}  k^{*}_{m} \; ]
\label{3.3b}
\end{eqnarray}

\noindent where $\bar{\delta} ^{c}_{b}$ is a~special Kronecker's symbol
\begin{eqnarray}
\bar{ \delta} ^{c}_{b} =
\left \{ \begin{array}{l}
                0 ,\; \;  if \;\;  c \neq  b \; ,\\
                +1 ,\;\;  if \;\;  c = b = 0  \; , \\
                -1 ,\;\;  if \;\;  c = b = 1,2,3 \; .
\end{array} \right.
\nonumber
\end{eqnarray}

It is normal practice that some different  tetrads  are
used at examining the~Dirac equation on the~background of a~given
Riemanniann space-time. If there is  a~need  for  analysis  of
the~correlation between  solutions in such distinct tetrads,  then  it
is important  to know how to  calculate  the~corresponding  gauge
transformations over the spinor wave functions.

First,  the~need
for taking into account such gauge transformations  was especially
emphasized  by Fock  V.I. \cite{1929-Fock(3)}.
The first who were interested in
explicit expressions for such spinor matrices, were
Schr\"odinger \cite{1932-Schrodinger, 1938-Schrodinger(1), 1938-Schrodinger(2)}
 and  Pauli \cite{1939-Pauli}. Thus, Schr\"odinger
found the~matrix  relating  spinor  wave  functions  in Cartesian
and  spherical tetrads:
\begin{eqnarray}
x^{\alpha } = (x^{0},x^{1},x^{2},x^{3})\;, \qquad e^{\alpha }_{(a)} (x) = \delta ^{\alpha }_{a}
\; ,
\nonumber
\\
dS^{2} = [(dx^{0})^{2} - (dx^{1})^{2} - (dx^{2})^{2} - (dx^{3})^{2} ]\;,
\nonumber
\end{eqnarray}

\noindent and
\begin{eqnarray}
x^{'\alpha } = ( t , r , \theta  , \phi  )\; , \;\; dS^{2} =
[\; dt^{2}  -  dr^{2}  -   r^{2}
 (d\theta ^{2} +  \sin ^{2}\theta  d\phi ^{2})\; ]\; ,
\nonumber
\\
e^{\alpha'}_{(0)} = ( 1 , 0 , 0 , 0 ) \; , \qquad
e^{\alpha '}_{(1)} = ( 0 , 0 ,1/r, 0 )\; ,
\nonumber
\\
e^{\alpha '}_{(2)} = ( 0 ,0 , 0 , {1 \over r \;  \sin \theta})\;, \qquad
e^{\alpha '}_{(3)} = ( 0 , 1 , 0 , 0 )\; ;
\nonumber
\end{eqnarray}

\noindent the relevant  spinor matrix is
\begin{eqnarray}
B = \; \pm \;
\left | \begin{array}{cc}
       \cos \theta/2 \; e^{i\phi /2}   & \sin \theta/2  \; e^{-i\phi /2} \\[2mm]
       -\sin \theta /2 \; e^{i\phi /2}   & \cos \theta /2 \;  e^{-i\phi /2}
\end{array} \right | \; .
\label{3.4}
\end{eqnarray}

This basis of spherical tetrad will play a~substantial  role
in our work.

 Now, let us  reexamine  the~problem  of  free
electron in the external spherically symmetric gravitational field,
but centering upon  some facts  which  will  be  of  great
importance at extending that method to  an~electron-monopole
system.

In particular,  we  consider briefly a~question of separating the~angular variables in
the Dirac equation on the background of  a~spherically  symmetric
Riemannian space-time. As a~starting point we take a~flat space-time
model, so that an original equation (\ref{3.1}) being specified for  the
spherical tetrad  takes on the form
\begin{eqnarray}
\left [ i \;
\gamma ^{0} \; \partial_{t}    +
i \;  ( \gamma ^{3}  \partial_{r}  +  { \gamma ^{1}  \sigma ^{31}  +
\gamma ^{2}  \sigma ^{32} \over  r } )   +
{1 \over r} \; \Sigma_{\theta \phi}   -  m
\right ]   \Psi (x)  =   0 \; ,
\label{3.5a}
\end{eqnarray}

\noindent where
\begin{eqnarray}
\Sigma _{\theta ,\phi } \; =
 i\; \gamma ^{1} \partial _{\theta} \;+\;
\gamma ^{2} \;  {\;  i \partial _{\phi} \; + \;
 i\; \sigma ^{12} \over \sin \theta } \; .
\label{3.5b}
\end{eqnarray}

\noindent
We  specialize   the~electronic   wave   function   through
substitution
\begin{eqnarray}
\Psi _{\epsilon jm}(x) \;  = \; {e^{-i\epsilon t} \over r} \;
\left | \begin{array}{l}
        f_{1}(r) \; D_{-1/2} \\ f_{2}(r) \; D_{+1/2}  \\
        f_{3}(r) \; D_{-1/2} \\ f_{4}(r) \; D_{+1/2}
\end{array} \right | \; ;
\label{3.6}
\end{eqnarray}

\noindent Wigner functions are designated by
$D^{j}_{-m,\sigma }(\phi ,\theta ,0)  \equiv  D_{\sigma}$.
Using recursive formulas  \cite{1974-Warshalovich-Moskalev-Hersonskii}
\begin{eqnarray}
\partial_{\theta} \; D_{+1/2} \; =  a\; D_{-1/2}  - b \; D_{+3/2} \; ,
\nonumber
\\
\partial_{\theta} \;  D_{-1/2} \; =  b \; D_{-3/2} - a \; D_{+1/2} \; ,
\nonumber
\\
{- m - 1/2 \;  \cos \theta  \over  \sin \theta } \; D_{+1/2}\; =
- a \;  D_{-1/2} - b \;  D_{+3/2}  \; ,
\nonumber
\\
{- m + 1/2 \; \cos \theta \over \sin \theta}  \;  D_{-1/2} \; =
- b \; D_{-3/2}  - a \;  D_{+1/2} \; ,
\nonumber
\end{eqnarray}

\noindent where
$
a = (j + 1)/2 , \;
 \nu  = (j + 1/2)/2 \; ,
$ we find
\begin{eqnarray}
 \Sigma _{\theta ,\phi } \; \Psi _{\epsilon jm}(x) \;  = \; i\; \nu \;
{e^{-i\epsilon t } \over r} \;
\left | \begin{array}{r}
        - \; f_{4}(r) \; D_{-1/2}  \\  + \; f_{3}(r) \; D_{+1/2} \\
        + \; f_{2}(r) \; D_{-1/2}  \\  - \; f_{1}(r) \; D_{+1/2}
\end{array} \right |\; .
\label{3.7}
\end{eqnarray}

\noindent  Further one gets the following set of radial equations
\begin{eqnarray}
\epsilon   f_{3}   -  i  {d \over dr}  f_{3}   -
i {\nu \over r}  f_{4}  -  m  f_{1} =   0  \; ,
\qquad
\epsilon   f_{4}   +  i  {d \over dr} f_{4}   +
i {\nu \over r}  f_{3}  -  m  f_{2} =   0   \;  ,
\nonumber
\\
\epsilon   f_{1}   +  i  {d \over dr}  f_{1}  +
i {\nu \over r}  f_{2}  -  m  f_{3} =   0 \; ,
\qquad
\epsilon   f_{2}   -  i  {d \over dr} f_{2}   -
i {\nu \over r}  f_{1}  -  m  f_{4} =   0 \; .
\label{3.8}
\end{eqnarray}

The usual $P$-reflection symmetry operator  in  the  Cartesian
tetrad basis  is $\hat{\Pi}_{C.} \; = \; i \gamma ^{0} \otimes
\hat{P}$ (see in \cite{1974-Landau-Lifshitz}), or in a more detailed form
\begin{eqnarray}
  \hat{\Pi}_{C.}  = \left |
\begin{array}{cccc} 0 &  0 &  i &   0  \\ 0 &  0 &  0 &   i  \\ i &
          0 &  0 &   0  \\ 0 &  i &  0 &   0 \end{array} \right |
\; \otimes  \; \hat{P} \; ,
\qquad \hat{P} (\theta , \phi ) = (\pi  -
\theta, \; \phi+ \pi )
 \nonumber
 \end{eqnarray}

\noindent being subjected to translation into the spherical tetrad basis (see (\ref{3.4}))
\begin{eqnarray}
\hat{\Pi}_{sph.} = S(\theta ,\phi ) \;  \hat{\Pi}_{C.} \;
 S^{-1}(\theta ,\phi )
\nonumber
\end{eqnarray}
gives us the~result
\begin{eqnarray}
\hat{\Pi}_{sph.} \; \; =
\left | \begin{array}{cccc}
0 &  0 &  0 & -1   \\
0 &  0 & -1 &  0   \\
0 &  -1&  0 &  0   \\
-1&  0 &  0 &  0
\end{array} \right |
\; \otimes  \; \hat{P} \; =  \Pi_{sph.} \otimes  \; \hat{P}\; .
\label{3.9}
\end{eqnarray}

\noindent With the help of identity \cite{1974-Warshalovich-Moskalev-Hersonskii}
\begin{eqnarray}
\hat{P} \; D^{j}_{-m,\sigma } (\phi ,\theta ,0)  =
(-1)^{j} \; D^{j}_{-m,-\sigma} (\phi ,\theta ,0)\; ,
\nonumber
\end{eqnarray}

\noindent
from the equation on proper values
$
\hat{\Pi}_{sph.}\; \Psi _{jm}
= \; \Pi \; \Psi _{jm}\;
$
we get
\begin{eqnarray}
\Pi = \; \delta \;  (-1)^{j+1}\; , \;\; \delta  = \pm 1 \;, \qquad
f_{4} = \; \delta \;  f_{1}\;  , \qquad  f_{3} = \;\delta \; f_{2}
\label{3.10}
\end{eqnarray}

\noindent so that $\Psi _{\epsilon jm\delta }(x)$ looks
\begin{eqnarray}
\Psi (x)_{\epsilon jm\delta } \; = \; {e^{-i\epsilon t} \over  r } \;
\left | \begin{array}{r}
     f_{1}(r) \; D_{-1/2} \\[1mm]
     f_{2}(r) \; D_{+1/2}\\[1mm]
\delta \; f_{2}(r) \; D_{-1/2}   \\[1mm]
\delta \; f_{1}(r) \; D_{+1/2}
\end{array} \right |  .
\label{3.11}
\end{eqnarray}

\noindent Noting (\ref{3.10}), we  readily simplify the~system  (\ref{3.8}); it is  reduced  to a (no imaginary $i$) form:
\begin{eqnarray}
({d \over dr} \;+\;{\nu \over r}\;) \; f \; + \; ( \epsilon  \;+ \;
 \delta \; m )\; g \; = \;0 \;  ,
\nonumber
\\
({d \over dr} \; - \;{\nu \over r}\;)\; g  \;- \; ( \epsilon \; - \;
 \delta\;  m )\; f\; =\; 0 \; ,
\label{3.12a}
\end{eqnarray}

\noindent where instead of $f_{1}$  and $f_{2}$  we have employed their
linear combinations
\begin{eqnarray}
f \; = \; {f_{1} + f_{2} \over \sqrt{2}} \;, \;
g \; = \; {f_{1} - f_{2} \over i \sqrt{2}} \; ; \qquad
f_{1} ={ f + i g \over \sqrt{2}} \; , \;
f_{2} = { f - i g \over \sqrt{2}} \; .
\label{3.12b}
\end{eqnarray}

 It should be  noticed that the~above  simplification
$( \Psi _{\epsilon jm} \rightarrow  \Psi _{\epsilon jm\delta } )$  can
also  be   obtained   through   the~diagonalization of the
operator  $\hat{K}$  -- in Cartesian tetrad basis it is given in  \cite{1980-Berestetzkiy-Lifshitz-Pitaevskiy};
usually it is called the Johnson -- Lippmann operator
\cite{1950-Johnson-Lippmann(1)}; though the following (spherical tetrad-based)
form had been presented yet
in Pauli's paper \cite{1939-Pauli}; also see \cite{2005-Khachidze-Khelashvili(1), 2005-Khachidze-Khelashvili(2)}:
\begin{eqnarray}
\hat{K} \; = \;  - \gamma ^{0} \gamma ^{3} \; \Sigma _{\theta ,\phi } \;.
\nonumber
\end{eqnarray}

\noindent
Actually, from
$
\hat{K} \; \Psi _{\epsilon jm}  =
K \; \Psi _{\epsilon jm}
$
 we produce
\begin{eqnarray}
K = - \delta \; (j+1/2) \;  ,\qquad   \delta \; = \pm  1 \;, \qquad
f_{4} = \; \delta \; f_{1} \; , \qquad f_{3} = \; \delta \;  f_{2}  \; ,
\nonumber
\end{eqnarray}

\noindent which coincides    with (\ref{3.10}).

Everything established above for the flat space-time  model  can
be readily generalized  into  an arbitrary  curved  space-time  with
a spherical  metric $g_{\alpha \beta }(x)$
\begin{eqnarray}
dS^{2} =  e^{\nu }  dt^{2}  -  e^{\mu }  dr^{2}  -
r^{2}  (d\theta ^{2} + \sin ^{2}\theta  d\phi ^{2})\;  ,
\nonumber
\end{eqnarray}

\noindent and  its naturally corresponding diagonal tetrad
 $e^{\alpha }_{(a)}(x)$
\begin{eqnarray}
e^{\beta }_{(0)} = ( e^{-\nu /2}, 0 , 0 , 0) \; , \qquad
e^{\beta }_{(3)} = (0 , e^{-\mu /2}, 0 , 0 ) \; ,
\nonumber
\\
e^{\beta }_{(1)} = (0, 0 , {1 \over r} , 0) \; , \qquad
e^{\beta }_{(2)} = (0, 0 , 0 , {1 \over r \sin \theta } ) \; .
\nonumber
\end{eqnarray}

\noindent The  Dirac equation can be specified for an~arbitrary diagonal tetrad as follows
\begin{eqnarray}
 [\;i\; \gamma ^{a}\; ( e^{\beta }_{(a)} \; \partial_{\beta } \;
+ \;{1 \over 2}\; e^{\beta }_{(a);\beta } ) \;  - \;  m\;  ] \; \Psi(x) = 0 \; ,
\nonumber
\\
e^{\beta }_{(a);\beta } \; = \; { 1 \over \sqrt{ - \det g }}\; {\partial \over
\partial x^{\beta}}  \;  \sqrt{-\det g } \; e^{\beta }_{(a)}\; .
\nonumber
\end{eqnarray}

\noindent So, for the function $\Phi (x)$ defined by
\begin{eqnarray}
\Psi (t,r,\theta ,\phi ) = \exp (-{1 \over 4} (\nu +\mu )) \; {1 \over r} \;
\Phi  (t,r,\theta ,\phi )
\nonumber
\end{eqnarray}

\noindent
 we  produce the equation
\begin{eqnarray}
 [\; i \; \gamma ^{0} \; e^{-\nu /2} \; \partial_{t} \; +\;
i \; \gamma ^{3}\; e^{-\mu /2} \; \partial_{r} \; + \;
 {1 \over r} \; \Sigma _{\theta ,\phi }\;  -\;  m \;
 ] \; \Phi  (t,r,\theta ,\phi )= 0\; .
\label{3.13b}
\end{eqnarray}

\noindent
On comparing (\ref{3.13b}) with (\ref{3.5a}), it follows immediately       that
all the calculations carried out above for the flat space-time case are
still valid only with some evident modifications. Thus,
\begin{eqnarray}
\Phi _{jm\delta }(x)  =
\left | \begin{array}{rrrr}
  f_{1}(r,t) \; D_{-1/2}(\theta ,\phi ,0) \\[1mm]
  f_{2}(r,t) \; D_{+1/2}(\theta ,\phi ,0) \\[1mm]
  \delta \;  f_{2}(r,t) \; D_{-1/2}(\theta ,\phi ,0)   \\[1mm]
  \delta \;  f_{1}(r,t) \; D_{+1/2}(\theta ,\phi ,0)
\end{array} \right |
\label{3.14a}
\end{eqnarray}

\noindent and instead of (\ref{3.12a}) now we find
\begin{eqnarray}
( e^{-\mu /2} {\partial \over \partial r} + {\nu \over r} ) f \; + \;
( ie^{-\nu /2} {\partial \over \partial t}  +   \delta  m )\;  g  = 0 \; ,
\nonumber
\\
( e^{-\mu /2} {d \over dr} - {\nu \over r} ) g  \; - \;
 ( ie^{-\nu /2} {\partial \over \partial t}   - \delta   m )\; f\;   = 0 \; .
\label{3.14b}
\end{eqnarray}

\section{  About electron functions
                  in the  monopole field}

In  the~literature,  a particle-monopole   system   has
attracted a~lot of attention being in a sense a 'classical' problem:

\begin{quotation}

Dirac \cite{1931-Dirac},
Tamm \cite{1931-Tamm},
Gro\"{o}nblom\cite{1935-Groonblom},
Jordan \cite{1938-Jordan},
Fierz \cite{1944-Fierz},
Banderet \cite{1946-Banderet},
Harish-Chandra \cite{1948-Harish-Chandra},
Wilson \cite{1949-Wilson},
 Eldridge \cite{1949-Eldridge},
  Saha \cite{1949-Saha},
Johnson -- Lippmann \cite{1950-Johnson-Lippmann(2)},
Case \cite{1950-Case},
Ramsey \cite{1958-Ramsey},
 Eliezer and Roy \cite{1962-Eliezer-Roy},
Goldhaber \cite{1965-Goldhaber},
Schwinger \cite{1968-Schwinger, 1968-Schwinger},
 Dulock and  McIntosh \cite{1966-Dulock-McIntosh, 1967-McIntosh, 1967-McIntosh-Cisneros, 1970-McIntosh-Cisneros},
 Peres \cite{1967-Peres, 1968-Peres},
Zwanziger  \cite{1968-Zwanziger(1), 1968-Zwanziger(2)},
Harst \cite{1968-Harst},
Lipkin -- Weisberger -- Peshkin  \cite{1969-Lipkin-Weisberger-Peshkin},
collection of paper edited by  Bolotovskiy and Usachev \cite{1970-Bolotovskiy-Usachev},
Zwanziger \cite{1972-Zwanziger},
Barut \cite{1972-Barut, 1973-Barut},
Magne \cite{1975-Magne},
Schwinger \cite{1975-Schwinger},
Strachev and Tomilchik \cite{1975-Strachev-Tomilchik},
Boulware et al \cite{1976-Boulware et al},
 Schwinger et al \cite{1976-Schwinger},
Goldhaber \cite{1976-Goldhaber},
Wu and  Yang \cite{1976-Wu-Yang, 1977-Wu-Yang.},
Tomilchik  et al  \cite{1977-Tomilchik-Kurochkin-Tolkachev},
Callias \cite{1977-Callias},
Kazama and Yang \cite{1977-Kazama, 1977-Kazama-Yang},
Frenkel  and Hrask$\grave{o}$ \cite{1977-Frenkel-Hrasko},
Petry \cite{1977-Petry},
Margolin and  Tomilchik \cite{1977-Margolin-Tomilchik},
Kazama et al  \cite{1977-Kazama-Yang-Goldhaber},
Goldhaber \cite{1977-Goldhaber},
Ruck and Biedenharn \cite{1978-Ruck-Biedenharn},
Kazama \cite{1978-Kazama},
Friedman and Mayer \cite{1978-Friedman-Mayer},
 Goddard  and  Olive \cite{1978-Goddard-Olive},
Tolkachev and Tomil'chik \cite{1979-Tolkachev-Tomil'chik},
Rossi \cite{1979-Rossi},
Petry \cite{1980-Petry},
Jackiw and Manton \cite{1980-Jackiw-Manton, 1980-Jackiw},
Horvathy \cite{1981-Horvathy},
Barut \cite{1981-Barut},
 Hou Bo-Yu \cite{1981-Hou Bo-Yu},
Calucci \cite{1981-Calucci},
Callan \cite{1982-Callan(1), 1982-Callan(2)},
Wilczek \cite{1982-Wilczek},
Lipkin  and Peshkin \cite{1982-Lipkin-Peshkin},
Balachandran et al \cite{1983-Balachandran-Roy-Singh},
Horvathy \cite{1983-Horvathy},
Yamagishi \cite{1983-Yamagishi},
Tolkachev et al \cite{1983-Tolkachev-Tomilchik-Schnir(1), 1983-Tolkachev-Tomilchik-Schnir(2)},
Moreira  et al  \cite{1985-Moreira-Ritter-Santos},
Fuschich  et al  \cite{1985-Fuschich-Nikitin-Susloparov.},
Sch\"{a}fer  et al  \cite{1985-Schafer-Muller-Greiner},
Bose \cite{1986-Bose},
Lipkin and Peshkin \cite{1986-Lipkin-Peshkin},
Martinez \cite{1987-Martinez},
Feher \cite{1987-Feher},
Mladenov \cite{1988-Mladenov},
Tolkachev et al \cite{1988-Tolkachev-Tomil'chik-Shnir},
Gal'tsov and  Ershov \cite{1988-Gal'tsov-Ershov},
Red'kov \cite{1988-Red'kov(1), 1988-Red'kov(2)},
Savinkov  et al
\cite{1988-Savinkov(1), 1988-Savinkov(2), 1988-Savinkov-Shapiro, 1988-Savinkov-Ryzhov},
Frampton et al \cite{1989-Frampton--Zhang Jian-Zu--Qi Yong-Chang},
Tolkachev et al  \cite{1989-Tolkachev-Tomil'chi-Schnir},
Savinkov and Ryzhov \cite{1989-Savinkov-Ryzhov},
Stahlhofen \cite{1990-Stahlhofen},
Olsen  et al \cite{1990-Olsen-Osland-Wu},
Tolkachev \cite{1991-Tolkachev},
Ryzhov and Savinkov  \cite{1991-Ryzhov-Savinkov(1), 1992-Ryzhov-Savinkov},
Labelle  et al \cite{1991-Labelle-Mayrand-Vinet},
Shnir et al  \cite{1992-Shnir-Tolkachev-Tomil'chik},
Ivanov  and Savinkov \cite{1992-Ivanov-Savinkov},
Barut et al  \cite{1993-Barut-Shnir-Tolkachev},
\cite{1994-Weinberg},
Ren \cite{1994-Ren},
Bose \cite{1994-Bose},
Sitenko \cite{1996-Sitenko, 1999-Sitenko, 2001-Sitenko},  Karat -- Schulz \cite{1997-Karat-Schulz},
Tolkachev et al  \cite{1997-Tolkachev-Tomilchik-Schir},
Red'kov \cite{1998-Red'kov(1), 1998-Red'kov(2), 1998-Red'kov(3), 1998-Red'kov(4), 1998-Red'kov(5),
1998-Red'kov(6), 2002-Red'kov},
Tolkachev et al \cite{1999-Tolkachev-Aleynikov-Tomil'chik}, Haas \cite{2002-Haas},
Nesterov -- de la Cruz \cite{2002-Nesterov-de la Cruz},
Loinger \cite{2003-Loinger},
Tokarevskaya et al  \cite{2004-Tokarevskaya-Kisel-Red'kov(1), 2004-Tokarevskaya-Kisel-Red'kov(2)}.

\end{quotation}

 In particular,  the~various
properties  of  occurring  so-called   monopole   harmonics   were
investigated in  detail. Here, we are going to look into this
problem in the~context of generalized  Pauli-Schr\"odinger  formalism
reviewed in Sections 2-3: this technique provides us with  an ideal tool to
solve many of monopole-triggered  problems.

For our further purpose  it  will  be  convenient  to  use  the Abelian
monopole  potential in Schwinger's form \cite{1975-Schwinger}:
\begin{eqnarray}
A^{a}(x) = (A^{0}, \; A^{i}) =  [\; 0 \; , \; g\;
 {(\vec{r} \times \vec{n})\;(\vec{r} \; \vec{n}) \over
r \; (r^{2} - (\vec{r} \; \vec{n})^{2}) } \; ]
\label{4.1a}
\end{eqnarray}

\noindent after translating  the 4-vector potential  $A^{\alpha }$  to  the~spherical
coordinates  and specifying $\vec{n} = (0, 0 , +1 )$ ,  we get
\begin{eqnarray}
A_{0} = 0 \; , \;\; A_{r} = 0 \; , \;\; A_{\theta } = 0\; , \;\;
A_{\phi } = g\; \cos \theta  \; .
\label{4.1b}
\end{eqnarray}

\noindent Correspondingly,  the~Dirac  equation  in  this   electromagnetic
potential takes the form
\begin{eqnarray}
\left [ i \gamma ^{0} \partial _{t} + i \gamma ^{3} (\partial _{r} +
{1 \over r}) + {1 \over r}\; \Sigma ^{k}_{\theta ,\phi } \; -\;
 mc/\hbar  \right ]\; \Psi (x) = 0 \; ,
\label{4.2a}
\end{eqnarray}

\noindent where
\begin{eqnarray}
\Sigma ^{k}_{\theta ,\phi }  =
 i \gamma ^{1} \partial _{\theta}   +
\gamma ^{2} \;  { i \partial _{\phi } + (i\sigma ^{12} - k )
 \cos \theta \over  \sin  \theta}\; ,
 \label{4.2b}
\end{eqnarray}

\noindent and $k \equiv  eg/hc$.
As readily verified, the~wave  operator
  in  (\ref{4.2a}) commutes with the~following three ones
\begin{eqnarray}
J^{k}_{1} = l_{1} +  (i\sigma ^{12} - k)\; {
 \cos \phi  \over \sin \theta } \; ,
 \nonumber
 \\
J^{k}_{2} =  \; l_{2} +  (i\sigma ^{12} - k) \; {
\sin \phi  \over \sin \theta } \;  , \;\;
 J^{k}_{3} = l_{3}
\label{4.3a}
\end{eqnarray}

\noindent which  in  turn    obey   the~$SU(2)$   Lie   algebra.   Clearly,
this    monopole     situation     come  entirely  under
the~Schr\"{o}dinger-Pauli approach, so that our  further  work  will  be
a~matter of quite elementary calculations.

Corresponding to diagonalization of the
 $\vec{J}^{2}_{k}$ and $J^{k}_{3}$,
the~function $\Psi$ is to be taken as
\begin{eqnarray}
\Psi ^{k}_{\epsilon jm} (t,r,\theta ,\phi ) = {e^{-i\epsilon t} \over  r} \;
\left | \begin{array}{r}
       f_{1} \; D_{k-1/2}   \\   f_{2} \; D_{k+1/2}   \\
       f_{3} \; D_{k-1/2}   \\   f_{4} \; D_{k+1/2}
\end{array} \right |\; ;
\label{4.3b}
\end{eqnarray}

\noindent  $D_{\sigma } \equiv D^{j}_{-m,\sigma }(\phi ,\theta ,0)$.
Further, noting recursive relations \cite{1974-Warshalovich-Moskalev-Hersonskii}
\begin{eqnarray}
\partial_{\theta} \;  D_{k+1/2} = (+ a \; D_{k-1/2} - b \; D_{k+3/2} )
\; ,\;
\partial_{\theta} \;  D_{k-1/2} = (+ c \; D_{k-3/2} - a \; D_{k+1/2} )\; ,
\nonumber
\\
\sin^{-1} \theta \;  [\; -m -(k+1/2) \cos \theta \;] \; D_{k+1/2} =
(- a \; D_{k-1/2} - b \; D_{k+3/2} )  \; ,
\nonumber
\\
\sin^{-1} \theta \; [\; -m -(k-1/2)\cos \theta \;] \; D_{k-1/2} =
(- c \; D_{k-3/2} - a \; D_{k+1/2} ) \; ,
\nonumber
\\
a = {1 \over 2} \sqrt{(j + 1/2)^{2} - k^{2}} \; , \;
b = {1 \over 2} \sqrt{(j - k - 1/2)(j + k + 3/2)}\; , \;
\nonumber
\\
c = {1 \over 2} \sqrt{(j + k - 1/2)(j - k + 3/2)}
\nonumber
\end{eqnarray}

\noindent
we find how the $\Sigma ^{k}_{\theta ,\phi }$  acts on $\Psi $:
\begin{eqnarray}
\Sigma ^{k}_{\theta ,\phi } \; \Psi ^{k}_{\epsilon jm} =
 i \;\; \sqrt{(j + 1/2)^{2} - k^{2}} \;\; {e^{-i\epsilon t} \over r} \;
\left | \begin{array}{r}
  - f_{4} \; D_{k-1/2}  \\  + f_{3} \; D_{k+1/2} \\
  + f_{2} \; D_{k-1/2}  \\  - f_{1} \; D_{k+1/2}
\end{array} \right | \; ,
\label{4.4}
\end{eqnarray}

\noindent hereafter the factor $\sqrt{(j + 1/2)^{2}- k^{2}}$
 will be denoted by $\nu $. For the  $f_{i}(r)$  we establish four equations
\begin{eqnarray}
\epsilon \; f_{3} \; - \; i \; {d\over dr} \; f_{3}  \;- \;i\; {\nu \over r}\; f_{4}\;
 - \; m \; f_{1} = 0 \; ,
\nonumber
\\
 \epsilon \; f_{4} \; + \; i\;  {d \over dr}\;  f_{4}\; + \;
 i \;{\nu \over r} \; f_{3} \; - \; m \; f_{2} = 0 \; ,
\nonumber
\\
\epsilon \; f_{1} \; +\; i\; {d \over dr} \; f_{1} \;  + \;
i \; {\nu \over r} \; f_{2} \;- \;m\; f_{3} = 0 \; ,
\nonumber
\\
\epsilon \; f_{2}\; - \;i\; {d \over dr}\; f_{2}\; - \;
i \;{\nu \over r}\; f_{1}\; -\; m \;f_{4} = 0 \; .
\label{4.5}
\end{eqnarray}

As evidenced by analogy with  {\bf Sec. 3}   and  also  on
direct  calculation,   yet other     operator  can  be
diagonalized together with $\{ i \partial _{t} \;, \;  \vec{J}^{\;2}_{k}\;,
\;J^{k}_{3} \}$:  namely, a~generalized Dirac operator
\begin{eqnarray}
\hat{K} ^{k} \; = - \; i \; \gamma^{0} \; \gamma ^{3} \;
\Sigma ^{k}_{\theta ,\phi }   \; .
\label{4.6a}
\end{eqnarray}

\noindent From the  equation $\hat{K}^{k} \Psi _{\epsilon jm}  =
K \; \Psi _{\epsilon jm}$    we  can  produce  two
possible values for this~$K$  and the corresponding limitations on $f_{i}(r)$:
\begin{eqnarray}
K = - \delta \;  \sqrt{(j + 1/2)^{2}- k^{2}} \; : \qquad
f_{4} = \delta \;  f_{1} \; , \;\;\; f_{3} = \delta \; f_{2}
\label{4.6b}
\end{eqnarray}

\noindent the~system  (\ref{4.5}) is reduced to
\begin{eqnarray}
({d \over dr} + {\nu \over r}) f \; + \; (\epsilon  + \delta\;  m )\;  g = 0\; ,
\nonumber
\\
({d \over dr} - {\nu \over r}) g \; - \; (\epsilon  - \delta\;  m ) \; f = 0\;.
\label{4.7}
\end{eqnarray}

\noindent On direct comparing  (\ref{4.7})  with analogous system in  {\bf Sec. 3},
we can conclude that these two systems are formally similar apart  from
the difference  between $\nu  = j +1/2$  and $\nu  =
 \sqrt{(j + 1/2)^{2}- k^{2}}$.

Now let us pass over to quantization of  $k = eg/hc$  and $j$.  As
a~direct result  from   the~first  Pauli condition we derive
\begin{eqnarray}
{eg \over hc} = \pm 1/2 , \; \pm 1, \; \pm 3/2, \ldots
\label{4.8a}
\end{eqnarray}

\noindent which coincides with the Dirac's  quantization,  and  from
the~second Pauli condition it follows immediately that
\begin{eqnarray}
j = \mid k \mid  -1/2, \mid k \mid +1/2, \mid k \mid +3/2,\ldots
\label{4.8b}
\end{eqnarray}

The case of minimal allowable value $j_{min.}= \mid k \mid - 1/2$  must  be
separated out and looked  in a special way.  For example, let
$k = +1/2$, then to the minimal value $j = 0$ there corresponds the~wave
function in terms of solely $(t,r)$-dependent quantities
\begin{eqnarray}
\Psi ^{(j=0)}_{k = +1/2}(x) = { e^{-i\epsilon t} \over  r}
\left | \begin{array}{l}
           f_{1}(r)  \\   0  \\  f_{3}(r)  \\  0
\end{array} \right |   \; .
\label{4.9a}
\end{eqnarray}

\noindent At $k = - 1/2$, in an analogous way,  we have
\begin{eqnarray}
\Psi ^{(j =0)}_{k = -1/2}(x) =
{e^{-i\epsilon t} \over  r}
\left | \begin{array}{l}
   0  \\  f_{2}(r)  \\   0   \\  f_{4}(r)
\end{array} \right |       \;          .
\label{4.9b}
\end{eqnarray}

Thus, if $k = \pm  1/2$, then to the minimal allowed values
 $j_{\min }$
there correspond the function substitutions which do not depend at
all on the angular variables $(\theta ,\phi )$; at
 this point there exists some
formal analogy between  these  electron-monopole  states  and
$S$-states (with $l = 0 $) for a~boson field of spin zero:
$\Phi _{l=0} = \Phi (r,t)$. However, it would be unwise to attach too much
significance
to this formal coincidence  because that $(\theta ,\phi )$-independence
of $(e-g)$-states  is  not the~fact  invariant   under   tetrad   gauge
transformations. In contrast, the relation below (let $k = +1/2)$
\begin{eqnarray}
\Sigma^{+1/2}_{\theta ,\phi }  \Psi ^{(j=0)}_{k=+1/2} (x)  =
\gamma ^{2}  \cot \theta \; ( i \sigma ^{12} - 1/2 ) \Psi ^{(j =0)} _{k=+1/2} \equiv  0
\label{4.10a}
\end{eqnarray}

\noindent is invariant under any gauge transformations. The identity (4.10a)
holds because all the zeros in the
$\Psi ^{(j=0)}_{k=+1/2}$ are adjusted to the non-zeros in
$( i \sigma ^{12} - 1/2 )$;  the  non-vanishing  constituents  in $\Psi ^{(j=0)}_{k=+1/2}$
are canceled out by zeros in $( i \sigma ^{12}- 1/2 )$.
Correspondingly, the~matter equation (\ref{4.2a}) takes on the form
\begin{eqnarray}
\left [\;  i \; \gamma ^{0} \; \partial_{t} \; +
\; i\; \gamma ^{3} \; (\partial_{r}
\; + \;  {1 \over r}\;  )\; - \;  m  \; \right ] \; \Psi ^{(j=0)} = 0\; .
\label{4.10b}
\end{eqnarray}

It is readily  verified  that  both (4.9a)  and (\ref{4.9b}) representations are
directly extended to $(e-g)$-states  with $j = j_{\min }$ at all  the other
$k =\pm 1, \pm 3/2, \ldots $ Indeed,
\begin{eqnarray}
k= +1, +3/2, +2,\ldots : \qquad
\Psi ^{k > 0} _{j_{min.}} (x) = {e^{-i\epsilon t} \over r}
\left | \begin{array}{l}
   f_{1}(r) \; D_{k-1/2}  \\  0  \\  f_{3}(r) \;  D_{k-1/2} \\  0
\end{array} \right | \; ;
\nonumber
\\
k = -1, -3/2,-2,\ldots : \qquad
\Psi ^{k<0} _{j_{min.}} (x) = { e^{-i\epsilon t} \over  r}
\left | \begin{array}{l}
    0    \\   f_{2}(r) \; D_{k+1/2}  \\  0  \\ f_{4}(r) \; D_{k+1/2}
\end{array} \right | \qquad
\label{4.11b}
\end{eqnarray}

\noindent and, as can be shown,  the relation
$\Sigma _{\theta ,\phi } \Psi _{j_{\min }} = 0 $ still  holds.
For instance, let us consider in more detail  the case of positive $k$.
Using the recursive relations
\begin{eqnarray}
\partial _{\theta } D_{k-1/2} =
 { 1 \over 2} \sqrt{ 2k-1}  D_{k-3/2}\; ,
\nonumber
\\
\sin^{-1} \theta [ - m - (k-1/2) \cos \theta ]  D_{k - 1/2}  =
 - { 1 \over 2} \sqrt{ 2k -1} D_{k-3/2} \; ,
\nonumber
\end{eqnarray}

\noindent we get
\begin{eqnarray}
i\gamma ^{1} \; \partial _{\theta}
\left | \begin{array}{c}
       f_{1}(r) \; D_{k-1/2} \\  0  \\  f_{3}(r) \; D_{k-1/2}  \\  0
\end{array} \right | = {i\over 2} \sqrt{2k-1} \;
\left | \begin{array}{c}
     0  \\ - f_{3}(r) \; D_{k-3/2}  \\ 0  \\ + f_{1}(r)\; D_{k-3/2}
\end{array} \right | \; ;
\nonumber
\\
\gamma ^{2}  \; {i\partial _{\phi } + (i\sigma ^{12} - k) \cos \theta \over
\sin \theta} \;
\left | \begin{array}{c}
      f_{1}(r) \; D_{k-1/2} \\  0  \\  f_{3}(r) \; D_{k-1/2} \\ 0
\end{array} \right | \; = \;
{i \over 2} \sqrt{2k-1} \;
\left | \begin{array}{c}
    0 \\ +f_{3}(r) \; D_{k-3/2}  \\ 0  \\ -f_{1}(r) \; D_{k-3/2}
\end{array} \right |
\nonumber
\end{eqnarray}

\noindent in a~sequence, the identity
$\Sigma _{\theta ,\phi } \; \Psi _{j_{\min }} \equiv  0$  is proved.
The  case of negative $k$ can be considered in the same way.
Thus, at every $k$, the $j_{\min }$-state's equation  has  the  same
unique form
\begin{eqnarray}
\left [ \; i\; \gamma ^{0} \; \partial_{t} \; + \; i\gamma ^{3} \;
(\partial_{r}\; + \; {1 \over r}\; ) \;  - \;  mc/\hbar \;
\right ] \; \Psi _{j_{mi}} = 0
\label{4.11c}
\end{eqnarray}

\noindent which leads to the same unique radial system:

\vspace{2mm}
$
\underline{k = +1/2,+1,\ldots \;  }
$
\begin{eqnarray}
\epsilon \; f_{3} - i \; { d\over dr}  \; f_{3}  - m \; f_{1} = 0\; ,
\;\; \epsilon \; f_{1} + i \; { d \over dr} \; f_{1}  - m \; f_{3} = 0 \; ;
\nonumber
\end{eqnarray}

$
\underline{k = -1/2,-1,\ldots \; }
$
\begin{eqnarray}
\epsilon \; f_{4} + i \; { d\over dr}\; f_{4} - m \;f_{2} = 0 \; ,
\;\; \epsilon \; f_{2} - i \; { d\over dr}\; f_{2} - m \;f_{4} = 0 \; .
\label{4.12b}
\end{eqnarray}

\noindent These equations are equivalent respectively to

\vspace{3mm}
$
\underline{k = + 1/2,+ 1,\ldots  }
$
\begin{eqnarray}
( {d^{2} \over dr^{2}}  + \epsilon ^{2}  - m^{2}  )\; f_{1}  = 0 \;
, \;\; f_{3} =  { 1 \over m} ( \epsilon  +
i { d \over dr } ) \; f_{1}     \; ;
\nonumber
\end{eqnarray}

$
\underline{k = - 1/2, - 1,\ldots  }
$
\begin{eqnarray}
( {d^{2} \over dr^{2}}  + \epsilon ^{2}  - m^{2}  ) \; f_{4} = 0\;
 , \;\; f_{2} = {1 \over m}   ( \epsilon  + i {d \over dr}  )
 \; f_{4}
\label{4.13b}
\end{eqnarray}

\noindent which both end up  with the function
 \begin{eqnarray}
 f (t,r) = e^{  \pm  \sqrt{m^{2} - \epsilon ^{2}} \; r } \;,
 \nonumber
 \end{eqnarray}

\noindent
one of these  at $\epsilon \; < \; m$ looks as
\begin{eqnarray}
f(t,r)  = e^{-\sqrt{m^{2} -\epsilon ^{2}} \; r } \; .
\label{4.13c}
\end{eqnarray}

\noindent The function given by (\ref{4.13c}) which seems  to  be appropriate to describe   a~bound   state   in
the~electron-monopole system. It  should  be  emphasized  that
today the $j_{\min }$   bound  state  problem   remains   still  yet
a~question to understand.  In  particular,  the  important  question
is of finding a~physical and mathematical   criterion  on
selecting values for $\epsilon $: whether $\epsilon \; < \; m$, or
$\epsilon  = m$, or $\epsilon \; > \; m$;
and what value of $\epsilon $  is to be chosen  after specifying an~interval
above.

Now let us  proceed with studying the  properties  which
stem  from  the $\theta ,\phi $-dependence  of  the  wave   functions.
In particular, we restrict ourselves to  the $P$-parity
problem  in  the  presence  of  the  monopole.  This  problem  was
investigated in the literature both in Abelian and non-Abelian cases:
Frampton et al  \cite{1989-Frampton--Zhang Jian-Zu--Qi Yong-Chang},
Tolkachev et al \cite{1988-Tolkachev-Tomil'chik-Shnir, 1989-Tolkachev-Tomil'chi-Schnir,
1997-Tolkachev-Tomilchik-Schir, 1992-Shnir-Tolkachev-Tomil'chik, 1993-Barut-Shnir-Tolkachev}
Ryzov and Savinkov  et al \cite{1991-Ryzhov-Savinkov(1), 1992-Ryzhov-Savinkov, 1988-Savinkov(1),
1988-Savinkov(2), 1988-Savinkov-Shapiro, 1988-Savinkov-Ryzhov, 1989-Savinkov-Ryzhov,
1992-Ivanov-Savinkov}, Red'kov
 \cite{1988-Red'kov(1), 1988-Red'kov(2), 1998-Red'kov(1), 1998-Red'kov(2), 1998-Red'kov(3),
 1998-Red'kov(4), 1998-Red'kov(5), 1998-Red'kov(6), 2002-Red'kov},
so our first  step is to particularize some relevant facts in accordance with the
formalism and notation used in  the present treatment.

As evidenced by straightforward computation,  the  well-known
purely  geometrical  bispinor $P$-reflection  operator  does not
commute with the Hamiltonian $\hat{H}$  under  consideration. The same
conclusion  is also arrived at by  attempt to solve  directly  the
proper value equation
\begin{eqnarray}
\;\hat{\Pi}_{sph.} \; \Psi ^{k} _{\epsilon jm} =
 \Pi \; \Psi ^{k}_{\epsilon jm}\;
\nonumber
\end{eqnarray}

\noindent which leads to
\begin{eqnarray}
(-1)^{j+1} \;
\left | \begin{array}{l}
   f_{4} \; D_{-k-1/2} \\   f_{3} \; D_{-k+1/2} \\
   f_{2} \; D_{-k-1/2} \\   f_{1} \; D_{-k+1/2}
\end{array} \right | \; = \;  P \;
\left | \begin{array}{l}
   f_{1} \; D_{k-1/2}   \\  f_{2} \; D_{k+1/2} \\
   f_{3} \; D_{k-1/2}   \\  f_{4} D_{k+1/2}
\end{array} \right  |
\nonumber
\end{eqnarray}

\noindent the  latter  matrix  relation  is  satisfied  only  by  the~trivial
substitution $f_{i}= 0$   for  all  $i$.  The~relation above
indicates how a~required discrete transformation can be  constructed
(further we will denote it as $\hat{N}_{sph.}$)
\begin{eqnarray}
\hat{N}_{sph.} \;  =    \; \hat{\pi } \otimes
\hat{\Pi}_{sph.} \; , \qquad  \; \hat{\Pi}_{sph.} = \Pi_{sph.} \otimes  \hat{P}
\label{4.14}
\end{eqnarray}

\noindent where $\hat{\pi }$  is a~special discrete operator changing
$k$  into $-k\;$ :
\begin{eqnarray}
\hat{\pi } \; F( k ) \;  = \; F (- k )\;.
\nonumber
\end{eqnarray}

\noindent
From
the equation
\begin{eqnarray}
\hat{N}_{sph.}\; \Psi ^{k}_{\epsilon jm} =
N \Psi ^{k}_{\epsilon jm}
\nonumber
\end{eqnarray}

\noindent   it follows
\begin{eqnarray}
N = \; \mu \; (-1)^{j+1} \;, \;  \mu  = \pm  1 \; ,  \qquad
f_{4} = \mu \; f_{1} ,\;\; f_{3} = \mu \; f_{2} \; .
\label{4.15a}
\end{eqnarray}

\noindent These relations are compatible with the~above  radial system --
 eqs. (\ref{4.5}) transform    into
\begin{eqnarray}
({d \over dr} + {\nu \over r})  f + (\epsilon  + \mu \;  m ) g = 0\; ,
\nonumber
\\
({d \over dr} - {\nu \over r}) g -  (\epsilon  - \mu \;  m ) f = 0 \; ,
\label{4.15b}
\end{eqnarray}

\noindent
 $f(r)$ and $g(r)$  are  already
used combinations from $f_{1}(r)$  and $f_{2}(r)$ -- see (\ref{3.12b}).

Here some additional remarks must be done. Everything just  said  about  diagonalizing  the
$\hat{N}_{sph.}$ is applied only to the cases when $j > j_{\min }$.
As  regards the~lower value of $j$, the situation turns out to be very
specific and unexpected. Actually, let $j  = 0$ then from equation
$\hat{N}_{sph.} \Psi ^{(j=0)} = N \Psi ^{(j=0)}$,
considering  the cases $k = + 1/2 $ and $-1/2$,  we get respectively
\begin{eqnarray}
\left | \begin{array}{r}
         0  \\  - f_{3} \\  0  \\ - f_{4}
\end{array} \right |  =  N
\left | \begin{array}{r}
          f_{1}\\  0 \\ f_{3} \\  0
\end{array} \right |   \;\;  , \;\;
\left | \begin{array}{r}
    -f_{4} \\  0  \\ -f_{2} \\  0
\end{array} \right | =  N
\left | \begin{array}{r}
  0 \\ f_{2} \\  0  \\ f_{4}
\end{array}  \right |     \; .
\nonumber
\end{eqnarray}

\noindent Evidently they both  have no solutions,
excluding trivially null ones (and therefore being of~no interest). Moreover,
as  may  be easily seen, in both cases a~function $\Phi (x)$,  defined by
$\hat{N}_{sph.} \; \Psi ^{(j =0)}  \equiv  \Phi (x)$,
lies outside a~fixed totality of states that are only valid as
possible  quantum  states of the system under consideration.
At greater values of this $k$, we come  to  analogous  relations:
the~equation $\hat{N}_{sph.} \; \Psi _{j_{min.}} = N \;
\Psi _{j_{min.}}$  leads to (at positive $k$ and negative $k$ respectively)
\begin{eqnarray}
(-1)^{j+1} \hspace{-2mm}
\left | \begin{array}{l}
        0  \\  f_{3} D_{k+1/2}   \\ 0 \\  f_{1} D_{k+1/2}
\end{array} \right | =  N
\left | \begin{array}{l}
         f_{1} D_{k-1/2}  \\   0 \\  f_{3} D_{k-1/2} \\  0
\end{array} \right |  \; ;
\nonumber
\\
(-1)^{j+1} \hspace{-2mm}
\left | \begin{array}{l}
   f_{4} D_{k-1/2} \\ 0  \\  f_{2} D_{k-1/2}  \\ 0
\end{array} \right |  = N
\left | \begin{array}{l}
  0 \\ f_{2} D_{k+1/2}  \\  0  \\ f_{4} D_{k+1/2}
\end{array}  \right |
\nonumber
\end{eqnarray}

\noindent and the above arguments may be repeated again.

In turn, as regards the operator $\hat{K}^{k}$, for the $j_{\min }$  states  we
get
$\hat{K}^{k} \; \Psi _{j_{min.}} =  0$;
that is, this state represents the  proper  function  of  the $\hat{K}$
with the null  proper  value.  So,  application  of  this $\hat{K}$
instead of  the $\hat{N}$   has  an  advantage  of  avoiding  the
paradoxical and puzzling situation when $\hat{N}_{sph.} \;
\Psi ^{(j_{min})} \not\in  \{ \Psi  \}$.
In a sense, this second alternative (the use  of $\hat{K}^{k}$ instead
of $\hat{N}$ at separating the variables and  constructing  the  complete
set of mutually commuting operators) gives us a possibility not to
attach great significance to the monopole discrete operator $\hat{N}$  but
to focus our  attention  solely  on  the  continual  operator $\hat{K}^{k}$.

\section{ Discrete symmetry in external monopole field and selection rules}

 It is known that the  quantum  mechanics,  when  dealing  with  some
specific operator  $\hat{A}$, implies    its  self-conjugacy
property:
$
< \Psi  \mid  \hat{A}\;\Phi >\;=\;< \hat{A}\; \Psi \mid \Phi >.
$
For  example,  the~usual  bispinor $P$-reflection
presents evidently a~self-conjugate one,  since  one  has
\begin{eqnarray}
<\Psi (\vec{r}) \mid  \gamma ^{0}  \hat{P} \Phi (\vec{r})> \; =
\int \tilde {\Psi } ^{*}(\vec{r})  \Phi (-\vec{r}) \;dV  \; ,
\nonumber
\\
< \gamma ^{0}  \hat{P}  \Psi (\vec{r}) \mid \Phi (\vec{r})>\; =
\int \tilde{\Psi }^{*}(- \vec{r}) \; \Phi (\vec{r}) \; dV   \; .
\label{b}
\end{eqnarray}

\noindent
The~$\Psi $ with over symbol $\sim$  denotes a~transposed  column-function,
that is, a~row-function; and   the~asterisk  $*$   designates  the
operation of complex conjugation.

In the presence of the external monopole field,  the~whole
situation is completely different from the above. Indeed
\begin{eqnarray}
<\psi ^{+eg}(\vec{r}) \mid  \hat{N} \; \Phi ^{+eg}(\vec{r}) >\; =
\int (\tilde {\Psi }^{+eg}(\vec{r}))^{*} \; \Phi ^{-eg}(-\vec{r}) dV
\; ,
\nonumber
\\
< \hat{N} \; \Psi ^{+eg}(\vec{r}) \mid \;\Phi ^{+eg}(\vec{r})>\; =
\int (\tilde  {\Psi }^{-eg}(\vec{r}))^{*} \; \Phi ^{+eg}(-\vec{r}) \;
dV \;
\label{b}
\end{eqnarray}

\noindent it is evident that  right-handed sides  of  these  two
equalities vary in sign at $eg$ parameter; thereby it follows  that
the~discrete operator $\hat{N}$  does not possess
a self-adjoint one.

In this connection, one must take notice  of  the  manner  in
which  the $eg$   parameter  enters   the   radial    system   for
$f_{1},\ldots, f_{4}$ : it  occurs through
 $\nu  = \sqrt {(j+1/2)^2 - \kappa^2 }$.
The latter leads to independence on $\kappa$'s sign.  Therefore,
the two distinct systems   with  the  characteristics $+eg$   and
$-eg$ respectively have their radial systems exactly identical:
\begin{eqnarray}
F^{+eg}_{s=1/2}(f_{1},\ldots, f_{4}) =
F^{-eg}_{s=1/2}(f_{1},\ldots, f_{4}) \; .
\label{5.1}
\end{eqnarray}

As an~illustration to manifestations of the non-self-adjointness  property
of  the $N$-operator,  let  us  consider a~question  concerning $P$-parity
selection rules  in  presence  of  the~monopole.   Here
Though   there  exists a seemingly appropriate operator
\begin{eqnarray}
\hat{N} = \hat{\pi } \otimes  \Pi_{sph.} \otimes
\hat{P} \; , \qquad \hat{\pi } \; \Psi ^{+eg}_{\epsilon jm\mu }(\vec{r}) =
\Psi ^{-eg}_{\epsilon jm\mu }(\vec{r}) \; ,
\nonumber
\\
\hat{N} \;  \Psi ^{eg}_{\epsilon jm\mu }(x) =
\mu \; (-1)^{j+1} \; \Psi ^{eg}_{\epsilon jm\mu }(x)
\label{5.2}
\end{eqnarray}

\noindent
but this does not allow us to  obtain
any $N$-parity selection rules.
Let us  consider  this  question  in
more detail.
A matrix element for some physical observable $\hat{G}^{0}(x)$   is  to
be
\begin{eqnarray}
\int \bar{\Psi}^{eg}_{\epsilon jm\mu }(\vec{r}) \;
\hat{G}^{0}(\vec{r})\;\Psi ^{eg}_{\epsilon j'm'\mu'}(\vec{r})\; dV
\equiv \int r^{2}dr \int f( \vec{r} ) \; d\Omega  \; .
\label{5.3}
\end{eqnarray}

\noindent
First we  examine the case $eg = 0,$ in order  to  compare  it
with the situation at $eg \neq  0$. Let us relate $f(-\vec{r})$
with $f(\vec{r})$. Considering the equality  (and the same  with
$j'm' \delta ')$
\begin{eqnarray}
\Psi ^{0}_{\epsilon jm\delta }(-\vec{r} ) =
\Pi_{sph.} \; \delta \; (-1)^{j+1} \;
  \Psi ^{0}_{\epsilon jm\delta }(\vec{r} )
\label{5.4a}
\end{eqnarray}

\noindent we  get
\begin{eqnarray}
f^{0}(-\vec{r}) = \delta  \; \delta '\; (-1)^{j+j'+1}
\bar{\Psi}^{0}_{\epsilon jm\delta }(\vec{r})  \;
 \left [\; \Pi_{sph.} \; \hat{G}^{0}(-\vec{r}) \;
  \Pi_{sph.} \; \right ]
\; \Psi ^{0}_{\epsilon j'm'\delta '}(\vec{r}) \; .
\nonumber
\end{eqnarray}

\noindent If $\hat{G}^{0}(\vec{r})$  obeys  the equation
\begin{eqnarray}
\Pi_{sph.} \; \hat{G}^{0}(-\vec{r}) \;
 \Pi_{sph.}  =
 \omega ^{0} \; \hat{G}^{0}(\vec{r} )
\label{5.4b}
\end{eqnarray}

\noindent here $\omega ^{0}$ defined to be $+1$   or   $-1$  relates  to
the  scalar  and pseudo scalar, respectively, then $f(\vec{r})$
can be brought to
\begin{eqnarray}
f^{0}(-\vec{r}) = \omega \;  \delta \; \delta '\; (-1)^{j+j'+1} \;
f^{0}(\vec{r}) \; .
\nonumber
\end{eqnarray}

The latter generates the well known $P$-parity selection rules:
\begin{eqnarray}
\int \bar{\Psi}^{0}_{\epsilon jm\mu }(r) \; \hat{G}^{0}(r) \;
\Psi ^{0}_{\epsilon j'm'\mu'}(r) \;dV
=
\nonumber
\\
=
\left [\; 1 \;+ \; \omega \;  \delta \;  \delta '\; (-1)^{j+j'+1} \; \right ]
\; \int r^2 \;  dr \; \int_{1/2} f^{0}(\vec{r}) \;  d\Omega
\label{5.4c}
\end{eqnarray}

\noindent where the $\theta,\phi$-integration is performed on a~half-sphere.

The~situation at $eg \neq  0$  is
completely different since here any  equality in the  form $(5.4a)$ does
not exist at all.  In other words, because of the absence any
correlation between $f^{eg}(\vec{r})$ and $f^{eg}(-\vec{r})$,
there is no selection rules on discrete quantum number $N$.
In  accordance  with  this,
for instance, an expectation value for the usual operator of space
coordinates $\vec{x}$  need not equal zero and one follows this
(see in  Tolkachev et al
\cite{1988-Tolkachev-Tomil'chik-Shnir,
1989-Tolkachev-Tomil'chi-Schnir,
1997-Tolkachev-Tomilchik-Schir, 1992-Shnir-Tolkachev-Tomil'chik, 1993-Barut-Shnir-Tolkachev},
Ryzhov and Savinkov et al
\cite{1991-Ryzhov-Savinkov(1), 1992-Ryzhov-Savinkov,
 1988-Savinkov(1), 1988-Savinkov(2), 1988-Savinkov-Shapiro,
1988-Savinkov-Ryzhov, 1989-Savinkov-Ryzhov, 1992-Ivanov-Savinkov}).

In the same time, from the above  it follows that there exist quite
definite correlations between $\Psi^{\pm eg}(-\vec{r})$ and
$\Psi^{\mp eg}(\vec{r})$:
\begin{eqnarray}
\Psi^{\pm eg}(-\vec{r}) = \Pi_{sph.}\;
\Psi^{\mp eg}(\vec{r}) \;\; ,
\nonumber
\\
f^{\pm eg}(-\vec{r}) =
 \omega \;  \delta \; \delta '\; (-1)^{j+j'+1} \;
f^{\mp eg}(\vec{r})  \; .
\label{5.5}
\end{eqnarray}

\noindent Those latter provide  certain indications that
in a non-Abelian (monopole-contained) model no problems with discrete
$P$-inversion-like symmetry might occur.

The above study has shown that the general outlook on
this matter which prescribes to consider a magnetic charge as
pseudo-scalar under $P$-reflection (just that interpretation is
implied by the use of additional $\pi$-transformation changing
$g$ into $-g$ and accompanying the ordinary $P$-reflection)
is not effective  one as  we
touch  relevant  selection rules.

\section{ Some technical facts on the Abelian monopole system}

Now let  us   consider
relationship between $D$-functions used above  and the
 spinor monopole harmonics. To this end one ought to
 perform two translations: from the spherical  tetrad  and
Weyl's spinor  frame in bispinor space  into
the~Cartesian  tetrad and the so-called Pauli's (bispinor) frame.
In the first place, it is convenient to accomplish those translations
for a~free electronic function; so as, in the second place, to follow this
pattern further in the monopole case.

So, subjecting the free electronic  function (spherical solution
from  Sec. 3)   to  the~local    gauge   transformation
associated with  the~tetrad change $e_{sph.} \rightarrow  e_{Cart.}$ :
\begin{eqnarray}
\Psi _{Cart.} =
\left | \begin{array}{cc}
     U^{-1}  & 0 \\ [2mm]0  & U^{-1}
\end{array} \right |   \Psi_{sph.}\; , \; U^{-1} =
\left | \begin{array}{lr}
  \cos \theta /2  e^{-i\phi /2}  &  - \sin \theta /2  e^{-i\phi /2}  \\
  \sin \theta /2  e^{+i\phi /2}  &    \cos \theta /2  e^{+i\phi /2}
\end{array} \right |
\nonumber
\end{eqnarray}

\noindent and further, taking the bispinor  frame  from  the~Weyl's one
to  the~Pauli's:
\begin{eqnarray}
\Psi ^{P.}_{Cart.} = \left | \begin{array}{c}
                 \varphi   \\ \xi
\end{array} \right  | \; , \qquad
\Psi _{Cart.} = \left | \begin{array}{c}
             \xi   \\ \eta
\end{array} \right | \; , \qquad
\varphi  = { \xi  + \eta  \over \sqrt{2}} ,\;\;
\chi     = { \xi -  \eta \over \sqrt{2}}
\nonumber
\end{eqnarray}

\noindent we get to
\begin{eqnarray}
\varphi  =  {f_{1} + f_{3} \over \sqrt{2} } \;
\left | \begin{array}{c}
    \cos \theta /2  e^{-i\phi /2}  \\
    \sin \theta /2  e^{+i\phi /2}
\end{array} \right |  D_{-1/2}  +
{ f_{2} + f_{4} \over \sqrt{2}}
\left | \begin{array}{c}
   -\sin \theta /2  e^{-i\phi /2} \\
    \cos \theta /2  e^{+i\phi /2}
\end{array} \right |   D_{+1/2} \;  ,
\nonumber
\\
\chi  =  {f_{1} - f_{3} \over \sqrt{2} } \;
\left | \begin{array}{c}
    \cos \theta /2  e^{-i\phi /2}  \\[2mm]
    \sin \theta /2   e^{+i\phi /2}
\end{array} \right |  D_{-1/2}  +
{ f_{2} - f_{4} \over \sqrt{2}}
\left | \begin{array}{c}
   -\sin \theta /2   e^{-i\phi /2} \\
    \cos \theta /2  e^{+i\phi /2}
\end{array} \right |   D_{+1/2}\; \; .
\nonumber
\label{6.1a}
\end{eqnarray}

\noindent Introducing special notation, $\chi _{+1/2}$   and $\chi _{-1/2}$,
for columns of   matrix $U^{-1}(\theta ,\phi )$, sometime they are termed as
helicity spinors:
\begin{eqnarray}
\chi _{+1/2} =
\left | \begin{array}{c}
    \cos \theta /2 \;  e^{-i\phi /2} \\[2mm]
    \sin \theta /2 \;  e^{+i\phi /2}
\end{array} \right | \; , \qquad
\chi _{-1/2} =
\left | \begin{array}{c}
          -\sin \theta /2  \; e^{-i\phi /2}  \\[2mm]
           \cos \theta /2  \; e^{+i\phi /2}
\end{array} \right | \; ,
\nonumber
\end{eqnarray}

\noindent previous formulas can be rewritten in the form
\begin{eqnarray}
\varphi  =  {f_{1} + f_{3} \over \sqrt{2} } \; \chi _{+1/2}
 \; D_{-1/2} \; +  \;
{ f_{2} + f_{4} \over \sqrt{2}} \chi _{-1/2}
 \; D_{+1/2} \;  ,
\nonumber
\\
\chi  =  {f_{1} - f_{3} \over \sqrt{2} } \;\chi _{+1/2}
 \; D_{-1/2} \; +  \;
{ f_{2} - f_{4} \over \sqrt{2}} \chi _{-1/2}
  \; D_{+1/2}\; \; .
\label{6.1c}
\end{eqnarray}

Further, for the~above solutions with fixed proper  values  of
 the operator  $\hat{\Pi}_{sph}\;$ -- see (\ref{3.10}):

\vspace{3mm}

$
\underline{\Pi = (-1)^{j+1} \; ,}
$
\begin{eqnarray}
\Psi _{Cart.} =
{e^{-i\epsilon t} \over r \sqrt{2}} \;
\left | \begin{array}{c}
(f_{1} + f_{2})\; (\; \chi _{+1/2} \; D_{-1/2}\; +\; \chi _{-1/2} \;D_{+1/2}\; ) \\[2mm]
(f_{1} - f_{2})\; (\; \chi _{+1/2} \; D_{-1/2} \;-\; \chi _{-1/2} \; D_{+1/2}\;)
\end{array} \right | \; ,
\label{6.2a}
\end{eqnarray}

$
\underline{\Pi =   (-1)^{j}\; , }
$
\begin{eqnarray}
\Psi _{Cart.} =
{ e^{-i\epsilon t} \over  r \sqrt{2}} \;
\left | \begin{array}{c}
(f_{1} - f_{2})\; (\; \chi _{+1/2} \; D_{-1/2}\; -\; \chi _{-1/2} \; D_{+1/2}\;) \\[2mm]
(f_{1} + f_{2})\; ( \chi _{+1/2} \; D_{-1/2} \;+\; \chi _{-1/2} \; D_{+1/2}\; )
\end{array} \right |\; .
\label{6.2b}
\end{eqnarray}

Now, using the  known extensions  for  spherical  spinors
$\Omega ^{j\pm 1/2}_{jm}(\theta ,\phi )$  in terms of
$\chi _{\pm 1/2}$  and  $D$-functions \cite{1974-Warshalovich-Moskalev-Hersonskii}:
\begin{eqnarray}
\Omega ^{j+1/2}_{jm} = (-1)^{m+1/2}   \sqrt{(2j+1)/8\pi} \;
(   \chi _{-1/2}  D_{+1/2}  +  \chi _{+1/2}  D_{-1/2}  ) \; ,
\nonumber
\\
\Omega ^{j-1/2}_{jm} = (-1)^{m+1/2} \sqrt{(2j+1)/8\pi} \;
(
 \chi _{-1/2}  D_{+1/2}  -   \chi _{+1/2}  D_{-1/2} )\; ,
\label{6.3a}
\end{eqnarray}

\noindent we      arrive   at   the~common   representation   of
the~spinor spherical solutions \cite{1980-Berestetzkiy-Lifshitz-Pitaevskiy}
\begin{eqnarray}
\Pi = (-1)^{j+1}\; , \qquad \Psi_{Cart.}=
{e^{-i\epsilon t} \over  r} \;
\left | \begin{array}{r}
   + f(r) \; \Omega ^{j+1/2}_{jm} (\theta ,\phi ) \\[2mm]
   - i\;g(r)\; \Omega ^{j-1/2}_{jm}(\theta ,\phi )
\end{array} \right |   \;    ;
\nonumber
\\
\Pi = (-1)^{j}\; , \qquad \Psi_{Cart.} =
{e^{-i\epsilon t} \over r} \;
\left | \begin{array}{r}
-i\; g(r) \; \Omega ^{j-1/2}_{jm} (\theta ,\phi ) \\[2mm]
     f(r) \; \Omega ^{j+1/2}_{jm}(\theta ,\phi )
\end{array} \right |    \;   .
\label{6.3b}
\end{eqnarray}

The monopole situation  can  be  considered in the same  way.
As a~result,  we  produce  the~following  representation  of
the~monopole-electron  functions  in  terms  of   `new'   angular
harmonics
\begin{eqnarray}
N = (-1)^{j+1}\; : \qquad \Psi _{Cart.} = {e^{-i\epsilon t} \over r} \;
\left | \begin{array}{r}
 + f(r) \; \xi ^{(1)}_{jmk} (\theta ,\phi ) \\[2mm]
-i \; g(r) \; \xi ^{(2)}_{jmk}(\theta ,\phi )
\end{array} \right |    \;       ;
\nonumber
\\
N = (-1)^{j} \; : \qquad \Psi _{Cart.} ={ e^{-i\epsilon t} \over  r} \;
\left | \begin{array}{r}
-i \; g(r)\; \xi ^{(1)}_{jmk}(\theta ,\phi ) \\[2mm]
+f(r) \; \xi ^{(2)}_{jmk}(\theta ,\phi )
\end{array} \right |  \; .
\label{6.4b}
\end{eqnarray}

\noindent Here, the two column functions $\xi ^{(1)}_{jmk} (\theta ,\phi )$
and   $\xi  ^{(2)}_{jmk} (\theta ,\phi )$ denote  special  combinations
of $\chi _{\pm 1/2}(\theta ,\phi )$ and
$D_{-m,eg/hc\pm 1/2}(\phi ,\theta ,0)$:
\begin{eqnarray}
\xi ^{(1)}_{jmk} =   \chi _{-1/2}\; D_{k+1/2}\; +  \;
 \chi _{+1/2} \;D_{k-1/2} \;  ,
\nonumber
\\
\xi ^{(2)}_{jmk} =  \chi _{-1/2} \; D_{k+1/2} \; - \;
\chi _{+1/2} \; D_{k-1/2}\; ;
\label{6.5}
\end{eqnarray}

\noindent compare them with  analogous  extensions (6.3a)   for
$\Omega ^{j\pm 1/2}_{jm}(\theta ,\phi )$.

These
2-component  functions
 $\xi ^{(1)}_{jmk}(\theta ,\phi)$  and $\xi ^{(2)}_{jmk}(\theta ,\phi)$
just  provide  what is called spinor  monopole   harmonics. It
should be useful to write out the detailed explicit form of these
generalized harmonics.  Given  the  known  expressions  for $\chi $-  and
$D$-functions, the formulas (\ref{6.5}) yield  the following
\begin{eqnarray}
\xi ^{(1,2)}_{jmk} =
e^{{\rm im}\phi }
\left | \begin{array}{r}
-\sin \theta /2   e^{-i\phi /2}  \\[2mm]
 \cos \theta /2   e^{+i\phi /2}
\end{array} \right | \;
 d^{j}_{-m,k+1/2} (\cos \theta )
 \nonumber
 \\
   \pm
  e^{im\phi } \;
\left | \begin{array}{c}
\cos \theta /2  e^{-i\phi /2} \\[2mm]
\sin \theta /2  e^{+i\phi /2}
\end{array} \right |
d^{j}_{-m,k-1/2} (\cos \theta ) \; ;
\label{6.6}
\end{eqnarray}

\noindent here, the signs  $+ \; (plus)$  and  $- \; (minus)$ refer  to
$\xi ^{(1)}$ and $\xi ^{(2)}$ respectively.
When $k=0$ from (6.6) it follow relations for  a pure fermion case in absence
of monopole potential.

Above, at translating the electron-monopole functions into the
Cartesian tetrad and Pauli's spin frame, we had overlooked the case
of minimal $j$. Returning to it, on  straightforward  calculation
we find (for $k\; <\; 0$ and $k\; > \; 0$ , respectively)

\vspace{3mm}

$
\underline{\mbox{positive }\;\; \kappa :}
$
\begin{eqnarray}
\Psi ^{Cart.}_{j_{min.}} \; = \; {e^{-i\epsilon t} \over \sqrt{2} r} \;
\left | \begin{array}{c}
   ( f_{1} +  f_{3}) \;  \chi _{+1/2} \\[2mm]
   ( f_{1} -  f_{3}) \;  \chi _{+1/2}
\end{array} \right | \;
D^{\mid k \mid -1/2} _{-m,k-1/2} (\theta ,\phi ,0) \; ;
\label{6.7a}
\end{eqnarray}
$
\underline{\mbox{negative} \;\; \kappa :}
$
\begin{eqnarray}
\Psi ^{Cart.}_{j_{min.}} =  {e^{-i\epsilon t} \over \sqrt{2} r} \;
\left  | \begin{array}{c}
   ( f_{2} +  f_{4}) \;  \chi _{-1/2} \\[2mm]
   ( f_{2} -  f_{4}) \;  \chi _{-1/2}
\end{array} \right | \;
D^{\mid k \mid -1/2} _{-m,k+1/2} (\theta ,\phi ,0)  \; .
\label{6.7b}
\end{eqnarray}

Concluding note: one can equally work  whether in  terms  of  monopole  harmonics
$\xi ^{(1,2)}(\theta ,\phi )$ or directly in terms of $D$-functions,
but  the  latter
alternative has an~advantage over the former because of
the~straightforward  access  to the 'unlimited'
$D$-function
apparatus; instead  of proving
and  producing   just disguised  old  results.

Now we pass on to another subject   and  take  up  demonstrating
how the major facts  obtained  so  far  are  extended  to
a~curved background geometry (of spherical symmetry).
All above,  the~flat space monopole potential
 $A_{\phi } = g \cos \theta $  preserves
its  simple form at changing the~flat  space  model  into  a~curved one
of spherical symmetry
\begin{eqnarray}
A_{\phi } = g \cos \theta  \;\; \Longrightarrow \;\;
 F_{\theta \phi } = - F_{\phi \theta } = - g \sin \theta
\nonumber
\end{eqnarray}

\noindent
and the general covariant Maxwell equation in such a curved space
yields
\begin{eqnarray}
{1\over \sqrt{-g} } {\partial \over \partial x^{\alpha }}
\sqrt{-g} \; F^{\alpha \beta } = 0 \qquad \Longrightarrow
\nonumber
\\
{\partial \over \partial \theta } \;   e^{\nu +\mu } r^{2} \;
\sin \theta  \; {- g \sin \theta  \over  r^{4} \sin ^{2}\theta  } \equiv  0 \; ,  \qquad (\theta \neq 0, \pi) \; .
\nonumber
\end{eqnarray}

So,  the  monopole
potential  (for a~curved background geometry) is given  again as
$A_{\phi } = g \cos \theta$. In a~sequence, the~problem of electron in
external monopole field (in a~curved  background)  remains,  in a~whole,
unchanged. There are only some new features  brought  about  by
curvature, but they do not affect the~$(\theta ,\phi )$-aspects of
 the~problem.
For instance, consider  the case of $j_{min}$ at $k>0$ (
the  case  $j_{min}\; ,\; k<0$  can   be
considered in  the  same  way):
\begin{eqnarray}
\kappa = +1, +3/2, +2,\ldots \; , \qquad
\Psi ^{k>0}_{j_{min.}}(x) = {1 \over r} \;
\left | \begin{array}{l}
    f_{1}(r,t) \; D_{k-1/2} \\  0  \\  f_{3}(r,t) \; D_{k-1/2} \\  0
\end{array} \right | \; ;
\label{6.8}
\end{eqnarray}

\noindent from that  it follows
\begin{eqnarray}
i e^{-\nu /2} \partial _{t}  f_{1} +
i e^{-\mu /2} \partial_{r}  f_{1} - m f_{3} = 0\;  ,
\nonumber
\\
i  e^{-\nu /2} \partial _{t} f_{3}  -
i  e^{-\mu /2} \partial_{r} f_{3}  -  m f_{1} = 0\; ,
\label{6.9}
\end{eqnarray}

\noindent and further
\begin{eqnarray}
f_{3} = {i \over m} \;  ( \; e^{-\nu /2} \; \partial_{t} \; + \;
 e^{-\mu /2} \; \partial_{r} \;  ) \; f_{1}(r,t) \; ,
\nonumber
\\
\left [  ( e^{-\nu /2}\; \partial_{t}  -
 e^{-\mu /2}  \partial _{r}  ) \;
  ( e^{-\nu /2} \partial_{t}  +  e^{-\mu /2} \partial _{r}   )
   +   m^{2}  \right ]   f_{1} = 0  \; .
\label{6.10}
\end{eqnarray}

\noindent

Finally, let us consider the question  of  gauge  choice  for
description  of  the  monopole  potential.
From  general considerations we can conclude that, for the problems
considered  above,  it  was  not  basically   essential
whether to use the Schwinger's form of the~monopole  potential  or
to use any other form. Every possible choice  could  bring  about  some
technical incidental variation in a~corresponding description,
but this will not affect the applicability  of $D$-function
apparatus to the procedure of separating out the~variables
$\theta ,\phi $  in the electron-monopole system.

 For example,  in
the~Dirac gauge the~monopole potential is given by
\begin{eqnarray}
(A_{a})^{D.}= [ \; 0 \; , \; g\; { \vec{n} \times \vec{r} \over
 r \; ( r + \vec{n} \; \vec{r})} \; ]
\label{6.11a}
\end{eqnarray}

\noindent which after translating to spherical coordinates becomes
\begin{eqnarray}
A_{\alpha }^{D.}= (\;  A_{t} = 0, A_{r} = 0 ,
A_{\theta } = 0 , A_{\phi } = g (\cos \theta  - 1)\; ) \; .
\label{6.11b}
\end{eqnarray}

\noindent On comparing $A^{D.}_{\phi }$ with $A^{S.}_{\phi}$, it follows
immediately  that we can relate these electron-monopole  pictures
 by  means of a simple gauge transformation:
\begin{eqnarray}
 S(\phi ) = e^{+ik\phi } \; , \qquad
\Psi ^{D.}(x) \; = \; S(\phi ) \; \Psi ^{S.}\; ,
\nonumber
\\
A^{D.}_{\beta }(x) \; =  \; A^{S.}_{\beta }(x) \; -  \;
i\; {\hbar c \over e}  \; S \; {\partial \over \partial x^{\beta}}  \; S^{-1} \; .
\label{6.11c}
\end{eqnarray}

\noindent Simultaneously translating the operators
 $\hat{J}^{k}_{j}, \; \hat{K}, \; \hat{N}$   from  S.-
to D.-gauge
\begin{eqnarray}
\hat{J}^{D.}_{j} = S \; J^{S.}_{j} \; S^{-1} \; , \qquad
\hat{K}^{D.} = S\; K^{S.} \; S^{-1} , \qquad
\hat{N}^{D.}= S \; \hat{N}^{S.} \; S^{-1}
\nonumber
\end{eqnarray}

\noindent we produce
\begin{eqnarray}
\hat{J}^{D.}_{1} =   l_{1} + {\cos \phi  \over \sin \theta } \;
 ( i \sigma ^{12} - k (1-\cos \theta ) )\; ,
\nonumber
\\
\hat{J}^{D.}_{2} =  l_{2} + {\sin \phi  \over \sin \theta }\;
 ( i \sigma ^{12} - k (1 - \cos \theta ))  \;, \qquad
 j^{D.}_{3}  = l_{3} - k   \; ,
\nonumber
\\
\hat{K}^{D.} = - i \; \gamma ^{0}\; \gamma ^{3}\;
  (\; i\gamma ^{1} \; \partial _{\theta } + \gamma ^{2} \;
{ i\partial _{\phi } + k + (i\sigma ^{12} - k )
\cos \theta \over \sin \theta} \;  )  \; ,
\nonumber
\\
\hat{N}^{D.} = e^{ik(2\phi + \pi )} \; \hat{N}^{S.} \;  .
\label{6.11d}
\end{eqnarray}

\noindent Thus, the explicit forms of the operators vary from  one  representation
to another, but their proper values remain unchanged;  any alterations
in operators and corresponding modifications  in  wave   functions
cancel out  each other  completely.  That is, as it certainly might expected,
the~complete set of proper values provides such a~description that is
invariant, by its implications, under  any possible $U(1)$ gauge transformations.

Now, let us  consider  else  one  variation  in $U(1)$  gauge,
namely, from  Schwinger's gauge \cite{1975-Schwinger} to  the~Wu-Yang's \cite{1975-Wu-Yang}-\cite{1976-Wu-Yang}.
In the~Wu-Yang  (hereafter, designated as (W-Y)-gauge,
the~monopole potential is characterized by two different respective
expressions  in  two complementary spatial regions
\begin{eqnarray}
0 \le  \theta  <  (\pi /2 + \epsilon ) \;\; \Longrightarrow \;\;
A^{(N)}_{\phi } = g (\cos \theta  - 1)  \; ,
\nonumber
\\
(\pi /2 - \epsilon ) <  \theta  \le  \pi \;\; \Longrightarrow \;\;
A_{\phi }(S) = g (\cos \theta  + 1) \; ,
\label{6.12a}
\end{eqnarray}

\noindent and the transition from  the $S.$-basis too $W-Y$'s  can be
obtained  by
\begin{eqnarray}
\Psi ^{S}(x)\; \Longrightarrow \; \Psi ^{W-Y}(x) \qquad \qquad
\nonumber
\\
 =
\left \{  \begin{array}{l}
\Psi ^{(N)}(x) \; = \; S^{(N)}(\phi ) \; \Psi ^{S.}(x)\;\; ,  \;
 S^{(N)}(\phi ) \; = \; e^{+ik\phi } \; ,\\[2mm]
\Psi ^{(S)}(x) \; = \; S^{(S)}(\phi ) \; \Psi ^{S.}(x) \;\; ,  \;\;
S^{(S)}(\phi ) \; = \; e^{-ik\phi} \; .
\end{array} \right.
\label{6.12b}
\end{eqnarray}

\noindent Correspondingly,
for the operators $\hat{J}^{k}_{j}, \; \hat{K},\; \hat{N}$   we     get
two different  forms in $N$- and $S$-regions, respectively:
\begin{eqnarray}
\hat{J}^{\pm }_{1} =  l_{1} + {\cos \phi  \over \sin \theta } \;
 (i\sigma ^{12} - k (1 \pm  \cos \theta )) \; ,
\nonumber
\\
\hat{J}^{\pm }_{2}\  =  l_{2} + {\sin \phi \over \sin \theta } \;
 ( i\sigma ^{12}  - k (1 \pm  \cos \theta ))  \; , \;\;
 j^{D.}_{3} = l_{3} \pm  k  \; ,
\nonumber
\\
\hat{K}^{\pm } = - i \; \gamma ^{0}\; \gamma ^{3}  \;
( \;  i \; \gamma ^{1} \partial _{\theta } + \gamma ^{2}\;
{ i\partial _{\phi } \mp k + (i\sigma ^{12} - k) \cos \theta  \over
\sin \theta }\;  ) \;  ,
\nonumber
\\
\hat{N}^{\pm } = \exp ( \mp i k (2\phi  +\pi ) )\; \hat{N}^{S.}
\label{6.12c}
\end{eqnarray}

\noindent where the over sign ($+$ or $-$ ) relates to $S.$-region,
 and  the lower one ($-$ or $+$, respectively) to $N.$-region.

It should be noted  that  only  the  Schwinger's $U(1)$  gauge,
in virtue of the relation $\hat{j}_{3} = - i \partial _{\phi } $, represents
analogue  of  the~Schr\"odinger's (tetrad) basis discussed in Sec.2,
whereas the~Dirac and Wu-Yang gauges are not. The explicit form
of the~third component of a~total conserved momentum
\begin{eqnarray}
J_{3} = - i \;\partial _{\phi } \equiv  J^{Schr.}_{3}
\nonumber
\end{eqnarray}

\noindent
can be regarded as a~determining characteristic,
which  specifies  this basis (and its possible generalizations).
The situations in  $S., \; D.$, and $W-Y$ gauges  are characterized by
\begin{eqnarray}
J^{S.}_{3} = l_{3} \; ; \qquad  J^{D.}_{3} =  l_{3} - k \; ; \qquad
J^{(N)}_{3}=  l_{3} - k \; , \;\; J^{(S)}_{3} =  l_{3} + k \; .
\label{6.13}
\end{eqnarray}

\vspace{5mm}
\begin{center}

{\bf PART III}

\end{center}

\vspace{5mm}

\section{The Dirac and Schwinger  gauges in isotopic space}

Together with  topological way of studying
monopole configurations, another approach to  mono\-poles
is possible: namely, which is based on manifestations
of mo\-no\-poles playing the role of external potentials. Moreover,
from the physical standpoint the latter method can be thought of  as
a more visualizing one in comparison with less obvious  topological language. So, the basic frame of the further
investigation is  analysis of particles in the external monopole
potentials; see also

Swank et al \cite{1975-Swank et al},
Jackiw and Rebbi \cite{1976-Jackiw-Rebbi', 1976-Jackiw-Rebbi},
Hasenfratz and Hooft \cite{1976-Hasenfratz-Hooft}
Callias \cite{1977-Callias},
Goddard and Olive \cite{1978-Goddard-Olive},
Jackiw and Manton \cite{1980-Jackiw-Manton},
Jackiw \cite{1980-Jackiw},
Proxhvatilov and  Franke \cite {1976-Proxhvatilov-Franke},
Rossi {1982-Rossi},
Blaer et al \cite{Blaer et al},
Tang Ju-Fei \cite{1982-Tang Ju-Fei},
Callan \cite{1982-Callan(1), 1982-Callan(2)},
Henneaux \cite{1982-Henneaux},
Farhi and D'Hoker \cite{1983-Farhi-D'Hoker},
Marciano and Muzinich \cite{1983-Marciano-Muzinich(1), 1983-Marciano-Muzinich(2)},
Din and  \cite{1983-Din-Roy},
Bhakuni et al \cite{1983-Bhakuni-Negi-Royput},
Tolkachev \cite{1984-Tolkachev},
Barut et al \cite{1993-Barut-Shnir-Tolkachev},
Red'kov \cite{1998-Red'kov(5), 1998-Red'kov(6), 1999-Red'kov(1), 1999-Red'kov(2)},
Volkov and Gal'tsov \cite{1999-Volkov-Gal'tsov},
Mezincescu \cite{2001-Mezincescu},
Tokarevskaya et al \cite{2004-Tokarevskaya-Kisel-Red'kov(1), 2004-Tokarevskaya-Kisel-Red'kov(2)},
Milton {2006-Milton},
Weinberg and Yi \cite{2007-Weinberg-Yi}.

It is well-known that the usual Abelian monopole potential
generates a certain non-Abelian potential being a solution of the
Yang-Mills (Y-M) equations. First, such a~specific non-Abelian
solution was found out in [19]. The procedure itself of that
embedding the Abelian monopole 4-vector $A_{\mu }(x)$
in the non-Abelian scheme:
$  A_{\mu }(x) \rightarrow  W^{(a)}_{\mu }(x)  \equiv
( 0 , 0 , A^{(3)}_{\mu }$ = $ A_{\mu }(x) )$
ensures automatically that   $W^{(a)}_{\mu }(x)$    will satisfy
the free Y-M equations. Thus, it may be readily verified that the
vector
      $A_{\mu }(x) = (0, 0, 0, A_{\phi } = g\cos\theta )$
obeys the Maxwell general covariant equations in every space-time
with  the  spherical symmetry:
\begin{eqnarray}
dS^{2}= e^{2\nu }dt^{2} - e^{2\mu }dr^{2} -
r^{2} (d\theta ^{2} +\sin^{2}\theta  d\phi ^{2}) \;,
\nonumber
\\
{1 \over \sqrt{-g} } \partial_{\alpha} \sqrt{-g} F^{\alpha \beta} = 0, \;
A_{\phi } =  g\cos\theta  \; , \;
F_{\theta \phi } = - g \sin\theta \; ;
\label{A}
\end{eqnarray}

\noindent
here we get essentially a single equation. One the same potential  $A_{\phi } =  g\cos\theta$ describes
the Abelian  monopole in arbitrary spherically-symmetric space-time.

In turn, the non-Abelian  strength tensor   $F^{a}_{\mu \nu }(x)$
associated with the $A^{(a)}_{\mu }$ above  has a  very  simple  isotopic
structure:  $ F^{(3)}_{\theta \phi } = - g \sin\theta $   and  all
other   $F^{(a)}_{\nu \mu }$    are  equal  to
zero. So, this substitution
$
F^{(a)}_{\nu \mu } =
( 0 , 0, F^{(3)}_{\theta \phi } = - g \sin \theta)
$
leads the Y-M equations to the single equation of  the  Abelian case.
 Thus, this monopole potential
may be interpreted as a trivially non-Abelian solution  of Y-M
equations.
Supposing that such a sub-potential is presented in the
well-known monopole solutions of t'Hooft-Polyakov,
 we will  establish explicitly that constituent structure.

The well-known form  of the monopole solution  (\ref{1.7})  may  be  taken
as a starting point
 The field $W^{(a)}_{\alpha }$  represents a~covariant vector
with the usual transformation law, and our first step is a~change of
variables in 3-space, so let us replace $x_{i}$  by the spherical coordinates
$(r, \theta , \phi )$. Thus, the given  potentials
$ (W^{(a)}_{\alpha })$
convert into
$ ( W^{(a)}_{t}, W^{(a)}_{r}$,
$W^{(a)}_{\theta }, W^{(a)}_{\phi})$.

Our  second  step  will be   a~special gauge transformation in the isotopic space.
A~required gauge matrix can be determined  by the condition
$$
(0_{ab} \Phi ^{b}(x) ) = ( 0 , 0 , r \Phi (r) ).
$$
This equation has a~set of solutions since the isotopic rotation
by every angle about the third axis  ( 0, 0 , 1 )  will not change
the~finishing vector $( 0, 0, r \Phi (r) )$.
We shall seek to fix such
an~ambiguity by deciding in favor of the  simplest  transformation
matrix. It will be convenient to utilize  the known group $SO(3.R)$
parametrization through the Gibbs  3-vector: see Fedorov \cite{1979-Fedorov}):
\begin{eqnarray}
O( {\bf c}) =  I  +2\; {  [   {\bf c} ^{\times } +
                    ( {\bf c}^{\times })^2 ] \over  1 +  {\bf c} ^{2} }
\;, \qquad
({\bf c} ^{\times})_{ij} = - \epsilon _{ijk} c_{k}\;.
\nonumber
\end{eqnarray}

\noindent
The simplest rotation above  is characterized by
\begin{eqnarray}
{\bf D} = O( {\bf c}) {\bf B} \;  ,   \qquad
{\bf c} = [ {\bf D} {\bf B} ] / ( {\bf B}+ {\bf D} ) {\bf B} \; ,
\nonumber
\\
{\bf B} = r \Phi \; \vec{n}_{\theta, \phi} \; ,  \;\;
{\bf D} = r \Phi \; ( 0 , 0 , 1 )  \; ,
\nonumber
\\
{\bf c} = {{ \sin \theta } \over {1 + \cos \theta } }
(  \sin \phi , - \cos \phi , 0 )   \; .
\label{A.5}
\end{eqnarray}

 Together with  varying  scalar  field $\Phi ^{a}(x)$  ,  the  vector
triplet $W^{(a)}_{\alpha }(x)$  is to be transformed from one isotopic
gauge to another under the  law  [...]
\begin{eqnarray}
W'^{(a)}_{\alpha }  =  O_{ab}({\bf c} (x)) \;  W^{(b)}_{\alpha }   +
{ 1\over e} \; \Delta_{ab}( {\bf c}(x))  \; {{ \partial c_{b}} \over {\partial x^{\alpha}}} \; ,
\nonumber
\\
 \Delta ({\bf c}) = - 2 \; { 1 + {\bf c}^{\times}  \over 1 + {\bf c}^2 }  \; .
\label{A.5'}
\end{eqnarray}

For definiteness let specify calculation the spherical space, in which we  need hyper spherical
coordinates  $y^{\alpha} = (x^{0}, \chi, \theta, \phi) $:
\begin{eqnarray}
x^{i} = 2 \tan { \chi \over 2}  \;  n_{i}  \; , \qquad
n_{i} = (\sin \theta \cos \phi \; , \; \sin \theta \sin \phi \; , \;
\cos \theta ) \; .
\nonumber
\end{eqnarray}

\noindent With the use of tensor law
$W^{b}_{\alpha} (y) = (\partial x^{\beta}  /   \partial y^{\alpha} ) \; W^{b}_{\beta} (x),
$
 starting from (\ref{1.7}), we obtain a  hyper spherical representation for the dyon substitution
 (note that $r = 2 \tan {\chi \over 2}$):
\begin{eqnarray}
\Phi^{b} =  \Phi (r )   \; r
\left | \begin{array}{c}
n_{1} \\ n_{2} \\ n_{3}
\end{array} \right | \; , \qquad
W^{b}_{0}  =  f(r)  \; r
\left | \begin{array}{c}
n_{1} \\ n_{2} \\ n_{3}
\end{array} \right | \; , \qquad
W^{b}_{\chi}  =  \left | \begin{array}{c}
0 \\ 0 \\ 0
\end{array} \right | \; ,
\nonumber
\\
W^{b}_{\theta}  =  K(r)   r^{2}
\left | \begin{array}{c}
-\sin \phi  \\ + \cos \phi \\  0
\end{array} \right | \; , \qquad
W^{b}_{\phi}  =  K(r)  r^{2}
\left | \begin{array}{c}
- \sin \theta \cos \theta \cos \phi  \\ - \sin \theta \cos \theta \sin \phi \\  0
\end{array} \right | \; ,
\label{A.7}
\end{eqnarray}

\noindent from which with the use of (\ref{A.5'}) we arrive at
\begin{eqnarray}
\Phi^{'b} =  \Phi (r )   \; r
\left | \begin{array}{c}
 0 \\  0 \\ 1
\end{array} \right |  , \;
W^{'b}_{0}  =  f(r )   \; r
\left | \begin{array}{c}
0 \\ 0 \\ 1
\end{array} \right |  , \; W^{'b}_{\chi}  =  \left | \begin{array}{c}
0 \\ 0 \\ 0
\end{array} \right |  ,
\nonumber
\\
 W'^{(b)} _{\theta } = (r^2 K(r) + 1 / e )
 \left | \begin{array}{c}
                      - \sin \phi  \\
                      + \cos \phi  \\
                           0
 \end{array} \right | ,
 \nonumber
 \\     W'^{(b)}_{ \phi} =
 \left | \begin{array}{cc}
            -(  r^{2}  K(r) + 1/e ) \sin \theta \cos \phi  \\
            -( r^{2}  K(r) + 1/e ) \sin \theta \sin \phi  \\
             {1 \over e } ( \cos \theta - 1 ) \; .
\end{array} \right |
\label{A.8}
\end{eqnarray}

\noindent   The factor $(r^{2} K + {1 / e})$
 vanishes when $K = - 1 / e r^{2}$. In other  words, only  the
delicate fitting of single proportional coefficient  results in the actual
formal  simplification  of the non-Abelian monopole potential (\ref{A.8}).

There exists a close link between $W^{(a)}_{\phi }$  from (\ref{A.8})  and  the
Dirac's expression for the Abelian monopole potential
(supposing $\vec{n} = (0,0,1)$)
\begin{eqnarray}
  {\bf A}^{D.} = g {{ [ {\bf n} {\bf r} ]} \over
{( r  + {\bf r} {\bf n} ) r}} \;\; ,  \;\;  \mbox{or} \;\;
A^{D.}_{\phi } =  g ( \cos \theta  - 1 ) \; ,
\label{A.9}
\end{eqnarray}

\noindent so that  $W^{triv.}_{(a) \alpha }(x)$   from
can be thought of  as a~result of embedding the Abelian  potential
in the non-Abelian gauge scheme:
$W^{(a)D.}_{\alpha }(x) = ( 0 , 0 , A^{D.}_{\alpha }(x) )$.
The quantity $W^{(a)D.}_{\alpha }(x)$  labelled with symbol $D.$  will be
named after its Abelian counterpart; that is, this potential can
be treated as relating to the Dirac's non-Abelian   gauge  in  the
isotopic space. In Abelian  case,  the  Dirac's  potential
$A^{D.}_{\alpha }(x)$ can  be converted into the Schwinger's  form:
\begin{eqnarray}
{\bf A}^{S.} =  g{{ [ {\bf r} {\bf n} ](  {\bf r} {\bf n} )}\over
{(r^2 - ({\bf r} {\bf n})^2 )}} \;\; , \;\;  \mbox{or} \;\;
A^{S.}_{\phi } = g  \cos \theta     \; .
\label{A.10}
\end{eqnarray}

\noindent It is  possible to draw an analogy between Abelian and
non-Abelian models. Thus, we may  introduce the Schwinger's
non-Abelian basis in the isotopic space:
\begin{eqnarray}
( \Phi ^{D.(a)} , W^{D.(a)}_{\alpha } )
       \qquad          \Longrightarrow     \qquad
( \Phi ^{S.(a)} , W^{S.(a)}_{\alpha } ) \; ,
\nonumber
\end{eqnarray}

\noindent with   $ {\bf c}' = ( 0 , 0 , - \tan \phi /2 )$.
Now an explicit form of the $\theta$ and$\phi$ components of the  monopole potential is given by
\begin{eqnarray}
                     W^{S.(a)}_{\theta }  =
\left | \begin{array}{c}
                      0  \\  r^{2} K  + 1 /e  \\ 0
\end{array} \right |  ;               \qquad
                     W^{S.(a)}_{\phi }  =
\left | \begin{array}{c}
                    -  ( r^{2}  K  + 1/e)   \\
                     0   \\  {1 \over e} \cos \theta
\end{array} \right |,
\label{A.11}
\end{eqnarray}

\noindent where the symbol $S.$  stands for the Schwinger's gauge.
Both $D.$- and $S.$- gauges are  unitary  ones  in  the  isotopic
space since   the  corresponding  scalar  fields
$
\Phi ^{D.}_{(a)}(x)
$
and   $\Phi ^{S.}_{(a)}(x)$   are   $x_{3}$-directional, but one of them
(Schwinger's) seems simpler than another (Dirac's).
 To the above-mentioned special monopole field ( (\ref{2.9b})
corresponds to the  $K(r) = - 1/er^{2}$, so that the relations  from (2)
turn out to be very simple and related to the Abelian potential
embedded into the non-Abelian scheme.

Let us determine the matrix $O({\bf c} '') = O({\bf c}~') O( {\bf c})$
relating the Cartesian gauge of isotopic space with Schwinger's
\begin{eqnarray}
\vec{c}~'' = (+ \tan\theta/2 \tan\phi/2 , - \tan\theta/2 , -\tan\phi/2)) \; ,
\nonumber
\\
O({\bf c}'') =  \left | \begin{array}{ccc}
\cos\theta \cos\phi  & \cos\theta \sin\phi & -\sin\theta \\
 -\sin\phi           &      \cos\phi       &    0        \\
\sin\theta \cos\phi  & \sin\theta \sin\phi &  \cos\theta
          \end{array} \right | \;  .
\nonumber
\end{eqnarray}

\noindent This matrix is also well-known in other context as a matrix
linking Cartesian and spherical tetrads in the flat space-time.

\section{ Dirac particles isotopic multiplet, separation of the variables in Sschr\"{o}dinger's tetrad
basis}

In this Section  we enter on analyzing the  isotopic doublet of
Dirac fermions in the external t'Hooft-Polyakov monopole field.
We are going to reexamine this problem, using
the general relativity  tetrad formalism.
Instead of the so-called monopole harmonics,  the  more
conventional  formalism of the Wigner's $D$-functions is used.

We will specify the case of spherical space $S_{3}$,  transition to Euclidean or Lobachewski  models is
achieved by a simple formal change (see below). In spherical coordinates the metric and  corresponding tetrad are
\begin{eqnarray}
dS^{2}  =  dt^{2}  - d\chi ^{2} - \sin  ^{2}\chi  (\; d\theta ^{2} +
\sin ^{2} \theta  d \phi ^{2}\; ) \;  \;  ,
\nonumber
\\
e^{\alpha }_{(0)} = (1, \; 0 ,\;  0 ,\;  0 )\; , \;\;
e^{\alpha }_{(1)} = ( 0 ,\; 0 ,\; \sin^{-1}\chi  ,\;  0 )\;  ,
\nonumber
\\
e^{\alpha }_{(2)} = (0,\; 0, \; 0, \;  \sin^{-1} \chi \sin^{-1} \theta  ) \; ,  \;
e^{\alpha }_{(3)} = ( 0 , 1 , 0 , 0 ) \; ,
\nonumber
\end{eqnarray}

\noindent and  the~Schwinger unitary
gauge of the~monopole  potentials, the  Dirac equation for an isotopic doublet
\begin{eqnarray}
[\; \gamma ^{\alpha }(x) \;  ( i \partial _{\alpha } \; +
\; \Gamma _{\alpha }(x)  +  e\; t^{a} \; W^{(a)}_{\alpha } \; ) \; - \;
 (m \; + \; \kappa \;  \Phi ^{(a)} t^{a}) \; ]\; \Psi (x) = 0  \; .
\nonumber
\end{eqnarray}

\noindent
takes  the form  (note that $r =2 \tan (\chi /2)$)
\begin{eqnarray}
 [\;\gamma ^{0} \; ( i \; \partial _{t} \; +\; e\;  r F(r) \; t^{3})\; + \;
i \gamma ^{3}\; ( \partial _{\chi} \;+ \; {1\over \tan \chi}) \; +\;
{1\over \sin \chi } \; \Sigma ^{S.}_{\theta ,\phi } \; +
\nonumber
\\
+ {{e r^{2}K + 1 } \over \sin \chi } \; (\gamma ^{1} \otimes  t^{2}\;  - \;
                              \gamma ^{2} \otimes  t^{1})\; - \;
 ( \; m  \; + \; \kappa \; r\; \Phi(r)\; t^{3} )\;   ] \; \Psi ^{S.} = 0 \;\;  ,
\nonumber
\\
\Sigma ^{S.}_{\theta ,\phi } =
 i \; \gamma ^{1} \; \partial _{\theta } \;+  \;
  \gamma ^{2}\; {{ i\partial _{\phi } \;+ \; (i \sigma ^{12} \; + \; t^{3})\cos \theta}
  \over {\sin \theta }}\; , \qquad t^{j} = (1/2)\; \sigma ^{j} \; .
\nonumber
\\
\label{3a}
\end{eqnarray}

\noindent
A characteristic feature of such a correlated choice of
frames in both these spaces is the explicit form of the total angular
momentum operator (the sum of orbital, spin, and isotopic ones)
\begin{eqnarray}
J^{S.}_{1} = l_{1} + {{(i \sigma ^{12} + t^{3}) \cos \phi } \over
{ \sin \theta }} \; ,
\nonumber
\\
J^{S.}_{2} =   l_{2} + {{(i \sigma ^{12} + t^{3}) \sin \phi }
\over { \sin \theta }} \; , \; \;    J^{S.}_{3} = l_{3}
\; ;
\label{3b}
\end{eqnarray}

\noindent
so that the present case entirely comes under the situation considered
by Pauli in [...]. The Pauli criterion allows here the following values
for $j: j = 0, 1, 2, 3,~\ldots $ The
$\theta ,\phi $-dependence of the  doublet wave function $\Psi _{jm}$
is to be built up in terms of  the Wigner $D$-functions:
 $D^{j}_{-m,\sigma } (\phi ,\theta ,0)$ ,
where the lower right index  $ \sigma $  takes the values from $(-1,0,+1)$,
which correlates with the explicit diagonal structure of the matrix
$ ( i\sigma ^{12} + t^{3} ) $ :
\begin{eqnarray}
\Psi _{\epsilon jm}(x) = {{ e^{-i \epsilon t}} \over \sin \chi } \;
 [\; { T_{+1/2} \otimes  F(\chi, \theta,\phi)} \; + \; { T_{-1/2} \otimes  G(\chi, \theta, \phi)} \; ]
\; ;
\label{4a}
\end{eqnarray}

\noindent here the fixed symbols $j$ and $(-m)$  in
${D}^{j}_{-m,\sigma } (\phi ,\theta ,0) $ are omitted and
\begin{eqnarray}
F = \left | \begin{array}{l}
            f_{1}(\chi) D_{-1} \\
            f_{2}(\chi) D_{ 0} \\
            f_{3}(\chi) D_{-1} \\
            f_{4}(\chi) D_{ 0}
            \end{array}
       \right |  , \;\;
G = \left | \begin{array}{l}
              g_{1}(\chi) D_{ 0} \\
              g_{2}(\chi) D_{+1} \\
              g_{3}(\chi) D_{ 0} \\
              g_{4}(\chi) D_{+1}
              \end{array} \right |  ,\;\;
              T_{+1/2}= \left | \begin{array}{c}
              1 \\ 0 \end{array} \right |,   \;\;   T_{-1/2} =
              \left | \begin{array}{c}
              0 \\ 1 \end{array} \right | \; ;
\nonumber
\end{eqnarray}

\noindent    throughout the paper the  factor  ${ e^{-i \epsilon t}} /  \sin \chi $ will be omitted.

Another essential feature of the given frame in  the

\begin{center}
({\em Lorentz}) $\otimes$ ({\em isotopic})-space
\end{center}

\noindent
is the appearance of the very simple expression (proportional to $er^{2} K +1 $) for
the term that mixes up together two distinct   components  of  the
isotopic doublet (see eq. (\ref{3a})).

An~important  case  in the~electron-monopole problem is  the~minimal value of quantum number $j$.
The~allowed values for $j$ are $0, 1, 2,\ldots$; the~case of $j = 0$
needs a~careful separate consideration. When $j=0$,  the  symbols
$D^{0}_{0,\pm 1}$
are  meaningless,  and  the~wave  function $\Psi
_{\epsilon 0}(x)$ is to be constructed as
\begin{eqnarray}
 \Psi _{\epsilon 0} =
 T_{+1/2} \otimes
             \left | \begin{array}{l}
  0 \\ f_{2}(\chi)  \\   0  \\  f_{4}(\chi)
             \end{array} \right | \; + \;
T_{-1/2} \otimes
             \left | \begin{array}{l}
  g_{1}(\chi)      \\   0   \\  g_{3}(\chi)  \\   0
             \end{array} \right |   \; .
\label{2.3}
\end{eqnarray}

\noindent
Using the~required recursive relations for Wigner functions

\vspace{3mm}
$(\nu = {\sqrt{j(j + 1)}}, \omega={\sqrt{(j - 1)(j + 2)}}, j
\neq 0) \; ,
$
\begin{eqnarray}
\partial _{\theta } D_{-1} = {1\over 2} (\omega  D_{-2} -
\nu D_{0}) \; , \qquad
{{m - \cos \theta  } \over{ \sin \theta }} D_{-1} =
{1\over 2} (\omega  D_{-2} + \nu  D_{0}) \; ,
\nonumber
\\
\partial _{\theta } D_{0} = {1\over 2} (\nu  D_{-1} - \nu D_{+1})
\; ,  \qquad
{m \over {\sin  \theta }} D_{0} =
{1\over 2} (\nu D_{-1} + \nu  D_{0}) \; ,
\nonumber
\\
\partial _{\theta } D_{+1} = {1\over 2} (\nu D_{0} - \omega D_{+2})
\; ,  \qquad
{{m + \cos \theta } \over{ \sin \theta }} D_{+1} =
{1\over 2} (\nu  D_{0} + \omega  D_{+2}) \; ,
\nonumber
\\
\label{2.4a}
\end{eqnarray}

\noindent we find
\begin{eqnarray}
\Sigma ^{S.}_{\theta ,\phi } \;\Psi ^{S.}_{jm}  = \; \nu \;
 [\; T_{+1/2} \otimes
             \left | \begin{array}{l}
-i f_{4} \; D_{-1} \\ +i f_{3} \; D_{0} \\ +i f_{2} \; D_{-1} \\ -i f_{1}\; D_{0}
             \end{array} \right | \;  + \;
T_{-1/2} \otimes
             \left | \begin{array}{l}
-i g_{4} \; D_{0} \\ +i g_{3} \; D_{+1} \\ +i g_{2} \; D_{0} \\ -i g_{1} \; D_{+1}
             \end{array} \right | \; ]  \; .
\label{2.4b}
\end{eqnarray}

\noindent
Further, let us write down the~expression for the term  mixing up
the~isotopic components
\begin{eqnarray}
{{e r^{2}K  + 1} \over \sin \chi } \;
(\gamma ^{1}\otimes  t^{2} \; - \; \gamma ^{2} \otimes  t^{1}) \; \Psi _{jm} =
\;{{e r^{2}K + 1 } \over 2 \sin \chi }
\nonumber
\\
 \times \; [ \; T_{+1/2} \otimes
             \left | \begin{array}{l}
   0 \\  +i g_{3} D_{0} \\   0   \\ -i g_{1} D_{0}
             \end{array} \right | +
T_{-1/2} \otimes
             \left | \begin{array}{l}
-i f_{4} D_{0} \\   0   \\ +i f_{2} D_{0} \\   0
             \end{array} \right | \; ] \; .
\label{2.5}
\end{eqnarray}

\noindent
After a  simple  calculation  one   finds  the~system  of  radial
equations (for shortness we  set
$W\equiv (e r^{2} K  + 1)/2 \; , \;
 \tilde{F} \equiv  e r  F/2 \; , \;
 \tilde{\Phi } \equiv \kappa r  \Phi /2\;$)
\begin{eqnarray}
(- i {d\over d\chi} + \epsilon + \tilde{F} ) f_{3} -
i{\nu \over \sin \chi} f_{4}  - ( m + \tilde{\Phi } ) f_{1} = 0 \; ,
\nonumber
\\
(+ i {d\over d \chi} + \epsilon  + \tilde{F}) f_{4} +
i{\nu \over \sin \chi } f_{3} + i{W \over \sin \chi } g_{3} - ( m + \tilde{\Phi }) f_{2} = 0 \; ,
\nonumber
\\
(+ i {d\over d \chi } + \epsilon  + \tilde{F} ) f_{1} +
i{\nu \over \sin \chi } f_{2} - ( m + \tilde{\Phi } ) f_{3} = 0 \; ,
\nonumber
\\
(- i {d\over d \chi } + \epsilon  + \tilde{F} ) f_{2} -
i{\nu \over \sin \chi } f_{1} - i{W\over  \sin \chi} g_{1} - ( m + \tilde{\Phi } ) f_{4} = 0 \; ,
\nonumber
\\
(- i{d\over d  \chi } + \epsilon  - \tilde{F} ) g_{3} - i{\nu \over \sin \chi } g_{4} -
i{W\over \sin \chi } f_{4} - ( m - \tilde{\Phi } ) g_{1} = 0 \; ,
\nonumber
\\
(+ i {d\over d  \chi } + \epsilon  - \tilde{F} ) g_{4} + i{\nu \over \sin \chi } g_{3}
    - ( m - \tilde{\Phi } ) g_{2} = 0 \; ,
\nonumber
\\
(+ i{d \over d  \chi } + \epsilon  - \tilde{F} ) g_{1} +
i{\nu \over \sin \chi } g_{2} + i{W\over \sin \chi } f_{2} - ( m - \tilde{\Phi } ) g_{3} = 0 \; ,
\nonumber
\\
(-i{d\over d \chi } + \epsilon  - \tilde{F} ) g_{2} -
i{\nu \over \sin \chi } g_{1} - ( m - \tilde{\Phi } ) g_{4} = 0  \; .
\label{2.6'}
\end{eqnarray}

\noindent When $j$ takes on  value $0$ (then
$\Sigma _{\theta ,\phi } \Psi _{\epsilon 0} \equiv  0$), the~radial system
is simpler:
\begin{eqnarray}
( + i {d\over d \chi} + \epsilon  + \tilde{F} ) f_{4} +
i{W\over \sin \chi } g_{3} - ( m + \tilde{\Phi } ) f_{2} = 0 \; ,
\nonumber
\\
( - i {d\over d \chi } + \epsilon  + \tilde{F} ) f_{2} -
i{W\over \sin \chi } g_{1} - ( m + \tilde{\Phi } ) f_{4} = 0 \; ,
\nonumber
\\
( - i{d\over d \chi } + \epsilon  - \tilde{F} ) g_{3} -
i{W\over \sin \chi } f_{4} - ( m - \tilde{\Phi } ) g_{1} = 0 \; ,
\nonumber
\\
( + i {d\over d \chi } + \epsilon  - \tilde{F} ) g_{1} +
i{W\over \sin \chi } f_{2} - ( m - \tilde{\Phi } ) g_{3} = 0  \; .
\label{2.7'}
\end{eqnarray}

\noindent Both    systems  (\ref{2.6'})   and   (\ref{2.7'})   are   sufficiently
complicated. To proceed further with a~situation like  that,  it  is
normal practice to have searched  a~suitable  operator  which  could
be diagonalized additionally. It is known that the~usual
$P$-inversion operator for a~bispinor field cannot be  completely
appropriate  for  this purpose  and  a~required quantity  is  to  be
constructed  as a~combination of  bispinor $P$-inversion operator and
a~certain discrete transformation  in  the~isotopic space.  Indeed,
considering that the~usual $P$-inversion operator for a~bispinor
field (in the~basis of Cartesian  tetrad, it is
$\hat{P}_{bisp.}^{Cart.} \otimes \hat{P} =
i \gamma^{0} \otimes \hat{P}$, where $\hat{P}$  causes the~usual
$P$-reflection
of  space coordinates)   is determined in the~given (spherical) basis as
\begin{eqnarray}
\hat{P}_{bisp.}^{sph.} \otimes \hat{P} =
\left | \begin{array}{rrrr}
             0  &  0  &  0  &  -1 \\
             0  &  0  & -1  &   0 \\
             0  & -1  &  0  &   0 \\
            -1  &  0  &  0  &   0
\end{array} \right | \otimes \hat{P} =
  - ( \gamma ^{5} \gamma ^{1} ) \otimes  \hat{P}
\nonumber
\end{eqnarray}

\noindent and  it acts upon the wave function $\Psi _{jm}(x)$ as follows
\begin{eqnarray}
(\hat{P}_{bisp.}^{sph.} \otimes \hat{P}) \; \Psi _{\epsilon jm}(x)
=
(-1)^{j+1} \;
\nonumber
\\
\times
 [\; T_{+1/2} \otimes
             \left | \begin{array}{l}
f_{4} \; D_{0} \\ f_{3} \; D_{+1} \\ f_{2} \; D_{0} \\ f_{1} \; D_{+1}
             \end{array} \right | \; + \;
T_{-1/2} \otimes
             \left | \begin{array}{l}
g_{4} \; D_{-1} \\ g_{3} \; D_{0} \\ g_{2} \; D_{-1} \\ g_{1} \; D_{0}
             \end{array} \right |\;  ] \; .
\label{2.8b}
\end{eqnarray}

\noindent The latter points the~way  towards  the~search  for
a~required discrete operator: it would  have the~structure
\begin{eqnarray}
\hat{N}^{S.}_{sph.} \equiv \hat{\pi}^{S.} \otimes
\hat{P}_{bisp.}^{sph.} \otimes \hat{P} \; , \qquad
\hat{\pi}^{S.} =  a  \sigma^{1}  +  b  \sigma^{2}  \; ,
\label{2.9}
\end{eqnarray}

\noindent so that $
\hat{\pi}^{S.} \; T_{\pm 1/2} = ( a  \pm  i b ) \; T_{\mp 1/2}\;$ .
The total multiplier at the~quantity
 $\hat{\pi }^{S.}$  is  not  material  for
separating the~variables, below one sets
$(\hat{\pi }^{S.})^{2} = ( a^{2} \; +\;  b^{2} ) = + 1$.
In the following we restrict ourselves to real valued $a$ and $b$ and use notation:
\begin{eqnarray}
a + i b = e^{iA}\;.
\nonumber
\end{eqnarray}

\noindent
From  the~equation
$\hat{N}^{S.A}_{sph.} \Psi _{jm}  = N_{A} \Psi _{jm}$  one   finds   two  proper
values $N_{A}$  and corresponding limitation on  the~functions $f_{i}(r)$
and $g_{i}(r)$:

\vspace{3mm}
$
N_{A} = \delta \; (-1)^{j+1} \; , \;\;  \delta = \pm \; 1 \; ,
$
\begin{eqnarray}
g_{1} = \delta \; e^{iA} \; f_{4} \; , \qquad
g_{2} = \delta \; e^{iA} \;  f_{3} \; ,
\nonumber
\\
g_{3} = \delta \; e^{iA} \; f_{2} \; ,\qquad
g_{4} = \delta \; e^{iA} \; f_{1}    \; .
\label{2.10a}
\end{eqnarray}

\noindent Taking into account the relations (\ref{2.10a}),  one  produces
the~equations
\begin{eqnarray}
(-i{d\over d \chi } + \epsilon  + \tilde{F} ) f_{3} -
{\nu \over \sin \chi } f_{4}  - ( m + \tilde{\Phi}) f_{1} = 0 \; ,
\nonumber
\\
(+i{d\over d \chi } + \epsilon  + \tilde{F} ) f_{4} +
{\nu \over \sin \chi } f_{3} + i{W \over \sin \chi } - \delta  e^{iA}  f_{2} -
( m  +  \tilde{\Phi} ) f_{2} = 0 \; ,
\nonumber
\\
(+i{d\over d \chi } + \epsilon  + \tilde{F} ) f_{1} +
{\nu \over \sin \chi } f_{2}   - ( m + \tilde {\Phi } ) f_{3} = 0 \; ,
\nonumber
\\
(-i{d\over d \chi } + \epsilon  + \tilde{F} ) f_{2} -
{\nu \over \sin \chi } f_{1} - i{W\over \sin \chi } \delta  e^{iA}  f_{4} -
( m + \tilde{\Phi } ) f_{4} = 0 \; ,
\nonumber
\\
(-i{d\over d \chi } + \epsilon  - \tilde{F} ) f_{2} -
{\nu \over \sin \chi } f_{1} - i {W\over \sin \chi } e^{-iA}  \delta  f_{4}-
( m - \tilde{\Phi } ) f_{4} = 0 \; ,
\nonumber
\\
(+i{d\over d \chi} + \epsilon  - \tilde{F} ) f_{1} +
{\nu \over \sin \chi } f_{2}   - ( m - \tilde{\Phi } ) f_{3} = 0 \; ,
\nonumber
\\
(+i{d\over d \chi } + \epsilon  - \tilde{F} ) f_{4} +
{\nu \over \sin \chi } f_{3} + i {W\over  \sin \chi } e^{-iA}  \delta  f_{2}-
( m - \tilde{\Phi } ) f_{2} = 0 \; ,
\nonumber
\\
(-i{d\over d \chi} + \epsilon  - \tilde{F} ) f_{3} -
{\nu \over \sin \chi } f_{4}  - ( m - \tilde{\Phi } ) f_{1} = 0  \;\;  .
\label{2.10b}
\end{eqnarray}

\noindent It is evident at once that the~system (\ref{2.10b})  would be compatible
with itself  provided that  $\tilde{F}(\chi) = 0$  and $\tilde{\Phi }(\chi) = 0$.
In other words,
the above-mentioned operator $\hat{N}^{S.}$ can be diagonalized on
the~functions $\Psi _{\epsilon jm}(x)$ if and only if $W^{(a)}_{t}= 0$
and $\kappa = 0$; below  we
suppose that these requirements  will be  satisfied.
Moreover, given this limitation satisfied, it is necessary to draw
distinction  between  two cases depending on expression for $W(r)$.

If $W(\chi) = 0$,  the~difference between $e^{iA} $ and $e^{-iA} $
in the~equations (\ref{2.10b})  is  not
essential in simplifying these equations (because
the~relevant terms just vanish). Thus, for the first case, the~system
(\ref{2.10b}) converts into

\vspace{3mm}
$\underline{ W(\chi) = 0 }, $
\begin{eqnarray}
( - i {d\over d \chi} + \epsilon  ) f_{3} - {\nu \over \sin \chi} f_{4} - m f_{1} = 0 \; ,
\nonumber
\\
( + i {d\over d \chi} + \epsilon  ) f_{4} + {\nu \over \sin \chi} f_{3} - m f_{2} = 0 \; ,
\nonumber
\\
( + i {d\over d \chi } + \epsilon  ) f_{1} + {\nu \over \sin \chi} f_{2} - m f_{3} = 0 \; ,
\nonumber
\\
( - i {d\over d \chi } + \epsilon  ) f_{2} - {\nu \over \sin \chi} f_{1} - m f_{4} = 0 \; .
\nonumber
\\
\label{2.11}
\end{eqnarray}

\noindent
There exists sharply distinct situation at $W \neq  0$. Here,
the~equations are  consistent  with  each  other  only  if
$e^{iA}  = e^{-iA}$;
therefore $e^{iA}  = a + i b =  \pm 1$ (for definiteness, let this parameter $a$
 be equal $+1$). The~corresponding set of radial equations, obtained from
(\ref{2.10b}), is

\vspace{3mm}
$\underline{W (\chi) \neq 0\; }, $
\begin{eqnarray}
( -i{d\over d \chi }
+ \epsilon ) f_{3} - {\nu \over \sin \chi} f_{4} - m f_{1} = 0 \; ,
\nonumber
\\
(+i{d\over d \chi } + \epsilon ) f_{4} + {\nu \over \sin \chi} f_{3} + i{W (\chi)\over \sin \chi }
\delta  f_{2} - m f_{2} = 0   \; ,
\nonumber
\\
( +i {d\over d \chi} + \epsilon
) f_{1} + {\nu \over \sin \chi } f_{2}  - m f_{3} = 0  \; ,
\nonumber
\\
 (- i
{d\over d \chi} + \epsilon ) f_{2} - {\nu \over \sin \chi} f_{1}  - i{W (\chi)\over \sin \chi}
 \delta  f_{4} - m f_{4} = 0        \; .
\nonumber
\\
\label{2.12}
\end{eqnarray}

The~case  $j = 0$ can be considered in the~same  way. Here
the $N_{A}$-symmetry produces
\begin{eqnarray}
N _{A}= - \; \delta  ,\;\; \delta  = \pm 1 \; : \qquad
g_{1} \; = \delta \; e^{iA} \; f_{4} \;,\qquad
g_{3} \; = \delta \; e^{iA} \; f_{2} \; .
\label{2.13a}
\end{eqnarray}

\noindent Further, the~quantities  $\tilde{F}$  and $\tilde{\Phi }$
are to be equated  to zero; again there are two  possibilities depending
on $W$:
\begin{eqnarray}
\underline{j=0\; , \; W(\chi) = 0 \;} : \qquad
( i {d\over d \chi}  + \epsilon ) f_{4} - m f_{2}  = 0 \;  ,
\nonumber
\\
(-i {d\over d \chi} + \epsilon ) f_{2}  - m f_{4}  = 0 \; ;
\label{2.13b}
\end{eqnarray}
\begin{eqnarray}
\underline{j=0\;, \; W(\chi) \neq  0 \;}
 : \qquad  ( i {d\over d \chi } + \epsilon ) f_{4} - ( m - i {\delta \; W(\chi)\over \sin \chi}  ) f_{2} = 0 \; ,
\nonumber
\\
\;\;
(- i {d\over d \chi } + \epsilon ) f_{2} - ( m + i {\delta \; W (\chi)\over \sin \chi }  ) f_{4} = 0 \; .
\label{2.13c}
\end{eqnarray}

\noindent The explicit forms  of  the  wave  functions
 $\Psi _{\epsilon jm\delta }(x)$ and $\Psi _{\epsilon 0\delta }(x)$ are
as follows:

\vspace{3mm}
the case $\; W(\chi) \neq  0 , \; j > 0 \; $,
\begin{eqnarray}
\Psi _{\epsilon jm}(x) =
 T_{+1/2} \otimes
             \left | \begin{array}{l}
f_{1} \; D_{-1} \\ f_{2} \; D_{0} \\ f_{3} \; D_{-1} \\ f_{4} \; D_{0}
             \end{array} \right |  \; + \;
\delta \; T_{-1/2} \otimes
             \left | \begin{array}{l}
f_{4} \; D_{0} \\ f_{3} \; D_{+1} \\ f_{2} \; D_{0} \\ f_{1} \; D_{+1}
             \end{array} \right | \; ;
\label{2.14a}
\end{eqnarray}

the case $W(\chi) \neq  0 , \;\; j = 0 \;$,
\begin{eqnarray}
\Psi _{\epsilon 0} =
 T_{+1/2} \otimes
             \left | \begin{array}{l}
  0 \\ f_{2}(r)  \\   0  \\  f_{4}(r)
             \end{array} \right |  \; + \;
\delta \; T_{-1/2} \otimes
             \left | \begin{array}{l}
  f_{4}(r)      \\   0   \\  f_{2}(r)  \\   0
             \end{array} \right | \; ;
\nonumber
\\
\label{2.14b}
\end{eqnarray}

\noindent when $W = 0$, the term  $\delta \; T_{-1/2}$  is to be changed to
 $\delta \;  e^{iA} \; T_{-1/2}$.

In the end of this Section let us specify explicit  form of $W(\chi)/ \sin \chi$,
in this point we consider all three model, $S_{3}, H_{3}, E_{3}$:

\begin{eqnarray}
in \;\; S_{3}-space, \qquad
 {W (\chi ) \over \sin \chi } = {er^{2}K +1 \over 2 \sin \chi} =  {1 \over 2} a f_{1} (a\chi + b) \; ,
 \;\; \chi \in [ 0, \pi ]\; ;
\nonumber
\\in \;\; H_{3}-space, \qquad
 {W (\chi ) \over \mbox{sh}\;  \chi } = {er^{2}K +1 \over 2 \mbox{sh}\;  \chi } =  {1 \over 2} a f_{1} (a\chi + b) \; ,
\;\;  \chi \in [0, + \infty ) \; ;
\nonumber
\\
in \;\; E_{3}-space, \qquad
 {W (r) \over r } = {er^{2}K +1 \over 2 r } =  {1 \over 2} a f_{1} (a r + b) \; , \;\;
 r \in [ 0 , + \infty ) \; .
\nonumber
\\
\label{2.15}
\end{eqnarray}

According to see (\ref{2.7}) we have three different possibilities to choose $f_{1}$:
\begin{eqnarray}
 f_{1}  = \pm \; {A \over
\sin \;(A r + B) } \;, \;
\pm \; {A \over
\mbox{sh}\; (A r + B) } \; , \;
\pm
\; {A \over Ar +B } \;.
\nonumber
\end{eqnarray}

\noindent
One may feel that among the above monopole solutions in   models $E_{3}, H_{3},S_{3}$
there exist  three  ones
which can be  naturally associated
with respective geometries. The situation can be illustrated  by the schema
$$
\left. \begin{array}{lccc}
  &  E_{3}  &  H_{3}  &  S_{3}  \\
(ar+b)  &  *  &  -   &  -  \\
\mbox{sh}\; (ar+b)  & - &  *  & -  \\
\sin (ar+b)  &  -  & -  &  *
\end{array}  \right.
$$

\noindent
It should be noted that the known non-singular BPS-solution in the flat
Minkowski space  can be understood as a result of somewhat artificial
combining the Minkowski space background with a possibility naturally
linked up with the Lobachevsky geometry.

\section{Analysis of the  case of
               singular monopole  field}

Now, some added aspects of the~simplest monopole are examined more closely.
The~system of radial equations, specified for this potential, is basically
simpler than in general case, so that the~whole problem including
the~radial functions can be carried out to its complete conclusion.
Actually,  the equation (\ref{2.11}) admits of some
further simplifications owing to diagonalyzing the  operator
$\hat{K}_{\theta ,\phi }=
- i \gamma ^{0} \gamma ^{5} \Sigma _{\theta ,\phi }$.
From the~equation
$
\hat{K}_{\theta ,\phi } \Psi _{jm}  =
\lambda  \Psi _{jm}$ , it
follows that $\lambda  = - \mu \;  {\sqrt{j(j + 1}}), \; \mu  = \pm 1$  and
\begin{eqnarray}
f_{4} = \mu \;  f_{1} , \qquad f_{3} = \mu \; f_{2} ,\qquad
g_{4} = \mu \;  g_{1} , \qquad g_{3} = \mu \;  g_{2} \; .
\label{3.1'}
\end{eqnarray}

\noindent Correspondingly, the system (\ref{2.11}) yields
\begin{eqnarray}
(+ i {d\over d \chi}  + \epsilon ) f_{1} + i {\nu \over \sin \chi} f _{2} -
\mu \; m \; f_{2}  = 0 \;  ,
\nonumber
\\
(- i {d\over d \chi } +  \epsilon ) f_{2}  - i {\nu \over \sin \chi} f _{1} -
\mu \;  m \; f_{1}  = 0  \; .
\label{3.2a'}
\end{eqnarray}

\noindent The wave function with quantum  numbers
$(\epsilon , j, m, \delta , \mu )$  has  the~form
\begin{eqnarray}
\Psi _{\epsilon jm\delta\mu}^{A}(x) =  T_{+1/2} \otimes
             \left | \begin{array}{rl}
f_{1} & D_{-1} \\ f_{2} & D_{0} \\ \mu &  f_{2}  D_{-1} \\ \mu &  f_{1}  D_{0}
             \end{array} \right | \;  + \;
e^{iA}  \mu   \delta   T_{-1/2} \otimes
             \left | \begin{array}{rl}
f_{1} &   D_{0} \\ f_{2} &  D_{+1} \\ \mu &  f_{2}  D_{0} \\ \mu &  f_{1}  D_{+1}
             \end{array} \right | \;  .
\nonumber
\\
\label{3.2b'}
\end{eqnarray}

Let us  relate the  non-Abelian  functions (\ref{3.2b})  and  (\ref{2.14b})  with the~wave functions satisfying  the~Dirac
equation  in  the~Abelian monopole potential. Those latter were investigated
by many authors in the case of flat space; below we  will use the~notation
according to [...]).

At $ j > j_{\min }$ these Abelian functions are described as follows (the factor $e^{-i\epsilon t} / \sin \chi$ is omited)
\begin{eqnarray}
\Phi^{(eg)} _{jm\mu}  =
                              \left | \begin{array}{r}
          f_{1}(\chi) \; D^{j}_{-m,eg -1/2}  \\
          f_{2}(\chi) \; D^{j}_{-m,eg +1/2} \\
 \mu  \;          f_{2}(\chi) \; D^{j}_{-m,eg -1/2} \\
 \mu  \;          f_{1}(\chi) \; D^{j}_{-m,eg +1/2}
                               \end{array} \right |
\label{3.3a'}
\end{eqnarray}

\noindent For the minimal values $\;j = j_{min.} = \mid eg\mid -1/2 $,
they are
\begin{eqnarray}
eg = + 1/2, + 1, + 3/2, ... \qquad
\Phi ^{(eg)}_{\epsilon 0} =
                              \left | \begin{array}{l}
          f_{1}(\chi) \; D^{j}_{-m,eg-1/2} \\
                            0                         \\
          f_{3}(\chi) \; D^{j}_{-m,eg-1/2} \\
                            0
                               \end{array} \right |  \;  ;
\nonumber
\\\\
eg = - 1/2, - 1, - 3/2,...\qquad
\Phi ^{(eg)}_{\epsilon 0} =
                              \left | \begin{array}{l}
                             0                    \\
          f_{2}(\chi) \; D^{j}_{-m,eg+1/2} \\
                             0                     \\
          f_{4}(\chi) \; D^{j}_{-m,eg+1/2}
                               \end{array} \right |   \;  .
\label{3.3'}
\end{eqnarray}

\noindent On comparing the formulas (\ref{3.2b})  and  (\ref{2.14b})    with
these  Abelian  fermion-monopole functions,
the~following expansions  can be easily  found (respectively, for
$j > 0$ and $j=0$ cases):
\begin{eqnarray}
\Psi ^{A \delta \mu }_{\epsilon jm}(x) \; = \;
 T_{+1/2} \otimes  \Phi ^{eg=-1/2}_{\epsilon jm\mu }(x)
  \;\; + \;\; \mu \; \delta \;  e^{iA}\;   T_{-1/2} \otimes
 \Phi ^{eg =+1/2}_{\epsilon jm\mu }(x)  \; ,
\nonumber
\\[3mm]
\Psi ^{A }_{\epsilon 0\delta } (x) \;= \;
 T_{+1/2} \otimes  \Phi ^{eg =-1/2}_{\epsilon 0}(x) \; \; + \;\;
\delta \; e^{iA} \;\;  T_{-1/2} \otimes  \Phi ^{eg =+1/2}_{\epsilon 0}(x)
\;  .
\label{3.4''}
\end{eqnarray}

 In connection  with the formulas (\ref{3.4''}), one additional remark
should be given.  Though,  as evidenced by (3.4a,b),  definite  close
relationships between the~non-Abelian doublet wave functions and
Abelian fermion-monopole
functions can be explicitly discerned, in reality,  the~non-Abelian
situation is intrinsically non-monopole-like (non-singular one).
Indeed,  in the~non-Abelian case, the~totality
of possible transformations (upon the~relevant wave functions) which bear
the~gauge status are very different from ones that there are
in the~purely Abelian theory. In  a~consequence of this, the~non-Abelian
fermion doublet wave functions  can be readily transformed, by carrying
out some special gauge transformations in Lorentzian and isotopic
spaces together, into the~form where they are single-valued
functions of spatial points.  In the~Abelian monopole situation,
the~analogous particle-monopole functions can by no means
be translated  to any single-valued ones.

\section{ Free parameter and $N_{A}$-parity selection rules}

Now we proceed with analyzing the~totality of the~discrete
operators $\hat{N}_{A}$, which all are suitable  for  separation  of  variables.
What is the~meaning of the~parameter $A$? In other words,  how
can this $A$ manifest itself and why does such an~unexpected  ambiguity
exist?  We remember that the $A$   fixes  up  one  of  the~complete
set of operators  $\{\; i\; \partial _{t} ,\; \vec{J}^{2} ,
\; J_{3} ,\;  \hat{N}_{A} ,\; \hat{K} \;\}$,
and correspondingly this~$A$ also labels all  basic
wave functions.
It is obvious, that this parameter  $A$  can manifest itself in
matrix elements of physical  quantities.

As  a~simple  example  let  us  consider  a~new   form of
the~above-mentioned selection rules depending on the~$A$-parameter. Now,
the~matrix element examined is
\begin{eqnarray}
\int \bar{\Psi} ^{A}_{\epsilon JM\delta \mu }(x) \; \hat{G}(x)\;
     \Psi ^{A}_{\epsilon J'M'\delta'\mu'}(x) \;  dV \; \equiv   \;
\int r^{2} dr \int f^{A}(\vec{x}) \; d \Omega
\nonumber
\end{eqnarray}

\noindent then
\begin{eqnarray}
f^{A}(-\vec{x} ) = \; \delta \;  \delta'\; (-1)^{J+J'} \;
                \bar {\Psi}^{A}_{\epsilon JM\delta \mu }(x) \; \times
\nonumber
\\
\left [\;(a^{*} \sigma ^{1} + b^{*} \sigma ^{2}) \otimes \hat{P}_{bisp.}
         \hat{G}(-\vec{x}) \;
 (a \sigma ^{1} + b \sigma ^{2}) \otimes \hat{P}_{bisp.} \; \right ]  \;
\Psi ^{A}_{\epsilon J'M'\delta'\mu'} (\vec{x}) \; .
\nonumber
\\
\label{9.2a}
\end{eqnarray}

\noindent If this $\hat{G}$  obeys the condition
\begin{eqnarray}
\left [\;(a^{*} \sigma ^{1} + b^{*} \sigma ^{2}) \otimes
\hat{P}_{bisp.}\;] \; \hat{G}(-\vec{x}) \; [\; (a \sigma ^{2} + b \sigma ^{1})
 \otimes \hat{P}_{bisp.}\; \right ]\; = \; \Omega ^{A} \; \hat{G}(\vec{x})
\label{9.2b}
\end{eqnarray}

\noindent which is equivalent to
\begin{eqnarray}
\left ( \begin{array}{cc}
 e^{i(A-A^{*})}  \; \hat{g}_{22} (-\vec{x})     &   \;
 e^{-i(A+A^{*})} \; \hat{g}_{21}(-\vec{x})  \\
 e^{i(A+A^{*})}  \; \hat{g}_{12} (-\vec{x})  &      \;
 e^{-i(A-A^{*})} \; \hat{g}_{11}(-\vec{x})
\end{array}  \right ) \otimes
\nonumber
\\
\left [ \; \hat{P}_{bisp.} \; \hat{G}^{0}(-\vec{x})\; \hat{P}_{bisp.} \;
 \right ] \;
 = \;
\Omega^{A}  \left ( \begin{array}{cc}
 \hat{g}_{11}(\vec{x})     &
 \hat{g}_{12}(\vec{x})     \\
 \hat{g}_{21} (\vec{x}) &
 \hat{g}_{22}(\vec{x})
\end{array}  \right ) \otimes  \hat{G}(\vec{x})
\label{9.2c}
\end{eqnarray}

\noindent where $\Omega ^{A} = + 1$ or  $-1$,
 then the relationship (\ref{9.2a}) comes to
\begin{eqnarray}
f^{A}(-\vec{x})\;  = \;  \Omega ^{A}\; \delta \;  \delta'\; (-1)^{J+J'} \;
 f^{A}(\vec{x}) \; .
\label{9.2d}
\end{eqnarray}

\noindent Taking into account (\ref{9.2d}), we bring the matrix element's
 integral above to the~form
\begin{eqnarray}
\int \bar{\Psi}^{A}_{\epsilon JM\delta \mu }(x) \; \hat{G}(x) \;
\Phi ^{A}_{\epsilon J'M'\delta'\mu'}(x) \;dV \;
\nonumber
\\
= \;
\left [\; 1\; + \; \Omega ^{A} \; \delta \; \delta'\; (-1)^{J+J'} \; \right ] \;
\int_{V_{1/2}} f^{A}( \vec{x} ) \;dV
\label{9.3a}
\end{eqnarray}

\noindent where the~integration in the~right-hand side is done on the~half-space.  This
expansion  provides the following selection rules:
\begin{eqnarray}
M E \equiv  0  \qquad \Longleftrightarrow  \qquad
\left [\; 1\; +\; \Omega^{A} \; \delta \; \delta' \; (-1)^{J+J'}\; \right ]
 \; =\; 0 \;\; .
\label{9.3b}
\end{eqnarray}

\noindent  $\Omega^{A}$  involves
its own particular limitations on composite scalars or pseudoscalars
because it implies definite configuration of
their isotopic parts, obtained by delicate  fitting all
the~quantities $\hat{g}_{ij}$.  Therefore, each of those  $A$  will
 generate  its own distinctive selection rules.

\section{Parameter $A$ and additional isotopic symmetry}

Where does this $A$-ambiguity come from and what is the meaning
of this parameter $A$? To proceed further with this problem, one is
to realize that the~all different values for $A$  lead to  the  same
whole functional space; each fixed value for $A$  governs  only
the~basis states $\Psi ^{A}_{\epsilon JM\delta \mu }(x)$ associated with $A$.
Connection  between any two sets
of functions $\{ \Psi(x)  \}^{A}$ and $\{ \Psi(x) \}^{A'=0}$
is characterized by
\begin{eqnarray}
\Psi ^{A\; S.}_{\epsilon JM\delta \mu } =
U_{S.}(A) \;\Psi ^{A'=0, \;S.}_{\epsilon JM\delta \mu }(x) \; , \qquad
U _{S.}(A) =
\left | \begin{array}{cc}
 1  & 0 \\
  0  &  e ^{iA}
\end{array}   \right | \otimes  I    \; .
\label{6.1a}
\end{eqnarray}

\noindent It is readily verified that the operator $\hat{N}^{S.}_{A}$
(depending on $A$) can be obtained  from the~operator $\hat{N}^{S.}_{A'=0}$  as follows
\begin{eqnarray}
\hat{N}^{S.}_{A} \;  = \; U_{S.}(A) \;\; \hat{N}^{S.} \; U^{-1}_{S.}(A) \; .
\label{6.1b}
\end{eqnarray}

\noindent The matrix $U^{S.}_{A}$  is so simple only  in  the~Schwinger
basis; after translating that into Cartesian one
\begin{eqnarray}
\Psi ^{A\;C.}_{\epsilon JM\delta \mu } (x)  =
U^{C.} _{A} \; \Psi ^{A'=0,\; C.}_{\epsilon JM\delta \mu }(x)  \; ,
\nonumber
\end{eqnarray}

\noindent it becomes
\begin{eqnarray}
U^{C.}_{A}=
                       \left | \begin{array}{cc}
( e^{iA}   \sin ^{2} \theta /2 \; + \;  \cos ^{2}\theta /2 ) &
{1\over 2} \; ( 1 - e^{iA} ) \sin \theta \; e^{-i\phi } \\[2mm]
{1\over 2} \; (1 -e^{iA}  ) \; \sin \theta \; e^{+i\phi }  &
(\sin ^{2}\theta /2 \; +\;  e^{iA}  \cos ^{2} \theta /2 )
\end{array} \right | \otimes I \; .
\label{6.1c}
\end{eqnarray}

The transformation $U^{C.}_{A}$ can be brought to the~form
\begin{eqnarray}
 U ^{C.}_{A}\; = \;{1 + e^{iA} \over 2} \; + \; {1 - e^{iA} \over 2} \;\vec{\sigma} \;
        \vec{n}_{\theta ,\phi }    \; .
\nonumber
\end{eqnarray}

\noindent Separating out the factor $e^{iA/2}$  in the right-hand  side  of
this formula, we can rewrite the $U^{C.}$  in the~form
\begin{eqnarray}
U^{C.}_{A} \; = \; e^{iA/2} \; \exp ( -\; i \; {A\over 2}\; \vec{\sigma }\;
 \vec{n}_{\theta ,\phi } )
\label{6.6}
\end{eqnarray}

\noindent where the second factor lies in the~(local) spinor representation
of  the   rotation group $SO(3.R)$.
This matrix  provides a~very  special
transformation upon the~isotopic fermion doublet and can be thought
of  as  an~analogue of the~Abelian chiral symmetry transformation.
This symmetry leads  to the $A$-ambiguity (6.5) and
permits to choose an~arbitrary reflection operator from the~totality
$\{ \hat{N}_{A} \}$.

Let us add some  generalities.  As  well  known,  when
analyzing any Lie group problems (or  their  algebra's)  there
indeed exists a~concept of equivalent representations:
 $U \; M_{k} \; U^{-1} \; = \; M'_{k} \;$ and $\;  M_{k} \; \sim \; M'_{k}$.
In this context,  the two  sets  of
operators $\{ J^{S.}_{i}, \; \hat{N}^{S.} \}$  and
$\{ J^{S.}_{i}, \; \hat{N}^{S.}_{A} \}$  provide   basically  just the same
representation of the $O(3.R)$-algebra
\begin{eqnarray}
\{ J^{S.}_{i}, \; \hat{N}^{S.}_{A} \} \; = \; U_{S.}(A) \;
\{ J^{S.}_{i}, \; \hat{N}^{S.} \} \; U^{-1}_{S.}(A) \; .
\label{6.2a}
\end{eqnarray}

\noindent The totally different situation  occurs  in the context of the use
of those two operator sets as physical observables concerning the  system
with the~fixed Hamiltonian
\begin{eqnarray}
\{ \vec{J}^{2}_{S.}, \; J^{S.}_{3}, \; \hat{N}^{S.} \}^{\hat{H}}
\qquad and \qquad
\{ \vec{J}^{2}_{S.}, \; J^{S.}_{3}, \; \hat{N}^{S.}_{A} \}^{\hat{H}} \; .
\label{6.2b}
\end{eqnarray}

\noindent Actually, in this case the two operator sets  represent different
observables at the  same  physical  system:  both  of  them  are
followed by the~same Hamiltonian $\hat{H}$  and also  lead  to  the~same
functional space, changing  only its basis vectors
 $\{ \Psi _{\epsilon JM\delta \mu }(x) \}^{A}$.
Moreover,
in the~quantum mechanics it  seems always possible to  relate
two  arbitrary  complete  sets  of  operators  by   some   unitary
transformation:
\begin{eqnarray}
\{ \hat{X}_{\mu } , \; \mu  = 1, \ldots \}^{\hat{H}}
 \;\;  \Longrightarrow  \;\;
\{ \hat{Y}_{\mu } , \; \mu  = 1, \ldots \}^{\hat{H}}    \; ,
\{ \Phi _{x_{1} \ldots x_{s}} \}  \;\;  \Longrightarrow  \;\;
\{ \Phi _{y_{1} \ldots y_{s}} \} \; .
\nonumber
\end{eqnarray}

\noindent But arbitrary transformations $U$ cannot generate, through
converting
$$
U \; \{ \hat{X}_{\mu } \} \; U^{-1} \; = \; \hat{Y}_{\mu },
$$
 a~new
complete  set  of  variables;
instead, only some Hamiltonian symmetry's operations are suitable for
this: $U \; \hat{H} \; U^{-1} = H$.

In this connection, we may recall  a more  familiar  situation
for Dirac massless field. The wave equation for  this  system
has the~form
\begin{eqnarray}
i \bar{\sigma}^{\alpha}(x)\; (\partial _{\alpha } +
\bar{\Sigma}_{\alpha })\; \xi (x) = 0 \;\; , \qquad
i \sigma ^{\alpha }(x) \; (\partial _{\alpha }  + \Sigma _{\alpha })\; \eta (x) = 0 \; .
\label{6.3a}
\end{eqnarray}

\noindent If the function $\Phi (x) = ( \xi (x) , \; \eta (x) )$ is subjected
to the~transformation
\begin{eqnarray}
\left | \begin{array}{c}
\xi'(x)  \\ \eta'(x)
\end{array}  \right | =
\left | \begin{array}{cc}
I  & 0  \\ 0  &  z\;I \end{array} \right |=
\left | \begin{array}{c}
\xi (x) \\ \eta (x) \end{array} \right |\; ,
\label{6.3b}
\end{eqnarray}

\noindent where $z$ is an arbitrary complex number,  then  the  new  function
$\Phi'(x) = ( \xi'(x), \eta'(x) )$ satisfies again the equation in the form
(\ref{6.3a}). This manifests the Dirac massless  field's  symmetry  with
respect to  the~transformation
\begin{eqnarray}
\hat{H}' = U \; \hat{H} \;  U^{-1}  = \hat{H} , \qquad
\Phi'(x) = U \; \Phi (x)   \; .
\label{6.3c}
\end{eqnarray}

\noindent The existence of the~symmetry raises the~question  as  to
whether this symmetry affects  determination of  complete set  of
diagonalized  operators   and  constructing spherical wave solutions.
These  solutions,  conformed  to  diagonalizing   the~usual
bispinor $P$-inversion operator  are as in  (\ref{i.2}).
In the same time, other spherical solutions,  together
with corresponding diagonalized discrete operator, can be produced:
\begin{eqnarray}
\Phi^{z}_{\epsilon jm \mu } = {e^{-i\epsilon t}\over r}
\left | \begin{array}{r}
                  f_{1} \; D^{j}_{-m, -1/2} \\
                  f_{2} \; D^{j}_{-m, +1/2} \\
    z \; \mu \; f_{2} \; D^{j}_{-m, -1/2} \\
    z \; \mu \; f_{1} \; D^{j}_{-m, +1/2}
\end{array} \right | \;  , \;
U \;  (\; \hat{P}^{sph.}_{bisp.} \otimes \hat{P} \;  ) \; U^{-1}\;  =
\nonumber
\\
=
 [\; {1\over 2} \; ( z + {1\over z} )\; (- \gamma^{5} \gamma^{1}) \; + \;
  {1\over 2} \; ( z - {1\over z} ) \; (- \gamma^{1}) \;  ] \otimes  \hat{P} \; .
\label{6.4b}
\end{eqnarray}

\noindent Introducing  another  complex  variable $A$   instead  of
 the~parameter $z :\;  z = (\cos A + i \sin A) = e^{iA}$; so that
 the~operator from (6.4b) is  rewritten in the~form
\begin{eqnarray}
( \cos  A \; + \; i \sin  A \; \gamma ^{5} )  \;
(- \gamma ^{5} \; \gamma ^{1}) \otimes \hat{P} \;
\equiv e^{+iA \gamma ^{5}} \; \hat{P}^{sph.}_{bisp.} \otimes \hat{P}
\label{6.4c}
\end{eqnarray}

\noindent (\ref{6.3b}) may be expressed  as follows
\begin{eqnarray}
\Phi'(x) \; = \; e^{+iA/2} \; \mbox{exp}\; (+i \gamma ^{5} {A\over 2}) \; \Phi(x)\; .
\label{6.4d}
\end{eqnarray}

\noindent Those are Abelian analogues of
\begin{eqnarray}
\hat{N}^{C.}_{A} \; =\; (-i) \; \exp \left [\; - i \; A \; \vec{\sigma }\;
 \vec{n}_{\theta ,\phi } \right ]
\otimes \hat{P}_{bisp.} \otimes \hat{P}  \; ,
\label{6.5b}
\\
U^{C.} \; = \; e^{iA/2} \; \exp \left [ -\; i \; {A\over 2}\; \vec{\sigma }\;
 \vec{n}_{\theta ,\phi }\right ] .
\label{6.6}
\end{eqnarray}

\noindent
This symmetry leads  to the $A$-ambiguity (6.5) and
permits to choose an~arbitrary reflection operator from the~totality
$\{ \hat{N}_{A} \}$.

\subsection*{Acknowledgements}

The work was supported by the grand  of BRFFI (Belarusian Republican Foundation for
 Fundamental Research), No F09K -- 123.

The author wishes to thank the Organizers of the International Conference "Non-Euclidean Geometry and its  applications."
 (5 -- 9 July 2010, Kluj-Napoca (Kolozv\'{a}r), ROMANIA)   for
having given us the opportunity to talk on this subject, as well as HCAA-ESF
 (Harmonic and Complex Analysis and its Applications -- European Science Foundation Research Networking Programme)
 for partial support.

\end{document}